\documentclass[a4paper,11pt,twoside]{report}
 \author{Robbert-Jan Dikken}
 \title{AC quantum transport:\\Non-equilibrium in mesoscopic wires due to time-dependent fields}
 \usepackage{graphicx}
 \usepackage{a4wide}
 \usepackage{times}
 \usepackage{subfigure}
 \usepackage{amsfonts}
 \graphicspath{{picture/}}
 \newcommand{\dIdV}{\frac{\partial I}{\partial V}}

\begin{document}

  \pagestyle{empty}
  \begin{titlepage}
    \begin{center}

    \begin{minipage}{.75\textwidth}
	\begin{center}
    \LARGE \textbf{AC quantum transport:\\Non-equilibrium in mesoscopic wires due to time-dependent fields} \\
	\end{center}
	\end{minipage}

    \vspace*{2.5cm}
\normalsize
	 Robbert-Jan Dikken \\
	Delft University of Technology

    \vspace*{3.5cm}

    \begin{minipage}{1\textwidth}
\normalsize
	\begin{abstract}
A model is developed describing the energy distribution of quasi-particles in a quasi-one dimensional, normal metal wire, where the transport is diffusive, connected between equilibrium reservoirs. When an ac bias is applied to the wire by means of the reservoirs, the statistics of the charge carriers is influence by the formed non-equilibrium.

The proposed model is derived from Green function formalism. The quasi-particle energy distribution is calculated with a quantum diffusion equation including a collision term accounting for inelastic scattering. The ac bias, due to high frequency irradiation, drives the wire out of equilibrium. For coherent transport the photon absorption processes create multiple photon steps in the energy distribution, where the number of steps is dependent on the relation between the amplitude of the field $eV$ and the photon energy $\hbar \omega$. Furthermore we observe that for the slow field regime, $\omega \tau_D<1$, the photon absorption is highly time-dependent. In the fast field regime $\omega \tau_D > 1$ this time-dependency disappears and the photon steps in the distribution have a fixed value.

When the wire is extended, the transport becomes incoherent due to interaction processes, like electron-electron interaction and electron-phonon interaction. These interactions give rise to a redistribution of the quasi-particles with respect to the energy. We focused on the fast field regime and concluded that the strong interaction limit for both mechanisms gives the expected result. Strong electron-phonon interaction forces the distribution function on every position in the wire to become a Fermi function with the bath temperature, while strong electron-electron interaction causes an effective temperature profile across the wire and the distribution function on every position in the wire is a Fermi function with an effective temperature.

So the complicated interplay between the effect of photon absorption, diffusive transport and inelastic scattering on the  quasi-particle energy distribution seems to be accurately described by our model.
\end{abstract}
    \end{minipage}

   \end{center}
    \end{titlepage}

  \pagestyle{empty}
  \begin{titlepage}
    \sffamily
    \begin{center}

    \begin{tabular*}{1\textwidth}
    {@{\extracolsep{\fill}}lll}

    Delft University of Technology & Faculty of Applied Sciences & Physics of NanoElectronics\\
    & Kavli Institute of NanoScience & \\
    \end{tabular*}

    \vspace*{3.5cm}
    \LARGE \textbf{AC quantum transport:\\Non-equilibrium in mesoscopic wires due to time-dependent fields}

    \end{center}
    \normalsize
    \vfill

    \begin{figure}[h]
    \begin{minipage}{0.3\textwidth}
    \flushleft
    \includegraphics[width=1\textwidth]{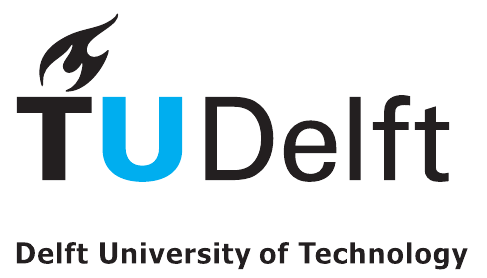}
    \end{minipage}
    \end{figure}

    \vspace{1.5cm}
    \flushleft
    \begin{tabular}{ll}
    Master's thesis by&: R.J. Dikken \\
    \\
    Research group &: Physics of NanoElectronics\\
    Groupleader and 1st Reviewer &: Prof. dr. ir. T.M. Klapwijk \\
    2nd Reviewer&: Dr. K.K. Berggren \\
    3rd Reviewer&: Dr. Y.M. Blanter\\
    Supervisors &: N. Vercruyssen,  MSc., H.L. Hortensius, MSc. \\
    \end{tabular}

    \normalfont
    \end{titlepage}
  \pagestyle{plain}

  \pagenumbering{Roman}

  
  \tableofcontents

	\begin{abstract}
A model is developed describing the energy distribution of quasi-particles in a quasi-one dimensional, normal metal wire, where the transport is diffusive, connected between equilibrium reservoirs. When an ac bias is applied to the wire by means of the reservoirs, the statistics of the charge carriers is influence by the formed non-equilibrium.

The proposed model is derived from Green function formalism. The quasi-particle energy distribution is calculated with a quantum diffusion equation including a collision term accounting for inelastic scattering. The ac bias, due to high frequency irradiation, drives the wire out of equilibrium. For coherent transport the photon absorption processes create multiple photon steps in the energy distribution, where the number of steps is dependent on the relation between the amplitude of the field $eV$ and the photon energy $\hbar \omega$. Furthermore we observe that for the slow field regime, $\omega \tau_D<1$, the photon absorption is highly time-dependent. In the fast field regime $\omega \tau_D > 1$ this time-dependency disappears and the photon steps in the distribution have a fixed value.

When the wire is extended, the transport becomes incoherent due to interaction processes, like electron-electron interaction and electron-phonon interaction. These interactions give rise to a redistribution of the quasi-particles with respect to the energy. We focused on the fast field regime and concluded that the strong interaction limit for both mechanisms gives the expected result. Strong electron-phonon interaction forces the distribution function on every position in the wire to become a Fermi function with the bath temperature, while strong electron-electron interaction causes an effective temperature profile across the wire and the distribution function on every position in the wire is a Fermi function with an effective temperature.

So the complicated interplay between the effect of photon absorption, diffusive transport and inelastic scattering on the  quasi-particle energy distribution seems to be accurately described by our model.
\end{abstract}  
  \pagestyle{plain}
  \include{exp}

  \pagenumbering{arabic}
  
  \chapter{\label{introduction}Introduction}

\section{\label{section}Non-equilibrium and mesoscopic systems}

The last decades the non-equilibrium in mesoscopic systems is intensively studied by a part of the nano-scientific community. 
Despite all the efforts the physics of this is still not fully understood due to the complexity of these systems. The systems have length scales between microscopic and macroscopic.  On one hand the system contains many particles, but on the other hand it can still exhibit quantum features. Because of the intermediate dimensions a specific approach is needed for calculating the physical properties. Pure quantum mechanics can not be used because the many particles complicate the quantum mechanical description in a horrible way and thermodynamics can not be used because of the significance of the quantum features in the system. Therefore often a quantum statistical approach is used which reveals the intriguing world of mesoscopic physics.

\begin{figure}[h!]
    \centering
    \includegraphics[width=0.85\textwidth]{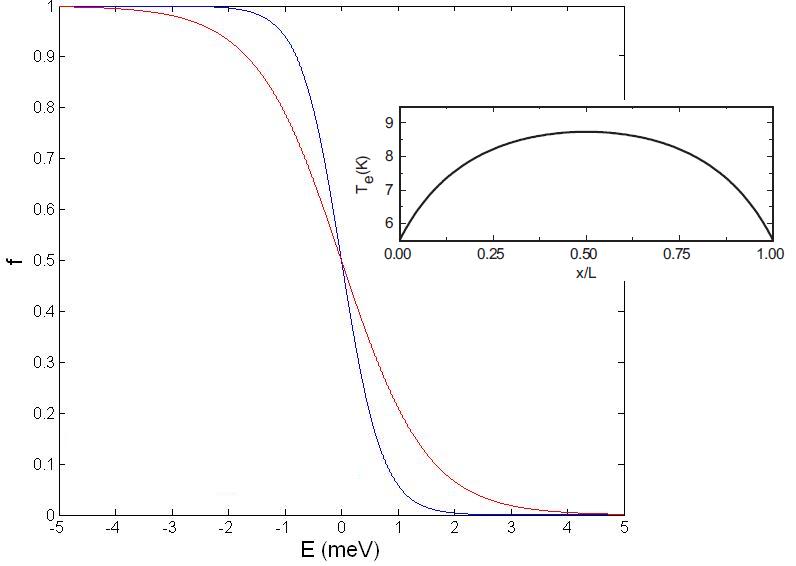}
    \caption{A bias voltage applied to a macroscopic wire causes local equilibrium in the wire and the electron statistics are described by Fermi functions with an effective temperature. The energy distribution in the reservoir and in the middle of the wire is given by the blue and red line, respectively. The effective temperature profile shows the effect of the potential difference across the wire \cite{hajenius}.}
    \label{figure:eff_temp}
\end{figure}

Before looking at mesoscopic systems, let's look at macroscopic and microscopic systems and the meaning of equilibrium and non-equilibrium in this context. Consider a macroscopic resistor $R$ placed between electron reservoirs at equilibrium, which means that the electrons in the reservoirs obey Fermi statistics and the electrons with energy $E$ are distributed according to a Fermi function, $f(E)=(e^{(E-\mu)/(k_bT)}+1)^{-1}$, where $k_b$ is the Boltzmann constant and $T$ the temperature. At zero temperature this Fermi function is just a step function at the chemical potential $\mu$ of the material. For energies lower than the chemical potential all energy levels are occupied and for higher energies all levels are empty. When the temperature is increased the electrons become thermally excited, creating holes for energies below chemical potential and electrons for higher energies. This can be seen as a quasi-equilibrium situation. When we look at the unexcited resistor between the reservoirs, we see that the electrons are at the same equilibrium, or quasi-equilibrium, as the reservoirs. Now when a dc voltage $V$ is applied on the reservoirs a current will flow from one reservoir through the resistor to the other reservoir by the relation $I=V/R$. The resistance on the flowing electrons due to impurities causes dissipation, heating the resistor. The statistics of the electrons in the resistor are no longer the same statistics as that of the reservoirs and becomes spatial dependent. The heating of the resistor causes a local equilibrium in the resistor and the electron energy distribution is described by an effective electron temperature \cite{effectivetemperature}. The applied power $P=V^2/R$ causes a temperature profile along the wire which is bounded by the temperature of the reservoirs. Such an effective temperature profile is shown in figure \ref{figure:eff_temp}. The temperature of the reservoirs is held at 4.2 K and the effective temperature is at maximum in the middle of the wire. Figure \ref{figure:eff_temp} also shows the local equilibrium distribution function at the boundary of the wire and in the middle of the wire. The effect of the dissipated energy is a thermal smearing around the Fermi energy.

For the opposite case, a microscopic system, the situation is completely different. A scatterer is placed between two reservoirs and a voltage is applied. In this microscopic situation it becomes more convenient to evaluate the transport using scattering theory \cite{nazarovblanter}, so we do not speak anymore of distribution functions inside the transport region. The electron approaches the scatterer as a Fermi particle with a wave function. Because of the wave-particle duality the electron can be transmitted or reflected with a certain probability by the scatterer, whereafter the electron leaves the scatterer as a Fermi particle with a certain wave function. The transport of the electrons through the scatterer depends on the properties of the scatterer. These properties are described by the transmission distribution, which gives the probability of finding a transport channel in the scatterer with a certain transmission probability. A bias voltage applied to the reservoirs will only create a potential difference across the structure and the transport depends on this potential difference and the transmission distribution of the scatterer. The number of electrons involved in the transport is a measure of the non-equilibrium.

So the non-equilibrium of macroscopic systems is described by the temperature and resistance of the object and the non-equilibrium of microscopic systems is revealed by scattering theory. Now the intermediate regime between macroscopic and microscopic: mesoscopic. In this research we will focus on a diffusive wire, which shows the most resemblance with the macroscopic situation where a resistor was evaluated. However, the general idea that the electron energy distribution inside the wire can be described by an effective temperature appears to breaks down. Pothier et al. studied the effect of a dc voltage on a diffusive wire between electron reservoirs \cite{pothier}. From this research it was concluded that the electron energy distribution obeys the time-independent Boltzmann equation when the driving term, i.e. the potential difference across the wire, is absorbed in the boundary conditions. 

\begin{equation}\label{collisionless boltzmann eq}
\frac{1}{\tau_D}\frac{d^2f(x,E)}{dx^2}+I_{coll}(x,E,{f})=0
\end{equation}

\begin{figure}[h!]
    \centering
    \includegraphics[width=0.85\textwidth]{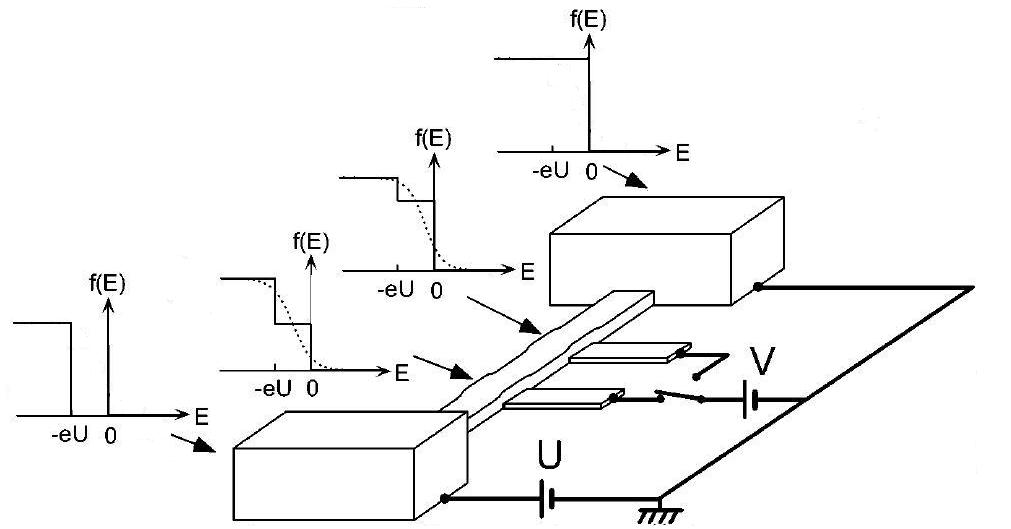}
    \caption{DC biased wire showing the spatial dependent superpositions of the equilibrium distribution functions of the reservoirs \cite{pothier}.}
    \label{figure:Pothier}
\end{figure}

In absence of inelastic interactions the collision integral vanishes and the solution is on every position in the wire a superposition of the boundary conditions which are the Fermi function of right reservoir and that of the left reservoir. When one reservoir is held at zero potential, the other reservoir is at maximum potential which shifts this Fermi function with $eU$. The superposition of these two distribution functions creates a two step function dependent on the position on the wire as shown in figure \ref{figure:Pothier}.

When inelastic interactions are involved the situation becomes a bit more complicated. The collision integral in the Boltzmann equation has to be evaluated. The energy that electrons gained from the electric field is redistributed during collisions on inelastic scatterers. These inelastic scattering processes are electron-electron and electron-phonon interactions. Depending on the characteristics of the diffusive wire and the dominant scattering processes, a relation for the energy relaxation time can be obtained which is self-consistently used in calculating the distribution function.

\section{\label{section}AC quantum transport}

So far only non-equilibrium of dc quantum transport is considered. The study of non-equilibrium of ac quantum transport in mesoscopic systems is interesting for better understanding of the physics of many-body systems and how small electronic devices respond to high frequency irradiation. Previous studies on ac quantum transport focused mainly on coherent structures, where the phase of electrons is preserved. Different examples of study objects of ac quantum transport are SIS junctions, quantum point contacts (QPC), quantum dots (QD) and resonant tunneling diodes (RTD). Tien and Gordon successfully constructed a theory describing the tunneling current between two superconducting films separated by an insulating layer biased with an ac voltage \cite{TienGordon}. The electrons involved in the transport can gain energy in discrete values from the ac field creating steps in the $I-V$ characteristics.  The success of their theory reached further than the SIS and was also successfully applied to the QPC, QD and RTD. 

Stimulated by the success of this theory for different structures Remco Schrijvers \cite{R.Schrijvers} tried to apply this theory to the reservoirs and use the Boltzmann equation to calculate the electron energy distribution in a diffusive wire excited by an ac voltage. The validity of this approach was a bit disappointing. The model was only valid for low frequencies in a wire without inelastic scattering. This was caused by the fact that Tien-Gordon theory assumes averaging over time and therefore the collision term of the Boltzmann equation can not be evaluated in a correct manner.

\begin{figure}[h!]
    \centering
    \includegraphics[width=0.8\textwidth]{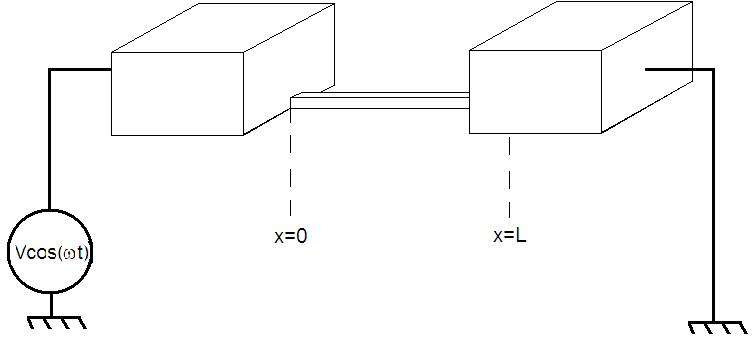}
    \caption{AC biased wire for which the non-equilibrium description is still unknown}
    \label{figure:ac biased wire}
\end{figure}

From the previous research on non-equilibrium due to time-dependent fields in diffusive wires it was concluded that the situation is still not completely understood. To avoid the deducted problems put forward by Remco Schrijvers, we derived from the Green function formalism a quantum diffusion equation for the electron energy distribution in a quasi-one dimensional diffusive wire subject to an oscillating electric field. The model is first derived for a coherent structure with elastic impurity scattering, whereafter this model is extended to account for inelastic scattering processes such as electron-electron and electron-phonon interactions.


\chapter{\label{chapter2}Phase coherent quantum transport}

\section{\label{scattering theory}Scattering theory}

\subsection{\label{transport}Transport}

Phase coherent quantum transport involves the transport of charge carriers where the phase of these charge carriers is preserved. Generally this means that scattering inside the structure is elastic, so that the energy of the charge carriers is not redistributed. Phase coherent transport of electrons in nanostructures is usually described with scattering theory. The nanostructure is defined as a scattering region between reservoirs and the wave function of the electrons subject to Hamiltonian $\hat{H}$ with potential $U(\textbf{r},t)$ obeys the Schrodinger equation

\begin{eqnarray}\label{Schrodinger equation}
i\hbar \frac{\partial \psi(\textbf{r},t)}{\partial t}=\hat{H}\psi(\textbf{r},t);    &      &   \hat{H} \equiv -\frac{\hbar^2}{2m}\nabla^2+U(\textbf{r},t).
\end{eqnarray}

The solution of the Schrodinger equation is a stationary space-dependent function multiplied by a time-dependent function dependent on the eigen energy $E$ of the Hamiltonian:

\begin{equation}\label{wave function}
\Psi(\textbf{r},t)=e^{-iEt/\hbar}\psi(\textbf{r}).
\end{equation}

The wave function $\psi(\textbf{r})$ obeys the time-independent Schrodinger equation $\hat{H}\psi(\textbf{r})=E\psi(\textbf{r})$. Due to the wave character of a charge carrier, an electron can contribute to the current through the scatterer between the reservoirs by either being reflected or being transmitted. The probability of being reflected or transmitted is dependent on the thickness and height of the barrier whereon the electron scatters. The potential difference across the structure is determined by the difference of the energy distribution of the two electron reservoirs. Landauer's result for the current through the scatterer between reservoirs is proportional to the integral over energy of the trace of the product of the transmission matrix $\hat{t}$ and its conjugate transpose $\hat{t}^+$ and the difference between the energy distribution of the left and right reservoir \cite{landauer}. An insightful derivation of the Landauer formula can be found in Ref \cite{nazarovblanter}.

\begin{equation}\label{Landauer eq}
I=\frac{2_se}{2\pi \hbar}\int^{\infty}_{0}Tr[\hat{t}^+\hat{t}][f_L(E)-f_R(E)]
\end{equation}

The factor $2_s$ accounts for the degeneracy of electrons with charge $e$. When a bias is applied to the reservoirs, creating across the structure a potential difference $V$, much smaller than the scale of energy dependence in the transmission eigenvalues $T_n$, equation \ref{Landauer eq} can be evaluated at the Fermi energy $\mu$. Introducing the conductance quantum $G_Q=2e^2/h$ gives for the current

\begin{equation}\label{Landauer at Ef}
I=G_QV\sum_{n}T_n(\mu).
\end{equation}

This expression for the current through a scattering structure clearly shows that the structure exists of different channels in which the electrons are transported with a certain probability from one reservoir to the other. The type of transport structure is characterized by the distribution of the transmission probabilities. This distribution is constructed by taking one specific nanostructure from an ensemble of identical design and counting the number of transmission eigenvalues of the transmission matrix in the interval of $T$ to $dT$. This is divided by the total number of nanostructures in the ensemble. For large enough ensembles, the result converges to $P(t)dT$, so that the transmission distribution is defined as $P(t)=\left\langle \sum_p \delta(T-T_p(E)) \right\rangle$. For very short structures, where the wavelength of the electron exceeds the length of the structure, the conductance quantization is prominent present and the distribution of the transmission probabilities is sharply peaked on certain values. When the length of the structure increases, the resistance due to defects in the system becomes dominant. The diffusive behavior of the electrons in the scatterer is random and for a diffusive scatterer the distribution of transmission probabilities is universal, i.e. independent on the details of the scatterer \cite{nazarovblanter}.

\begin{equation}\label{transmission distribution}
\rho_D(T)=\frac{\left\langle G\right\rangle}{2G_Q}\frac{1}{T\sqrt{1-T}}
\end{equation}

Here $\left\langle G\right\rangle$ is the average conductance due to many scattering events. Now with the increasing dimensions of the structure the describing picture becomes more and more complicated due to the fact that more charge carriers are involved and inelastic scattering processes affect the energy of the charge carriers. Therefore one has to let go the idea that the energy of electrons is unchanged by the scattering events. Pure scattering theory can no longer describe in an effective way the transport. Quantum statistical mechanics provides a way out as we will see later on.	First we look at the statistical information of charge carriers that the noise due to finite transmission probabilities in scattering processes provides.

\begin{figure}[h!]
    \centering
    \includegraphics[width=0.60\textwidth]{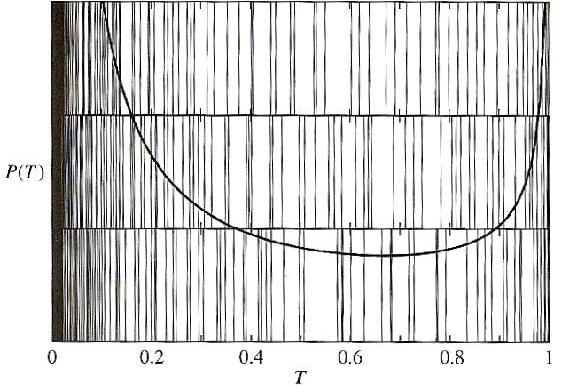}
    \caption{The transmission distribution of a diffusive wire for three different kind of disorder configurations \cite{nazarovblanter}.}
    \label{figure:Tdistribution}
\end{figure}

\subsection{\label{shot noise}Shot noise}

\begin{figure}[h!]
    \centering
    \includegraphics[width=0.8\textwidth]{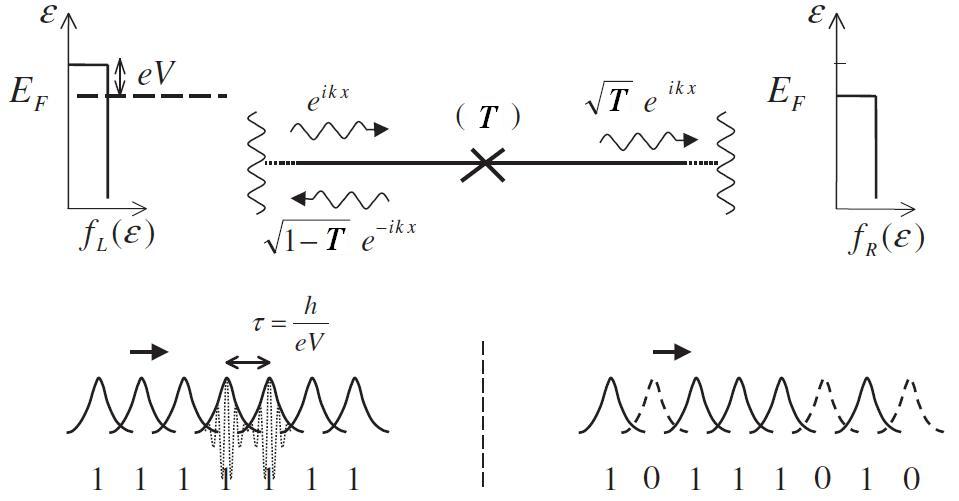}
    \caption{Shot noise arises when the wave packet of an electron is splitted due to a scattering event and the finite transmission probability $T$ causes fluctuations in the current \cite{glattli}.}
    \label{figure:shot_noise}
\end{figure}

A physical phenomenon that contains information about statistics of charge carriers in a mesoscopic conductor is shot noise. Shot noise is caused by the quantization of charge \cite{blanterbuttiker}. When a single incident charge in a state with occupation 1 scatters on some potential barrier it has a probability $R$ of being reflected and a probability $T=1-R$ of being transmitted. Figure \ref{figure:shot_noise} shows how the incoming wave packet of an electron scattering on a barrier with transmission probability $T$ is splitted and only a part of the initial wave packet is transmitted with a probability $T$, causing fluctuations in the current.

When the initial state is occupied by the distribution function $f$, an incident particle is reflected with probability $fR$ and transmitted with probability $fT$, so the averaged occupation of the reflected state is $\left\langle n_R\right\rangle=fR$ and the averaged occupation of the transmitted state is $\left\langle n_T\right\rangle=fT$. By looking at many scattering processes the fluctuations from the average occupation can be determined. For the incident state the average occupation is just the Fermi distribution $\left\langle n_{in}\right\rangle=f$, so that the mean squared fluctuations in the incident state vanishes: $\left\langle (f-\left\langle n_{in}\right\rangle)^2\right\rangle=0$. The fluctuations in the reflected and transmitted state have a finite value. The fluctuations are expressed as a deviation from the average so $\delta n_T=n_T-\left\langle n_T\right\rangle$ and $\delta n_R=n_R-\left\langle n_R\right\rangle$. When we use these identities to calculate the mean squares of the correlations between reflected and transmitted state and of the reflected and transmitted state itself we find:

\begin{eqnarray}\label{mean sq fluctuations}
\left\langle \delta n_T \delta n_T\right\rangle&=&-TRf^2\\
\left\langle (\delta n_T)^2\right\rangle&=&Tf(1-Tf)\\
\left\langle (\delta n_R)^2\right\rangle&=&Rf(1-Rf).
\end{eqnarray}

From these expressions we can distinguish two limits. One limit is given by full transparency and the other limit is given by full reflectance. Both limits have the same outcome in the fluctuations. In a situation where the occupation of the initial state is given by a Fermi distribution at zero temperature, the mean square fluctuations vanish. However, for finite temperature this is not the case. The mean square fluctuations does not vanish, but fluctuates like the incident state with occupation $f$.

These mean square fluctuations contribute in the current and from the current expressions derived in appendix \ref{appendix_shotnoise} the noise power can be obtained. When a multi-channel scatterer between two reservoirs is considered, the noise power can be evaluated at Fermi energy when the scale of energy dependence of the transmission coefficients is much larger than the thermal energy and the energy associated with the applied bias voltage on the reservoirs. The shot noise power is then \cite{blanterbuttiker}:

\begin{equation}\label{noise power3}
S=\frac{e^2}{\pi \hbar}[2k_bT\sum_{n} T_n^2+eVcoth\left(\frac{eV}{2k_bT}\right)\sum_n T_n(1-T_n)].
\end{equation}

As we will see later on, the shot noise for an ac bias has a bit different form than equation \ref{noise power3} and therefore the non-equilibrium due to the ac transport can be seen in the shot noise.

\section{\label{theory1_section1}Tien-Gordon theory}

\begin{figure}[h!]
    \centering
    \includegraphics[width=0.65\textwidth]{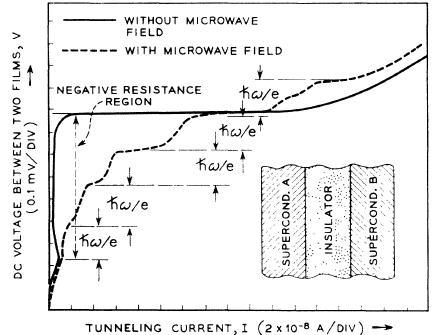}
    \caption{The by Dayem and Martin measured $I-V$ characteristic of an SIS junction biased with and without oscillating field \cite{TienGordon}.}
    \label{figure:Dayem_Martin}
\end{figure}

When we make the switch from dc quantum transport to ac quantum transport, the needed describing theoretical frameworks become a bit more sophisticated. Approximately five decades ago Dayem and Martin observed interactions of electrons with photons in the tunneling current between the superconducting films A and B separated by an insulating layer, when the structure was illuminated with microwave radiation, causing an ac bias across the junction \cite{dayem_martin}. Figure \ref{figure:Dayem_Martin} shows the clear difference between the $I-V$ characteristic with and without this oscillating electric field.

In order to explain these quantum interactions Tien and Gordon developed a describing theory for electric fields normal and parallel to the surface of the superconductor \cite{TienGordon}. Here we will only consider the case where the field is normal to the surface of the superconductor.

The potential difference between the superconductors A and B due to the electric field is given by $Vcos(\omega t)$, where the bias is applied to one reservoir and the other reservoir is held at zero potential. When no field is present the wave functions of the charge carriers of energy $E$ satisfy the unperturbed Hamiltonian $H_0$.

\begin{equation}\label{wavefunction}
\psi(x,y,z,t)=\psi_0(x,y,z)e^{-iEt/\hbar}
\end{equation}

The perturbed Hamiltonian due to the oscillating electric field is given by

\begin{equation}\label{hamiltonian}
H=H_0+eVcos(\omega t).
\end{equation}

This interaction Hamiltonian only effects the time-dependent part of the wave function given by equation \ref{wavefunction}. The new wave function under influence of the oscillating electric field becomes

\begin{eqnarray}\label{new wavefunction}
\nonumber
\psi(x,y,z,t)&=&\psi_0(x,y,z)e^{-\frac{i}{\hbar}\left[Et+\int^{t}_{0}eVcos(\omega t')dt'\right]}
\\
\nonumber
&=&\psi_0(x,y,z)e^{-iEt/\hbar}e^{\frac{eV}{\hbar \omega}sin{\omega t}}
\\
&=&\psi_0(x,y,z)e^{-iEt/\hbar}\sum^{\infty}_{n=-\infty}J_n\left(\frac{eV}{\hbar \omega}\right)e^{in\omega t}.
\end{eqnarray}

To come to the last line the identity $e^{zsin(\theta)}=\sum^{\infty}_{n=-\infty}J_n(z)e^{in\theta}$ is used, where $J_n(z)$ is the Bessel function giving the probability of the absorption of $n$ field quanta. The wave function in equation \ref{new wavefunction} is normalized, since $\left[\sum^{\infty}_{n=-\infty} J_n(z)\right]^2=1$. It appears that the wave function no longer has one energy variable. The energy variable is extended in a sum of multiples of the photon energy. This means that where a charge carrier in the situation without the oscillating field could only tunnel to a state with the same energy, now also could tunnel to states with energy $E\pm n\hbar \omega$. Basically the density of states of the superconductor is modulated by the electric field. The unperturbed density of states of the superconductor is $\rho(E)$. In the presence of the oscillating field the density of states $ \widetilde{\rho}(E)$ becomes

\begin{equation}\label{dos}
\widetilde{\rho}(E)=\sum^{\infty}_{n=-\infty}\rho(E+n\hbar \omega)J^{2}_n\left(\frac{eV}{\hbar \omega}\right).
\end{equation}

The tunnel current is calculated from the density of states. For an SIS junction biased with a dc voltage $V_0$ the tunnel current is

\begin{equation}\label{tunnelcurrent dc}
I_{AB}=C\int^{\infty}_{-\infty}\left[f(E-eV_0)-f(E)\right]\rho_A(E-eV_0)\rho_B(E)dE.
\end{equation}

Here $C$ is a proportionality constant depending on the junction resistance. When an additional ac voltage is applied to the SIS junction the tunnel current shows the multiple photon steps.

\begin{equation}\label{tunnelcurrent ac}
\widetilde{I}_{AB}=c\sum^{\infty}_{n=-\infty}J^{2}_n\left(\frac{eV}{\hbar \omega}\right)\int^{\infty}_{-\infty}\left[f(E-eV_0)-f(E+n\hbar \omega)\right]\rho_A(E-eV_0)\rho_B(E+n\hbar \omega)dE.
\end{equation}

\begin{figure}[h!]
    \centering
    \includegraphics[width=0.95\textwidth]{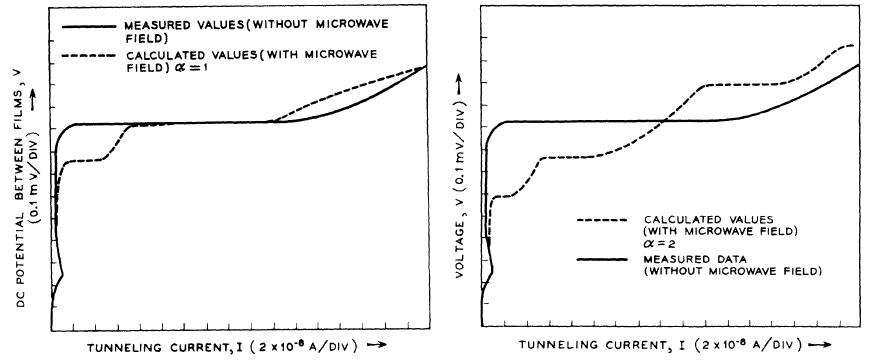}
    \caption{The measured $I-V$ characteristic of an SIS junction without oscillating electric field and the calculated $I-V$ characteristic with oscillating field for different ratios of $\frac{eV}{\hbar \omega}$ \cite{TienGordon}.}
    \label{figure:TienGordon}
\end{figure}

When the tunnel current is explicitly calculated it shows indeed the photon steps as measured by Dayem and Martin. Figure \ref{figure:TienGordon} shows the difference between the measured tunnel current without oscillating electric field given by the solid lines and the calculated tunnel current with oscillating electric field between two superconducting films for two different ratios of $\frac{eV}{\hbar \omega}$ given by the dashed lines.

\begin{figure}[h!]
    \centering
    \includegraphics[width=0.4\textwidth]{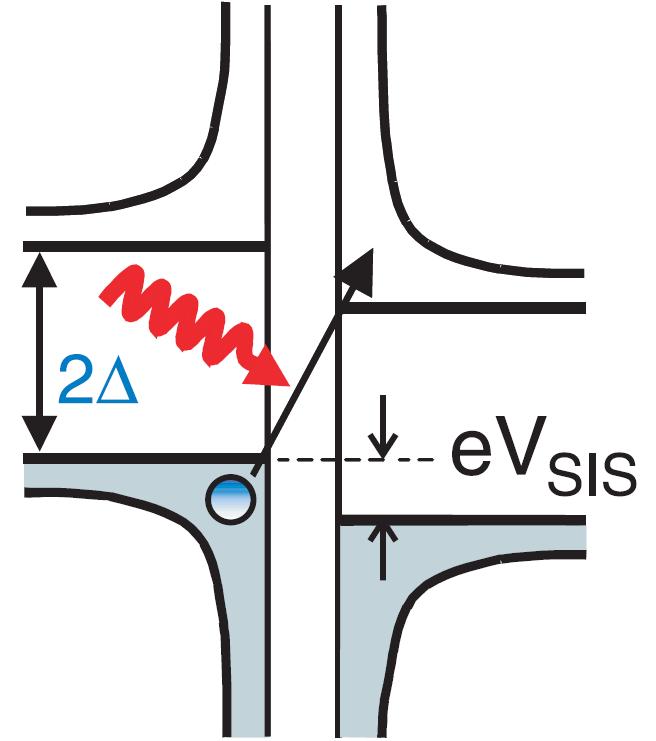}
    \caption{Photon-assisted transport in an SIS junction \cite{deblock}.}
    \label{figure:SISdiagram}
\end{figure}

The energy diagram of an ac biased SIS junction in figure \ref{figure:SISdiagram} shows explicitly how a photon assist the transport of an electron from the first superconductor through the insulating layer to the second superconductor. The gap in the density of states of the superconductor makes it impossible for an electron unaffected by the electric field to tunnel through the barrier to an unoccupied level in the second superconductor. The absorption of a photon can provide the required energy to make this possible.

As said in chapter \ref{introduction} photon-assisted transport is, besides in SIS junctions, also observed in other nano-electronic systems. We won't discuss all examples. Here we only have an additional look at the transport in a quantum dot illuminated with radiation, where the driving frequency exceeds the normal tunneling rate of electrons through the dot, since it provides great insight in the mechanism of photon-assisted transport. 

\begin{figure}[h!]
    \centering
    \includegraphics[width=0.5\textwidth]{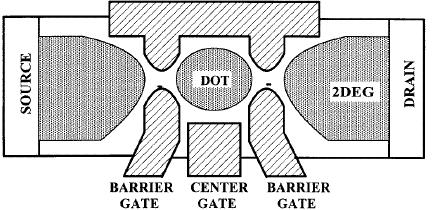}
    \caption{A schematic of a quantum dot \cite{PAT_QD_kouwenhoven}.}
    \label{figure:QD}
\end{figure}

A quantum dot is usually some island coupled by tunnel barriers to leads, the source and drain. The electronic properties of the island and the tunnel barriers can be controlled by gates. Figure \ref{figure:QD} shows this schematically.

The energy levels on the island are assumed to be discrete with a spacing $\Delta E$ while the energy spectrum of the leads is assumed to be a continuum. The radiation is coupled to the island by the gate \cite{ETinQD}. We will not go into detail about this, since we mainly want to focus on the transport from drain to source. The normal tunneling rates are modified by the radiation due to the modification of the wave function of the electrons given by equation \ref{new wavefunction}.

\begin{equation}\label{mod tunnelrate}
\widetilde{\Gamma}=\sum^{\infty}_{n=-\infty}J^2_n(z)\Gamma(E+n\hbar \omega)
\end{equation}

\begin{figure}[h!]
    \centering
    \includegraphics[width=0.5\textwidth]{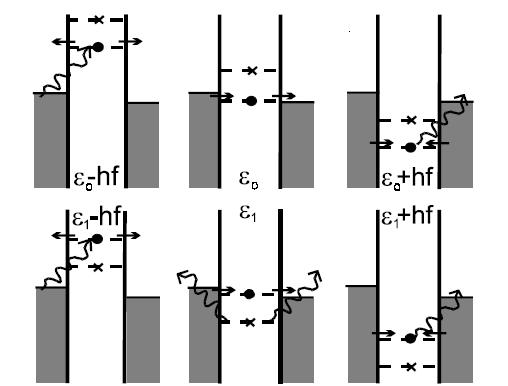}
    \caption{Tunneling processes in a quantum dot \cite{ETinQDkouwenhoven}.}
    \label{figure:QDtunneling}
\end{figure}

Here $z=e\widetilde{V}/\hbar \omega$ and $\widetilde{V}$ is the amplitude of the oscillation. The tunneling is assisted by the absorption of photon with energy $E+n\hbar \omega$ and emission of photons with energy $E-n\hbar \omega$. The possible tunneling processes in the dot with and without radiation are shown in figure \ref{figure:QDtunneling}. Only the upper energy diagram in the middle can contribute to a current through the dot without help of radiation. The remaining diagrams show the photon-assisted tunneling through the ground state $\epsilon_0$ and the first excited state $\epsilon_1$ of the dot. Electrons which normally do not have the right energy to tunnel to an unoccupied state can now absorb or emit a photon. This modifies their energy in such a way that tunneling becomes possible.

For both the SIS junction and the quantum dot the ac bias, due to radiation coupled on the structure, modulates the electronic properties making transport possible to energy states which are not accessible without the energy gain from the field. The photons from the field assist in the transport of charge carriers through the structure.

\section{\label{theory1_section1}Photon-assisted shot noise}

In section \ref{shot noise} the basic idea of shot noise in mesoscopic conductors for dc quantum transport is evaluated and we stated that the expression for the shot noise differs a bit for ac quantum transport. Here we will look how it differs and how this difference arises.

A general scatterer is placed between two reservoirs and an ac voltage is applied to the scatterer by the left reservoirs while the other reservoir is grounded. The transport of electrons can be divided into two regimes: transport of affected and unaffected electrons by the ac bias \cite{glattli}. The unaffected electrons do not contribute to the shot noise, because the number of emitted, unaffected electrons from the right reservoir is the same as that of the left reservoir. Since according to the Pauli exclusion principle both left and right outgoing states can only be occupied by one electron, the current cancels and so does the fluctuation in current. 

The affected electrons from the left reservoir can contribute to the shot noise. An electron with energy $\epsilon \leq \hbar \omega$ below the Fermi energy can get excited to an energy $\hbar \omega - \epsilon$. At energy $-\epsilon$ a hole is created. Since only the left reservoir can excite electrons in this way (the other reservoir is grounded), there is no counter current, so that this becomes the source of the fluctuations in the current. Now when also a dc voltage is applied to the scatterer, the shot noise expression becomes an extended version of equation \ref{noise power3} \cite{lesoviklevitov}, where the photon-assisted features are presented by the Bessel functions like in the tunnel current calculated by Tien and Gordon.

\begin{equation}\label{photassshotnoise}
S_I=4G_Qk_b T \sum_n T_n^2+2G\sum_n T_n(1-T_n) \sum_\pm \sum ^{\infty}_{l=0} J^2_l(\alpha)(eV \pm l \hbar \omega)coth\left(\frac{eV \pm l\hbar \omega}{2 k_b T}\right)
\end{equation}

\begin{figure}[h!]
    \centering
    \includegraphics[width=0.3\textwidth]{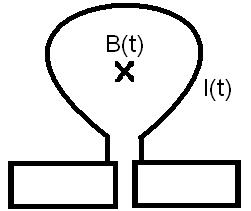}
    \caption{The schematic layout of the photon-assisted shot noise measurements.}
    \label{figure:shotnoisesetup}
\end{figure}

Here $\alpha=eV_{ac}/\hbar \omega$. For $V_{ac}=0$ the normal expression for shot noise is obtained. Now when we make the transition to a diffusive wire it appears that this description still holds. Schoelkopf et al. \cite{schoelkopf} investigated the photon-assisted shot noise experimentally for phase-coherent diffusive conductors and compared their results to the theoretical predictions for photon-assisted shot noise stated by Lesovik and Levitov \cite{lesoviklevitov}. The ac bias is applied on the conductor by bending the conductor between the reservoirs in a loop. A time-dependent magnetic field enters the loop, which induces a time-dependent electric field in the conductor. The situation is depicted in figure \ref{figure:shotnoisesetup}.

\begin{figure}[h!]
    \centering
    \includegraphics[width=0.5\textwidth]{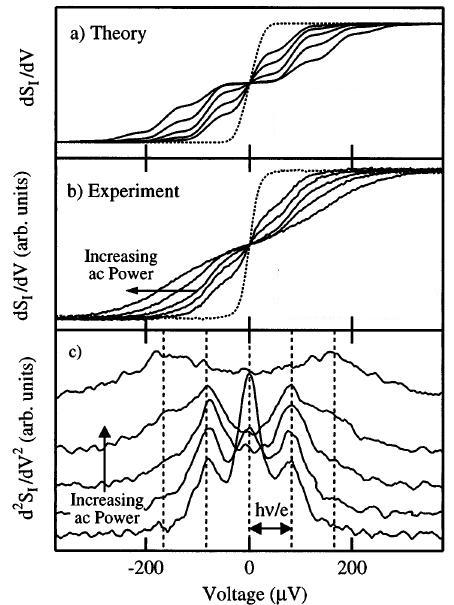}
    \caption{Photon steps in the shot noise both calculated and measured \cite{schoelkopf}.}
    \label{figure:photassshotnoise}
\end{figure}

Lesovik and Levitov predicted theoretically the photon steps in the noise power for an ac biased diffusive conductor where the phase of electrons is preserved. The experiment of Schoelkopf verifies this model. Figure \ref{figure:photassshotnoise} shows the experimental results and the expected results from equation \ref{photassshotnoise} of the differential noise power. The photon steps are not that clear in the first derivative of the noise power. The second derivative of the noise power however clearly shows at the expected energies the steps, indicating the photon-assisted mechanism in the shot noise.

The discrete steps in the shot noise shows the absorption of field quanta and give information about the statistics of the charge carriers in the diffusive wire. It reveals that the energy distribution of the charge carriers inside the wire is affected by the ac bias. This is a completely different point of view in comparison to the transport in the SIS junction and the quantum dot where the electronic properties of the reservoirs are affected by the ac bias. So apparently there arises some interesting physics in the diffusive wire. This is still a relatively simple model, where the electron transport is coherent, so that scattering theory still can be used to describe the transport. However, when the length of the diffusive wire is increased and not only diffusivity and photon absorption causes a change in statistics in the wire, but also inelastic scattering processes induce energy redistribution, scattering theory is no longer the most convenient describing theory. As said in section \ref{transport} we can proceed with quantum statistical theory to determine the statistics of the charge carriers described by the energy distribution function. In the next chapter we will evaluate the conditions for such an approach.
 

\chapter{\label{chapter3}Diffusive transport}

\section{\label{theorymetals_section1}Drude-Sommerfeld model}

The model that was proposed in the 1900s by Drude describes the transport properties of electrons in metals on a microscopic level from a classical point of view. The electronic properties of a metal are then described by a gas of electrons bouncing on heavier positive charged ions. Because of the higher mass of the ions, they are seen as static potentials and the collisions of the electrons on these ions are purely elastic. The electrons involved in the transport are assumed to be free. Between two scattering events no forces act on the electron. In a situation where no electric field is applied on the metal conductor, the average velocity due to different electrons cancels, as the electrons move in a variety of directions. When an electric field is applied the average velocity and thus the net current becomes finite. If $n$ electrons per unit volume with charge $-e$ move with the average velocity $\textbf{v}_{ave}$ and move in a time $dt$ a distance $\textbf{v}dt$, then the net charge passing through a cross-section $A$ is $-ne\textbf{v}_{ave}Adt$ \cite{ashcroftmermin} \cite{datta}. The current density becomes
 
\begin{equation}\label{current density}
 \textbf{j}=\frac{1}{A}\frac{dQ}{dt}=-ne\textbf{v}_{ave}.
\end{equation}
 
Now when an electron is considered at time zero with velocity $\textbf{v}_0$, the velocity that this electron can gain from the electric field in time $t$ is $-e\textbf{E}t/m$ following from Newton's laws of motion. The initial velocity $\textbf{v}_0$ of every electron does not contribute to the average velocity, due to the random collisions from which the electron emerges on time zero. From this it is also directly clear that the average time $t$ is the average time between collision $\tau$, so that the average velocity is $\textbf{v}_{ave}=-e\textbf{E}\tau/m$. Substituting this in the current density gives

\begin{equation}\label{current density2}
\textbf{j}=\frac{ne^2\tau}{m}\textbf{E}.
\end{equation}

Ohm's law is given by $\textbf{j}=\sigma \textbf{E}$, where $\sigma$ is the conductivity. Equating the current density of equation \ref{current density2} and from Ohm's law gives the final expression of the conductivity.

\begin{equation}\label{conductivity}
\sigma=\frac{ne^2\tau}{m}
\end{equation}

Based on the observation that metals conduct heat better than insulators the assumption was made that the electrons involved in the electric conduction also carry the thermal current. The original Drude model used the Maxwell-Boltzmann distribution to account for the probability of finding an electron with a certain energy and thus a certain velocity. However, the ratio between thermal and electric conductivity observed in experiments was not explained in this way. Then the Pauli exclusion principle was put forward, which stated that two fermions can never occupy the same state. From this the conclusion was drawn that the Maxwell-Boltzmann distribution had to be replaced by the Fermi-Dirac distribution. Sommerfeld exchanged the Maxwell-Boltzmann distribution by the Fermi-Dirac distribution in the classical electron gas of Drude. This modified the expression for the electronic velocity and gave the correct expression of the ratio between thermal and electric conductivity, the Wiedemann-Franz law \cite{ashcroftmermin}:

\begin{equation}\label{wiedemannfranz}
\frac{\kappa}{\sigma}=\frac{\pi^2}{3}\left(\frac{k_b}{e}\right)^2 T.
\end{equation}

The idea that electrons form a gas in a metal is sufficient for cases where no energy exchange is present in all processes involving the electrons. However, this is not always the situation. When the collisions of the electrons are no longer purely elastic and they cause energy exchange, the Drude-Sommerfeld model breaks down. Fortunately Landau's theory of Fermi liquids provides a strong replacement.

\section{\label{theorymetals_sections2}Landau theory of Fermi liquids}

As said in the previous section, at a certain stage the transport of electrons can no longer be explained in a electron gas model where the interactions are purely elastic. The effect of inelastic interactions becomes significant and the energy exchange processes initiate the break down of the electron gas concept. Instead one considers the transport of electrons in a  liquid model. This Fermi liquid model is developed by Lev Landau in 1956. The transport of one electron is affected by the surrounding electrons and its wave function is extremely complicated due to screening effects. It behaves however still very like a particle with a charge $e$. The screening can simply be seen as the modification of the relation between energy and wave vector, so $E(\textbf{k})=\hbar^2 k^2/2m^*$, where $m^*$ deviates from the free electron mass $m$. The electrons are defined as quasi-particles which are stable near the Fermi level, but lose their stability far from the Fermi level \cite{ziman}.

The domain of validity for excitations near the Fermi surface in the Landau theory of Fermi liquids has its origin in the assumed one-to-one correspondence between states of a non-interacting system and states of an interacting system when the interaction is adiabatically turned on. Since the lifetime of a quasi-particle is proportional to $(\epsilon-\epsilon_F)^{-2}$, the high energy quasi-particles are decayed before the interaction process is fully complete \cite{negeleorland}. The adiabatic continuation leads to the assumption that the excited states of the interacting system are labeled with the same quantum numbers as the excited states of the non-interacting system. The validity of adiabatic continuation from a non-interacting system to a interacting system can be shown by looking at the wave function. An example is given by a particle trapped in an one-dimensional potential $V(x,t)=V_0(t)h(x)$ \cite{bruusflensberg}. The wave function obeys the Schrodinger equation

\begin{equation}\label{Schrodingereq}
i\hbar \frac{\partial \psi(x,t)}{\partial t}=H(x,t)\psi(x,t)=\left(\frac{p^2}{2m}+V(x,t)\right)\psi(x,t).
\end{equation}

Now the potential changes slowly from initial value $V_{01}$ to a final value $V_{02}$. Because the potential varies slowly, the solution of the Schrodinger equation can be approximated by the solution of the static Schrodinger equation $H(x,t)\psi_{V_0(t)}(x)=E_{V_0(t)}\psi_{V_0(t)}(x)$ \cite{griffiths}. The adiabatic solution becomes

\begin{equation}\label{adiabaticwavefunction}
\psi_{adiabatic}(x,t) \approx \psi_{V_0(t)}(x) e^{-iE_{V_0(t)}t/\hbar}.
\end{equation}

By inserting equation \ref{adiabaticwavefunction} in equation \ref{Schrodingereq} the accuracy of equation \ref{adiabaticwavefunction} is obtained.

\begin{eqnarray}\label{accuracyadiab}
\nonumber
i\hbar \frac{\partial \psi_{adiabatic}(x,t)}{\partial t} & = & E_{V_0(t)}\psi_{adiabatic}(x,t)+i \hbar \left(\frac{\partial \psi_{adiabatic}(x,t)}{\partial V_0(t)}\right)\left(\frac{\partial V_0(t)}{\partial t}\right) \\
& = & H(x,t)\psi_{adiabatic}(x,t)
\end{eqnarray}

The adiabatic solution is a good approximation for the wave function in an one-dimensional potential $V(x,t)$ if the first term of equation \ref{accuracyadiab} dominates the second term, which is true if the rate of change of $V_0(t)$ is small enough. Then the solution for the new potential $V_0=V_{02}$ is found from the old value of the potential $V_0=V_{01}$ from which it adiabatically rises. This implies that when the excited state of the initial potential is a bound state, the excited state of the final potential is also a bound state. A transition from a bound state to an un-bound state will never occur from an adiabatic continuation, no matter how small the rate of change in $V(x,t)$, because one is a decaying function while the other is an oscillatory function.

As said the interactions cause a modification of the relation between energy and momentum of a particle. The total energy of an unperturbed electron system is given by the kinetic energy of the electrons \cite{kardar}.

\begin{equation}\label{tot energy unperturbed}
E=\hbar^2 \sum_{\textbf{k}}\frac{\textbf{k}^2}{2m}n(\textbf{k})
\end{equation}

Here $n(\textbf{k})$ is the occupation number of the state with momentum $\textbf{k}$. When a weak external field is coupled on the system, there will occur a change in occupation number and thus a change in total energy.

\begin{equation}\label{change of energy}
\delta E=\hbar^2 \sum_{\textbf{k}}\frac{\textbf{k}^2}{2m}\delta n(\textbf{k})
\end{equation}

If the system now is perturbed by a adiabatically turned on interaction, with interaction energy $g(\textbf{k},\textbf{k}')$ between states of wave vector $\textbf{k}$ and $\textbf{k}'$, the system is taken away from its ground state energy and a change of occupation numbers is induced. Therefore the change of energy is

\begin{equation}\label{adiab. turned on int.}
\delta E=\sum_{\textbf{k}}\epsilon^0_{\textbf{k}}\delta n(\textbf{k})+\frac{1}{2V}\sum_{\textbf{k},\textbf{k}'}g(\textbf{k},\textbf{k}')\delta n(\textbf{k}) \delta n(\textbf{k}').
\end{equation}

Due to the interaction the electron is no longer a pure particle, but it is a quasi-particle. It behaves still like a particle, but it arises from the interactions with its local environment. A quasi-particle with wave vector $\textbf{k}$ has an energy of

\begin{equation}\label{quasi particle energy}
\epsilon_{\textbf{k}}=\frac{\delta E}{\delta n(\textbf{k})}=\epsilon^0_{\textbf{k}}+\frac{1}{V}\sum_{\textbf{k},\textbf{k}'}g(\textbf{k},\textbf{k}') \delta n(\textbf{k}').
\end{equation}

In the above we have suppressed magnetic fields, so that spin dependency can be neglected, since $\epsilon(\textbf{k},\sigma)=\epsilon(\textbf{k})$ in absence of magnetic fields.

A fundamental parameter in the Landau theory of Fermi liquids is the effective mass. The interaction experienced by a quasi-particle changes its mass with respect to the mass in an environment free of interactions. The velocity and density of states at the Fermi surface can be calculated using this effective mass.

\begin{eqnarray}\label{effective mass intro}
v_F=\frac{p_F}{m^*}, & & N(0)=\frac{3Nm^*}{p^2_F}
\end{eqnarray}

The expressions for these quantities are similar to that of a non-interacting system which confirms the one-to-one correspondence between the states of a non-interacting system and an interacting system. So concluding this section, we can take interactions into account in calculating the electronic properties in quantum transport by considering the charge carriers being quasi-particles for low excited states. Therefore the total energy of the system is not the sum of the energy of the individual particles, but is function of the energy distribution among the quasi-particles. Also due to the one-to-one correspondence between the states of a non-interacting system and an interacting system, the energy distribution of the quasi-particles can be calculated from a diffusion equation, like the semi-classical Boltzmann equation.

\section{\label{theorymetals_section4}Transport in quasi-one dimensional metallic systems}

The Landau theory of Fermi liquids, discussed in the previous section, provides the justification of using a semi-classical Boltzmann equation to calculate the energy distribution of the quasi-particles in a diffusive wire. In this work we focus on a quasi-one dimensional metallic wire of mesoscopic dimensions where the transport of the quasi-particles is diffusive. We will first explain what we exactly understand when we talk about quasi-one dimensional, mesoscopic and diffusive. Then we discuss the non-equilibrium in such a system biased with a dc voltage by looking at the energy distribution of the quasi-particle involved in the transport.

Mesoscopic structures are defined by the relation between length scales defining the geometrics of the structure and defining microscopic processes in the structure.

The length scales defining the microscopic processes involving an quasi-particle are:
\begin{itemize}
	\item The Fermi wavelength $\lambda_F=2 \pi/k_F$, where $k_F$ is the Fermi wave vector,
	\item The elastic mean free path $l_e$, which is the average distance between elastic collisions on impurities for instance,
	\item The phase coherence length $l_{\phi}$, which is the distance that the phase of a quasi-particle is preserved,
	\item The energy relaxation length $l_E$, which is the distance the energy of a quasi-particle is preserved.
\end{itemize}

The length scales defining the geometrics of the structure are given by:
\begin{itemize}
	\item The length of the structure $L$,
	\item The cross-section of the structure $S$.
\end{itemize}

When the length of the structure is significantly larger than the cross-section it is more natural to talk about the structure as being a wire. The wire is said to be diffusive if the length $L$ of the wire is significantly larger than the elastic mean free path $l_e$ of an quasi-particle in the wire. The wire is quasi-one dimensional for $\lambda^2_F<<S<<l^2_e$ provided that the width and the thickness are of the same order of magnitude. When we want to be able to apply the Landau theory of Fermi liquids we are bound to at least quasi-one dimensional systems. For purely one dimensional systems the Landau theory of Fermi liquids is no longer valid. This has its origin in the nesting property of the Fermi surface, which means that a part of the Fermi surface can be matched onto an other part by a translation of $2k_F$. Therefore there arises a divergence in calculating physical properties. A more detailed explanation can be found in Ref. \cite{voit}.

\begin{figure}[h!]
    \centering
    \includegraphics[width=0.6\textwidth]{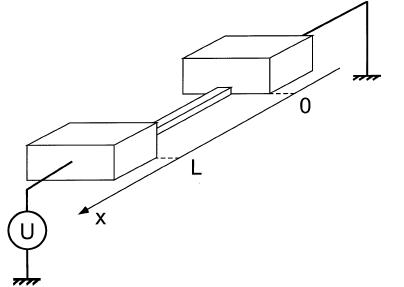}
    \caption{Diffusive wire biased with a potential difference $U$ \cite{pothier2}.}
    \label{figure:dcbiasedwire}
\end{figure}

Pothier et al. studied the quantum transport in dc biased diffusive wires by looking at the effect of the induced non-equilibrium on the quasi-particle energy distribution \cite{pothier}. The diffusive wire is placed between large electron reservoirs where the electron energy distribution is described by an equilibrium Fermi function. A dc voltage $U$ is applied on one reservoir while the other reservoir is held at zero potential, creating a potential difference $U$ over the wire.

The distribution function can be calculated by using semi-classical kinetic theory, which is the Boltzmann equation extended with an interaction term. For wires where the diffusion time $\tau_D$ is shorter than the relaxation time $\tau_E$ the transport is coherent and the distribution is described by a Boltzmann equation without interaction term. When the driving term $eU$ due to the potential difference $U$ across the wire is absorbed in the boundary conditions at the reservoirs, one Fermi function is unchanged, while the other is shifted by $eU$. This leads to the equation:

\begin{equation}\label{boltzmann}
\frac{\partial f(x,E)}{\partial t}+D\frac{\partial^2f(x,E)}{\partial x^2}=0
\end{equation}

The explicit boundary conditions for this equation are given by $f(0,E)=f_F(E)$ and $f(L,E)=f_F(E+eU)$, where $f_F(E)=(1+e^{E/k_bT})^{-1}$ is just the Fermi distribution. The stationary solution on every position in the wire is a superposition of the two boundary conditions.

\begin{equation}\label{f dcbiased wire}
f(x,E)=\left(1-\frac{x}{L}\right)f_F(E)+\frac{x}{L}f_F(E+eU)
\end{equation}

\begin{figure}[h!]
    \centering
    \includegraphics[width=0.6\textwidth]{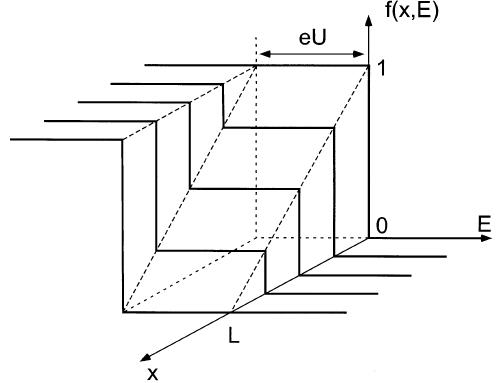}
    \caption{The two step distribution for a dc biased wire without interactions \cite{pothier2}.}
    \label{figure:fdcbias}
\end{figure}

Figure \ref{figure:fdcbias} shows the two step function of the electron energy distribution on every position in a dc biased wire with no interactions present.

If inelastic scattering is introduced the situation becomes a bit more sophisticated. Two main phase breaking mechanisms can be distinguished: electron-electron interactions and electron-phonon interactions, where the electrons are considered to be quasi-particles. We first consider electron-electron interactions and neglect electron-phonon interactions. Strong scattering induces a local equilibrium with temperature $T_e(x)$ and the distribution is described by 

\begin{equation}
f(x,E)=f_F(E-\mu(x),T_e(x))
\end{equation}

where $\mu(x)=-eU\frac{x}{L}$ \cite{pothier2}. The effective temperature $T_e(x)$ in a wire with cross-section $S$ and resistance $R$ is calculated from the heat equation \cite{pothier2}.

\begin{equation}\label{heat eq}
\frac{\partial}{\partial x}\left(\kappa \frac{\partial T_e}{\partial x}\right)+\frac{1}{SL}\frac{U^2}{R}=0
\end{equation}

The boundary conditions of this equation are $T_e(0)=T_e(L)=T$ and using the Wiedemann-Franz law (equation \ref{wiedemannfranz}) for the heat conductivity $\kappa$ the effective temperature is \cite{pothier2}

\begin{equation}\label{effectiveT}
T_e(x)=\sqrt{T^2+\frac{x}{L}\left(1-\frac{x}{L}\right)\frac{3}{\pi^2}\left(\frac{e}{k_b}\right)^2U^2}.
\end{equation}

\begin{figure}[h!]
    \centering
    \includegraphics[width=1\textwidth]{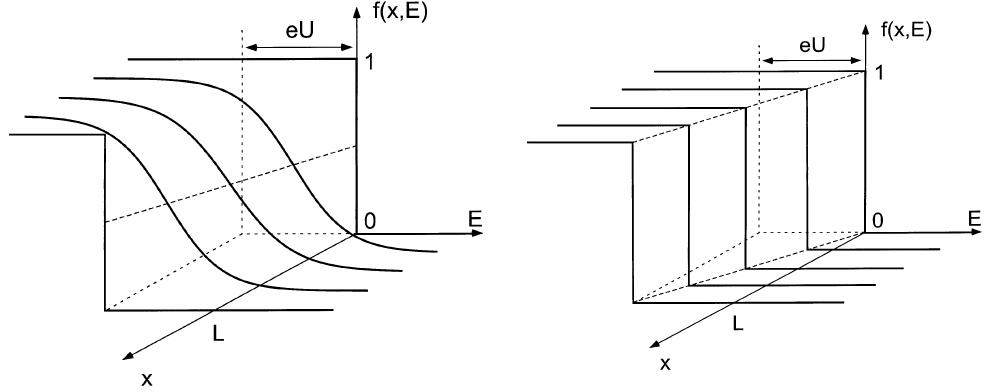}
    \caption{Left: Strong electron-electron scattering, right: strong electron-phonon scattering \cite{pothier2}.}
    \label{figure:scatteringdcbiasedwire}
\end{figure}

Now the electron-electron interactions are negligible and the electron-phonon scattering is the dominant phase breaking mechanism. For strong scattering the electrons thermalize with the temperature of the phonons. The distribution function is given by $f(x,E)=f_F(E-\mu(x),T)$ where $\mu=-eU\frac{x}{L}$ and $T$ is the phonon bath temperature \cite{pothier2}. The space dependence of the distribution functions is shown for both situations in figure \ref{figure:scatteringdcbiasedwire}.

For intermediate regimes where neither electron-electron scattering nor electron-phonon scattering is strong, but still present, the interaction term in the Boltzmann equation has to be evaluated. The interaction term can be calculated from the Fermi golden rule and the belonging kernel follows from a microscopic derivation \cite{huard}. We will come back to this later in chapter \ref{chapter4} where we calculate the interactions in a diffusive wire due to electron-electron scattering and electron-phonon scattering.

\section{\label{theorymetals_section3}Quantum corrections to the conductivity}

\begin{figure}[h!]
    \centering
    \includegraphics[width=0.75\textwidth]{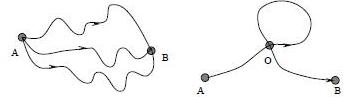}
    \caption{Feynman diagrams showing on the left classical trajectories and on the right trajectories resulting in weak localization \cite{Galperin}}
    \label{figure:weakloc}
\end{figure}

On quantum scale the conductance of a diffusive wire is not simply given by the Drude result of the conductivity in equation \ref{conductivity}. Because an electron has a wave-character, the electron is not localized. Therefore, when no phase-breaking processes are present, an electron can interfere with itself when it returns to a certain initial position after multiple elastic scattering events. This modification of the conductance is called localization and is depicted in figure \ref{figure:weakloc}.

The probability for an electron of passing between A and B is given by a classical probability and additionally an interference term

\begin{equation}\label{prob. selfcrossing}
W=\left|\sum_i{A_i}\right|^2=\sum_i{|A_i|^2}+\sum_{i\neq j}{A_i A^*_j}.
\end{equation}

The phase gained by an electron while traveling through the diffusive medium is $\Delta \phi=\hbar^{-1} \int^B_A \textbf{p} d \textbf{l}$. For most of the trajectories this phase gain will be much larger than one and therefore vanish in the interference term. The self-crossings have the same phase gain when the direction of the traveled trajectory is reversed, i.e. $\textbf{p}\rightarrow -\textbf{p}$ and $d\textbf{l} \rightarrow -d\textbf{l}$. This results in two paths and the probability of self-crossing is

\begin{equation}\label{prob. selfcrossing2}
W=|A_1+A_2|^2=|A_1|^2+|A_2|^2+2A_1 A^*_2=4|A_1|^2.
\end{equation}

The quantum interference doubles the result. So the probability of scattering is increased, which results in a decrease of conductance. To determine qualitatively the effect of weak localization on the  conductance we shall follow a heuristic derivation which can be found in Ref. \cite{Galperin}. The de Broglie wavelength $\lambda_F=2\pi/k_F$ of the electron determines the scattering cross-section on site $O$. In time $t$ it travels diffusively a distance $\sqrt{Dt}$, where $D$ is the diffusion coefficient. The interference volume in $d$ dimensions becomes $(Dt)^{d/2}r^{3-d}$, where $r$ is the thickness of the system. The electron has to enter the interference volume to experience interference, which occurs with a probability of $\frac{v_F \lambda_F^2 dt}{(Dt)^{d/2}r^{3-d}}$. This leads to a relative correction to the conductivity of 

\begin{equation}\label{corr.con.}
\frac{\Delta \sigma}{\sigma} \propto -\int^{\tau_{\phi}}_{\tau_e}\frac{v_F \lambda^2_F dt}{(Dt)^{d/2}r^{3-d}}.
\end{equation}

The phase coherence time in the upper limit of the integral shows the condition for phase preservation. Now when we focus on the one dimensional situation for our quasi-one dimensional wire, the evaluation of the integral gives

\begin{equation}\label{corr.con2}
\frac{\Delta \sigma}{\sigma} \propto -2\frac{v_F \lambda^2_F}{D^{1/2} r^{2}} (\sqrt{ \tau_{\phi}}-\sqrt{ \tau_e})=-2\frac{v_F \lambda^2_F}{D r^{2}} (l_{\phi}-l_e).
\end{equation}

For the last expression we used 

\begin{eqnarray}\label{id}
l_{\phi} \propto \sqrt{D \tau_{\phi}} , & l_e \propto v_F \tau_e , & D \propto v_F l_e.
\end{eqnarray}

If the elastic mean free path is much smaller than the phase coherence length we can neglect this term in the conductivity correction.

\begin{equation}\label{corr.con3}
\frac{\Delta \sigma}{\sigma} \propto -2\frac{v_F \lambda^2_F}{D r^{2}} l_{\phi}
\end{equation}

The Drude conductivity can be expressed in terms of the elastic mean free path and the Fermi momentum.

\begin{equation}\label{Drude}
\sigma \propto \frac{ne^2 \tau_e}{m} \propto \frac{n e^2 l_e}{p_F} \propto \frac{e^2 p^2_F l_e}{\hbar^3}
\end{equation}

Substituting this in the relative correction expression, where we use the identities \ref{id} and $\lambda_F \propto \hbar/p_F$ leads to

\begin{equation}\label{deltasigma}
\Delta \sigma \propto -2 \frac{e^2}{\hbar r^2}l_{\phi}.
\end{equation}

To get the correction to the conductance we introduce $\Delta G=\Delta \sigma r^2/L$ and arrive at

\begin{equation}\label{deltacon}
\Delta G \propto -2 \frac{e^2}{\hbar} \frac{l_{\phi}}{L}.
\end{equation}

This quantum correction to the conductance is known as weak localization and arises due to a self-crossing in the diffusive transport of an electron. However, when $l_{\phi}<<L$ the correction becomes negligible.

At zero temperature the phase of an electron is not broken $(l_{\phi} \rightarrow L)$, so that the correction to the conductance is no longer negligible \cite{Galperin}. This is known as strong, or Anderson, localization. When again the conductance is implemented, the expression for this correction is obtained \cite{R.Schrijvers}.

\begin{equation}\label{strongloc}
\frac{\Delta G}{G} \propto -2 \frac{e^2/\hbar}{\frac{e^2p^2_Fl_er^2}{\hbar^3 L}} = -2 \frac{\hbar^2 L}{p^2_Fl_er^2} \propto -2\frac{L}{l_e(r^2/\lambda^2_F} \propto -2\frac{L}{N_{\bot}l_e}
\end{equation}

The number of transverse channels available for conduction is determined by the ratio of Fermi wavelength and cross-section. Now the correction is negligible if $\Delta G/G<<1$, which is true for a large number of open conduction channels. Since we consider a diffusive wire, we can look at the distribution of transmission probabilities in equation \ref{transmission distribution}, and see that if the average conductance increases the number of open channels increases.

We can conclude that for our quasi-one dimensional diffusive wire, we can neglect the quantum correction to the conductance due to interference effects when we consider wires with length much larger than the phase coherence length and a conductance significantly larger than the conductance quantum.


\chapter{\label{chapter4}Photon absorption and other energy exchange processes in diffusive wires}

\section{\label{intro theory}Introduction}

The model proposed by Remco Schrijvers had the aim to describe the electron energy distribution in a diffusive wire subject to  high frequency irradiation with energy relaxation present inside the wire \cite{R.Schrijvers}. Unfortunately this aim was not fully achieved. The assumption was made that the path traveled by the electron inside the wire is of no influence to the energy distribution, so that Tien-Gordon theory could be applied to the reservoirs and the distribution inside the wire was described by the Boltzmann equation. However, this turned out to be incorrect since Tien-Gordon theory assumes averaging over time and therefore the collision integral can not be evaluated in the correct manner. Therefore a different approach is required.

A.V. Shytov developed a theoretical framework to calculate the electron energy distribution for wires where the phase coherence time and energy relaxation time exceeds the diffusion time, so that the transport is fully coherent. We derive from Green function formalism an equivalent model. The insight we gain from this derivation is helpfull in the extension of the theoretical framework of Shytov with a term accounting for inelastic scattering, breaking the phase of the electrons. Since Green function formalism is not basic knowledge, the most important parts for our derivation are first shortly explained.

\section{\label{theory_section2}Green function formalism}

The Green function formalism provides a strong calculation method which can be used to calculate a variety of properties of many-particle systems. In mathematics Green functions obey a inhomogeneous differential equation, where the inhomogeneity is singular. As we have seen in the previous chapters, the Schrodinger equation is the central equation in quantum mechanics. Since this is a differential equation the Green functions apply in describing many-body physics in both equilibrium and non-equilibrium situations. The basis of the formalism is the definition of the single-particle Green function by the wave function \cite{haug_jauho}.

\begin{equation}\label{single-particle Green function}
	G(x,t;x',t')=\frac{-i}{\hbar}\frac{\left\langle \psi_0|T[\psi_H(x,t)\psi_H^+(x',t')]|\psi_0\right\rangle}{\left\langle \psi_0|\psi_0\right\rangle}
\end{equation}

So the Green function is based on the wave function $\psi_0$ of the ground state of the system with Hamiltonian $H$ and the time-evolving wave function $\psi_H$ of the system which evolves like $e^{iHt/\hbar}\psi(t=0)e^{-iHt/\hbar}$. The time-ordening operator $T$ is defined in such a way that it always moves the operator with the earlier time-argument to the right.

\begin{equation}\label{time-ordening}
 T[A(t)B(t')]=\theta(t-t')A(t)B(t')\mp \theta(t'-t)A(t')B(t)
\end{equation}

The sign in the time-ordening is dependent on the nature of the considered particle. For fermions the sign is negative, so that the Pauli exclusion principle is not violated, and for bosons the sign is positive. In the following we shall only consider fermions. The equation of motion is now derived by differentiating the equation for the single particle Green function with respect to $t$.

\begin{equation}\label{eq. of motion}
 i\hbar \frac{\partial G(x,t;x',t')}{\partial t}=\delta(t-t')\frac{\left\langle \psi_0|[\psi_H(x,t),\psi_H^+(x',t')]_+|\psi_0\right\rangle}{\left\langle \psi_0|\psi_0\right\rangle}-\frac{i}{\hbar}\frac{\left\langle \psi_0|i\hbar\frac{\partial \psi_H(x,t)}{\partial t}\psi_H^+(x',t')|\psi_0\right\rangle}{\left\langle \psi_0|\psi_0\right\rangle}
\end{equation}

From second quantization it is know that the anticommutation of a wave function in the Heisenberg picture with its conjugate gives a delta-function, so that the first term on the right side of the equation of motion is a multiplication of a spatial and a temporal delta-function. For the second term we use the Heisenberg equation of motion $i\hbar \frac{\partial \psi_H}{\partial t}=[\psi_H,H]$. When we consider a particle free of interactions subject to a Hamiltonian $H=-\frac{\hbar^2}{2m}(-i\nabla-\frac{e}{\hbar}A(t))^2$, where the vector potential $A(t)$ representing an electric field is integrated in the momentum operator by principle of minimal substitution, the equation of motion for the Green function $G_0$ of a free particle becomes

\begin{equation}\label{eq. of motion2}
\left\{i\hbar \frac{\partial}{\partial t}-\frac{\hbar^2}{2m}\left(-i\nabla-\frac{e}{\hbar}A(t)\right)^2\right\}G_0(x,t;x',t')=\hbar \delta(t-t') \delta(x-x').
\end{equation}

\begin{figure}[h!]
    \centering
    \includegraphics[width=0.55\textwidth]{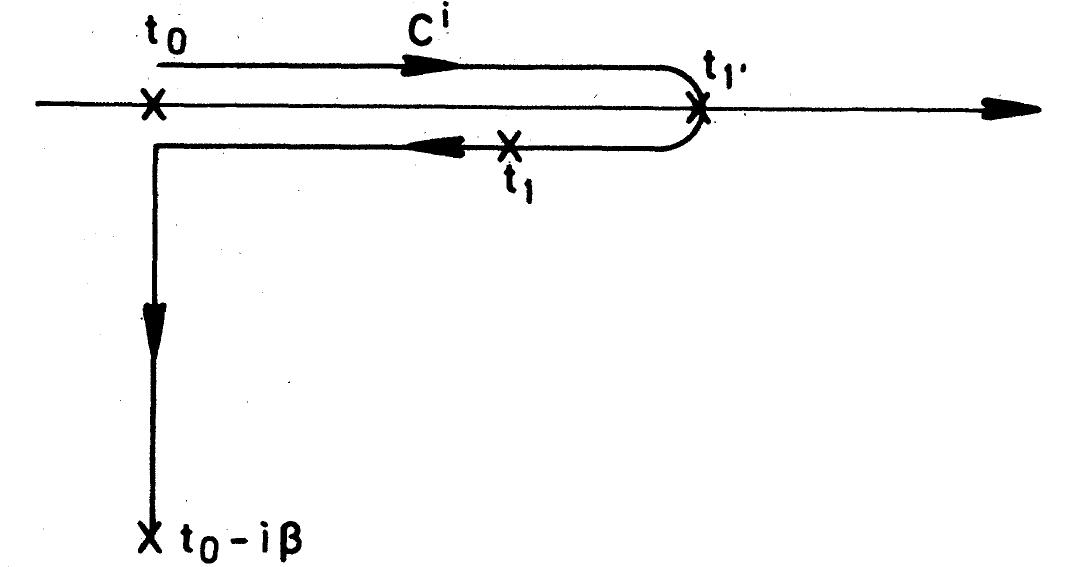}
    \caption{The contour on which the particle is defined in non-equilibrium \cite{rammersmith}.}
    \label{figure:complexcontour}
\end{figure}

Because the Hamiltonian is time-dependent in the vector potential we are already considering non-equilibrium. When now also a many-particle system is considered where the particles interact with eachother, the picture becomes a bit complicated. The wave functions, and thus the Green functions, are subject to both an external potential and an internal potential. To ease the calculations the operations are contour-ordered. This replaces the time-ordening operator $T$ in equation \ref{single-particle Green function} with the contour-ordening operator $T_C$ which has the same properties, only not in time, but on the defined contour. Because in non-equilibrium the final state does not have to return to the initial state the contour, on which the particle is defined, lies in the complex plane depicted in figure \ref{figure:complexcontour}. We won't go into detail on this, but a insightful derivation can be found in Ref.\cite{haug_jauho} and Ref.\cite{rammersmith}. 

The derivation in Ref.\cite{haug_jauho} and Ref.\cite{rammersmith} is an approach from non-equilibrium statistical mechanics and leads to the Dyson equation for the Green function which consists of the free particle Green function $G_0$ and a self energy term responsible for the interactions.

\begin{equation}\label{eq:Dysonwithinteractions}
		G(1,1')=G_{0}(1,1')+\frac{1}{\hbar}\int dx_2 \int dx_3 \int_C d\tau_2 \int_C d\tau_3 G_0(1,2)\Sigma(2,3)G(3,1')
\end{equation}

The complex contour integral in equation \ref{eq:Dysonwithinteractions} is rather impractical in calculations. Fortunately analytic continuation provides a method to replace the contour integrals by real time integrals. The Green function is defined by different Green functions on the contour, the lesser and greater Green function, the time-ordered and anti-time-ordered Green function and the advanced and retarded Green function, dependent on the position of the time coordinates of the Green function on the contour. When the initial time $t_0$ is set to infinity and the interactions are coupled adiabatically, the complex part of the contour depicted in figure \ref{figure:complexcontour} vanishes. By doing this one neglect initial correlations, but in many situations the interactions in the process of reaching a steady state will wash out these initial correlations. In highly transient situations it can however cause problems.

\begin{figure}[h!]
    \centering
    \includegraphics[width=0.5\textwidth]{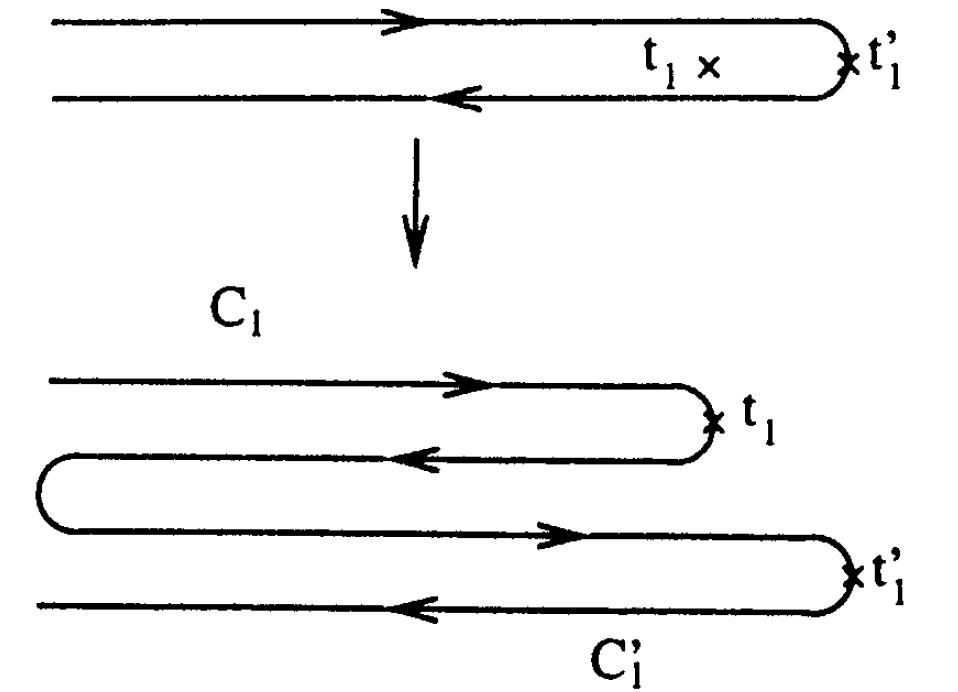}
    \caption{Deformation of the contour \cite{haug_jauho}.}
    \label{figure:deformC}
\end{figure}

When we consider the lesser Green function, which contains the information on the energy distribution, the first time coordinate is on the first half of the contour and the second time coordinate on the second half. The contour can be deformed to form two contours in the limit of initial time going to infinity as indicated in figure \ref{figure:deformC}.

When we look at the product $C(t_1,t_{1'})=\int_C d\tau A(t_1, \tau) B(\tau,t_{1'})$, the lesser function becomes on the new deformed contour $C^<(t_1,t_{1'})=\int_{C_1} d\tau A(t_1, \tau) B^<(\tau,t_{1'})+\int_{C_2} d\tau A^<(t_1, \tau) B(\tau,t_{1'})$. The integration on the first contour can run from $-\infty$ to $t_1$ and from $t_1$ to $+\infty$ and on the second contour from $-\infty$ to $t_{1'}$ and from $t_{1'}$ to $+\infty$. By doing this all functions can be expressed in lesser functions (for $t_1<t_{1'}$) and greater functions (for $t_1>t_{1'}$) and when the relations $G^a(1,1')=\theta(t_{1'}-t_1)[G^<(1,1')-G^>(1,1')]$ and $G^r(1,1')=\theta(t_{1}-t_{1'})[G^>(1,1')-G^<(1,1')]$ are used Langreth's result for analytic continuation is obtained \cite{langreth}.

\begin{equation}\label{langreth}
C^<(t_1,t_{1'})=\int^{+\infty}_{-\infty} dt[A^r(t_1,t)B^<(t,t_{1'})+A^<(t_1,t)B^a(t,t_{1'})]
\end{equation}

In the next section we shall derive from a simplified Dyson equation a quantum diffusion equation. In the subsequent section equation \ref{langreth} is used to derive from the complete Dyson equation a quantum diffusion equation with an interaction term accounting for inelactic scattering.

\section{\label{theory_section}Quantum diffusion equation for the distribution function}

To calculate the energy distribution function of electrons in a mesoscopic wire biased with an ac voltage induced by THz radiation on the reservoirs whereon the wire is coupled, we derive a quantum diffusion equation from the Dyson equation. First we derive an equation for a situation where inelastic interactions are neglected by neglecting the self energy term in the Dyson equation and introduce instead an elastic interaction term which will lead to a relaxation time approximation to account for the diffusivity of the system \cite{kamenev}.

\begin{equation}\label{eq:Dyson}
		G(1,1')=G_{0}(1,1')+i I[G(1,1')]
\end{equation}

Here $G(1,1')$ is the non-equilibrium Green function of a particle at coordinates $x_1$ and $t_1$ provided that the particle arises from the coordinates $x_{1'}$ and $t_{1'}$ defined by equation \ref{single-particle Green function}. $G_0(1,1')$ is the Green function of a free particle given by equation \ref{eq. of motion2} and $I[G(1,1')]$ is the collision term for elastic impurity scattering. By substituting the equation \ref{eq. of motion2} in the Dyson equation we can obtain the differential form consisting of the two conjugate parts.

\begin{equation}\label{eq: diff Dyson1}
		\left\{i\frac{\partial}{\partial t_{1}}-\frac{\hbar}{2m}\left(-i\nabla_{1}-\frac{e}{\hbar}A_{1}\right)^{2}\right\}G(1,1')=\delta(x_1-x_{1'})\delta(t_{1}-t_{1'})+i I_1[G(1,1')]
\end{equation}

\begin{equation}\label{eq: diff Dyson2}
		\left\{-i\frac{\partial }{\partial t_{1'}}-\frac{\hbar}{2m}\left(i\nabla_{1'}-\frac{e}{\hbar}A_{1'}\right)^{2}\right\}G(1,1')=\delta(x_{1}-x_{1'})\delta(t_{1}-t_{1'})-i I_2[G(1,1')]
\end{equation}

These two conjugate parts are subtracted from each other where the two collision terms are re-defined in a single collision term which will later provide the relaxation time approximation for elastic impurity scattering.

\begin{eqnarray}\label{eq: difference Dyson1 and Dyson2}
\nonumber
		\left\{i\left(\frac{\partial}{\partial t_{1}}+\frac{\partial}{\partial t_{1'}}\right)-\frac{\hbar}{2m}\left[\left(-i\nabla_{1}-\frac{e}{\hbar}A_{1}\right)^{2}-\left(i\nabla_{1'}-\frac{e}{\hbar}A_{1'}\right)^{2}\right]\right\}G(1,1')\\
		=\frac{i}{2} I_{coll}[G(1,1')]
\end{eqnarray}

Now the quadratic terms are expanded and we can use the fact that the vector potential is taken only time-dependent, so that according to commutation rules the operation $\nabla A$ is equivalent to $A \nabla$. 

\begin{eqnarray}\label{quadratic exp}
\nonumber	\left\{i\left(\frac{\partial}{\partial t_{1}}+\frac{\partial}{\partial t_{1'}}\right)+\frac{\hbar}{2m}\left[\left(\nabla_{1}^{2}-\nabla_{1'}^{2}\right)-2i\frac{e}{\hbar}\left(\nabla_{1} A_{1}+\nabla_{1'} A_{1'}\right)-\frac{e^{2}}{\hbar^{2}}\left(A_{1}^{2}-A_{1'}^{2}\right)\right]\right\}G(1,1')\\
=\frac{i}{2}I_{coll}[G(1,1')] & ~
\end{eqnarray}

For reasons of convenience we will proceed with this equation expressed in Wigner coordinates defined like:

\begin{eqnarray}\label{Wigner coordinates}
T &=& \frac{t_{1}+t_{1'}}{2}, \\
t &=& t_{1}-t_{1'},\\
R &=& \frac{r_{1} + r_{1'}}{2},\\
r &=& r_{1} - r_{1'}.  
\end{eqnarray}

To introduce the Wigner coordinates the quadratic parts of equation \ref{quadratic exp} has to be expanded. The summation and difference of the vector potential can be replaced by an representive symbols: $A_+(t)=A(t_1)+A(t_{1'})$ and $A_-(t)=A(t_1)-A(t_{1'})$. Because we are interested in the distribution function we proceed with the lesser Green function in the equations. The interaction term is now just dependent on the lesser Green function. No analytic continuation procedures have to be followed, because in the end the interaction is given by a relaxation time approximation.

\begin{eqnarray}\label{eq2}
\nonumber
		\left\{\frac{i}{2}\frac{\partial}{\partial T}+\frac{\hbar}{2m}\left[\nabla_R\nabla_r-i\frac{e}{\hbar}(\nabla_r A_- + \frac{1}{2}\nabla_R A_+)-\frac{e^2}{2\hbar^2}A_+A_-\right]\right\}G^{<}(r,R,t,T)\\
		=\frac{i}{2}I_{coll}[G^<]
\end{eqnarray}

Here we make the transition to proceed with the distribution function in a momentum representation of equation \ref{eq2}.

\begin{eqnarray}\label{eq3}
\nonumber
		\left\{\frac{i}{2}\frac{\partial}{\partial T}+\frac{\hbar}{2m}\left[\nabla_R\nabla_r-i\frac{e}{\hbar}(\nabla_r A_- + \frac{1}{2}\nabla_R A_+)-\frac{e^2}{2\hbar^2}A_+A_-\right]\right\}\int\frac{dp'}{2\pi^3}e^{ip'r/\hbar}f(p',R,t,T)\\
		=\frac{i}{2}I_{coll}[\int\frac{dp'}{2\pi^3}e^{ip'r/\hbar}f(p',R,t,T)]
\end{eqnarray}

The terms containing $\nabla_r$ operate first on the integral, so that the operator is replaced by $ip'/\hbar$, and the terms are rearranged.

\begin{eqnarray}\label{eq4}
\nonumber
\int\frac{dp'}{2\pi^3}\left\{\frac{\partial}{\partial T}+\frac{(p'-\frac{eA_+}{2})}{m}\left[\nabla_R-i\frac{e}{\hbar}A_-\right]\right\}e^{ip'r/\hbar}f(p',R,t,T)\\
=I_{coll}[\int\frac{dp'}{2\pi^3}e^{ip'r/\hbar}f(p',R,t,T)]
\end{eqnarray}

Then the equation is multiplied by $e^{-ipr/\hbar}$ and a Fourier transform is performed by integrating over all $r$. 

\begin{eqnarray}\label{fourier1}
\nonumber
\int dp' \int\frac{dr}{2\pi^3}\left\{\frac{\partial}{\partial T}+\frac{(p'-\frac{eA_+}{2})}{m}\left[\nabla_R-i\frac{e}{\hbar}A_-\right]\right\}e^{i(p'-p)r/\hbar}f(p',R,t,T)\\
=I_{coll}[\int\ dp' \int \frac{dr}{2\pi^3}e^{i(p'-p)r/\hbar}f(p',R,t,T)]
\end{eqnarray}

The Fourier transform in $r$ of the exponent creates the delta function $\delta(p'-p)$ and the integral over $p'$ forces by means of the delta function all $p'$ to $p$.

\begin{equation}\label{shifting p}
\left\{\frac{\partial}{\partial T}+\frac{(p-\frac{eA_+}{2})}{m}\left[\nabla_R-i\frac{e}{\hbar}A_-\right]\right\}f(p,R,t,T)=I_{coll}[f(p,R,t,T)]
\end{equation}

The sum of the vector potential on time $t_1$ and $t_{1'}$ modulates the momentum of the charge carrier. This is a second order effect so that the term in front of the momentum part of the equation above can be replaced by the velocity of the charge carrier. The vector potential is defined as $A(t)=U/(L\omega)cos(\omega t)$. The difference term in the vector potential is then expressed in the Wigner coordinates.

\begin{eqnarray}
\nonumber
A_-(t,T) &=& \frac{U}{L\omega}(cos(\omega(T+t/2))-cos(\omega(T-t/2)))\\
\nonumber
&=&-2\frac{U}{L\omega}sin(\omega T)sin(\omega t/2)
\\
&=&-\frac{U}{iL\omega}sin(\omega T)(e^{i\omega t/2}-e^{-i\omega t/2})
\end{eqnarray}

This vector potential is substituted in equation \ref{shifting p} and the same procedure is followed for an energy representation as previous done for the momentum representation. A Fourier transform in $t$ is performed and this is integrated over $E'$. For the terms without the vector potential this operation is trivial since it just replaces the variable $t$ in the distribution function by $E$. For the part containing the vector potential the situation is a bit more subtle and essential in the understanding of the absorption of energy quanta of the field by electrons. Therefore this is explicitly shown.

\begin{eqnarray}\label{photon absorption}
\nonumber
& &sin(\omega T)\int dt e^{-iEt/\hbar}(e^{i\omega t/2}-e^{-i\omega t/2})\int dE'e^{iE't/\hbar}f(p,R,E',T)  \\
\nonumber
&=&sin(\omega T)(\int dE'\int dt e^{-i(E'-E/+\omega \hbar/2)t/\hbar}f(p,R,E',T)-\int dE'\int dt e^{-i(E'-E/-\omega  \hbar/2)t/\hbar}f(p,R,E',T)  \\
\nonumber
&=&sin(\omega T)(\int dE'\delta(E'-E+\omega \hbar/2)f(p,R,E',T)-\int dE'\int dt \delta(E'-E-\omega \hbar/2)f(p,R,E',T)  \\
\nonumber
&=&sin(\omega T)\left[f(p,R,E-\omega \hbar/2,T)-f(p,R,E+\omega \hbar/2,T)\right]  \\
\nonumber
&=&-sin(\omega T)\left[f(p,R,E+\omega \hbar/2,T)-f(p,R,E-\omega \hbar/2,T)\right] \\
&=&-\omega sin(\omega T)D_{\omega}f(p,R,E,T)  
\end{eqnarray}

When this is substituted in the kinetic equation and the operator $\nabla_R$ is replaced by a derivative with respect to the one-dimensional space coordinate $x$ we arrive at a form from which we can go to a diffusion equation. 

\begin{equation}\label{kinetic}
\left\{\frac{\partial}{\partial T}+v\left[\frac{\partial}{\partial x}-\frac{eU}{\hbar L}sin(\omega T)D_{\omega}\right]\right\}f(p,x,E,T)=I_{coll}[f]
\end{equation}

The distribution function can be divided in an odd and an even part with respect to $p$. 

\begin{eqnarray}\label{even odd}
f_e(p,R,E,T)=\frac{f(p,R,E,T)+f(-p,R,E,T)}{2}\\
f_o(p,R,E,T)=\frac{f(p,R,E,T)-f(-p,R,E,T)}{2}
\end{eqnarray}

Because the field is considered to be uniaxially symmetric the even part of the distribution function only depends on the absolute value of $p$, so that the even part of the distribution function is the distribution function as function of energy only: $f_e(p,x,E,T)=f(x,E,T)$. First the kinetic equation is transformed into two equation for positive and negative momentum.

\begin{equation}\label{even odd 1}
\left\{\frac{\partial}{\partial T}+v \left[\frac{\partial}{\partial x}-\frac{eU}{\hbar L}sin(\omega T)D_{\omega}\right]\right\}f(p,x,E,T)=I[f(p,x,E,T)]
\end{equation}

\begin{equation}\label{even odd 2}
\left\{\frac{\partial}{\partial T}-v\left[\frac{\partial}{\partial x}-\frac{eU}{\hbar L}sin(\omega T)D_{\omega}\right]\right\}f(-p,x,E,T)=I[f(-p,x,E,T)]
\end{equation}

The equations are added and subtracted from each other and divided by 2.

\begin{eqnarray}\label{even odd add}
\nonumber
\frac{\partial}{\partial T}(f(p,x,E,T)+f(-p,x,E,T))/2 & ~ &\\
\nonumber
+v\left[\frac{\partial}{\partial x}-\frac{eU}{\hbar L}sin(\omega T)D_{\omega}\right](f(p,x,E,T)-f(-p,x,E,T))/2\\
=I[(f(p,x,E,T)+f(-p,x,E,T))/2]
\end{eqnarray}

\begin{eqnarray}\label{even odd subtr}
\nonumber
\frac{\partial}{\partial T}(f(p,x,E,T)-f(-p,x,E,T))/2 & ~ &\\
\nonumber
+v\left[\frac{\partial}{\partial x}-\frac{eU}{\hbar L}sin(\omega T)D_{\omega}\right](f(p,x,E,T)+f(-p,x,E,T))/2\\
=I[(f(p,x,E,T)-f(-p,x,E,T))/2]
\end{eqnarray}

The identities of the even and odd part of the distribution function can be implemented and the even part is changed to the distribution function as function of energy only.

\begin{equation}\label{even coll}
\frac{\partial}{\partial T}f(x,E,T)+v\left[\frac{\partial}{\partial x}-\frac{eU}{\hbar L}sin(\omega T)D_{\omega}\right]f_o(p,x,E,T)=I[f(x,E,T)]
\end{equation}

\begin{equation}\label{odd coll}
\frac{\partial}{\partial T}f_o(p,x,E,T)+v\left[\frac{\partial}{\partial x}-\frac{eU}{\hbar L}sin(\omega T)D_{\omega}\right]f(x,E,T)=I[f_o(p,x,E,T)]
\end{equation}

Now if we only consider inelastic impurity scattering we only have an collision integral acting on the odd part of the distribution function. The impurity scattering can only change the momentum of a charge carrier but can not change the energy. When for this collision integral the relaxation time approximation $I[f_o(p,x,E,T)]=-f_o(p,x,E,T)/\tau_{im}$ is used and the impurity time is considered to be small the time derivative of equation \ref{odd coll} can be neglected. Then $f_o(p,x,E,T)$ is just a function of the momentum part times f(x,E,T) times $-\tau_{im}$. When this is substituted in equation \ref{even coll} we arrive at the final form of the quantum diffusion equation, where we take $D=v^2 \tau_{im}$ the diffusion constant.

\begin{equation}\label{quantum diffusion eq}
\left\{\frac{\partial}{\partial T}-D\left[\frac{\partial}{\partial x}-\frac{eU}{\hbar L}sin(\omega T)D_{\omega}\right]^2\right\}f(x,E,T)=0
\end{equation}

A.V. Shytov also studied the energy distribution of electrons in a diffusive, coherent wire. The equation he used to calculate the distribution function is equivalent to that derived above.

\section{\label{limitcases}Limit situations for the simple quantum diffusion equation}

The quantum diffusion equation \ref{quantum diffusion eq} can be solved analytically for certain limit situations \cite{shytov}. Therefore it is convenient to express the equation in dimensionless parameters.

\begin{eqnarray}\label{dimension degradat}
t \rightarrow t \omega, & x \rightarrow x/L, & E \rightarrow E/eV
\end{eqnarray}

When we introduce the diffusion time for an electron in the wire $\tau_D=L^2/D$ equation \ref{quantum diffusion eq} becomes

\begin{equation}\label{dimensionless qde}
\left\{\frac{\partial}{\partial t}-\frac{1}{\omega \tau_D}\left[\frac{\partial}{\partial x}-sin(t)D_{\omega}\right]^2\right\}f(t,E,x)=0
\end{equation}

This differential equation has for the initial and boundary conditions a Fermi distribution

\begin{eqnarray}\label{initial boundary conditions}
f(t=0,E,x)=n_F(E)\\
f(t,E,x=0)=n_F(E)\\
f(t,E,x=1)=n_F(E)
\end{eqnarray}

The limit situations are defined by the ratio of the field frequency $\omega$ and the diffusion time $\tau_D$ and the ratio of the photon energy $\hbar \omega$ and the field energy $eV$.

\subsection{\label{slow field}Slow field limit}

For $\omega \tau_D << 1$ the field oscillates slowly with respect to the time that the electron travels diffusively through the wire. In equation \ref{dimensionless qde} the time derivative can be neglected and the solution is obtained by solving the spatial second order differential equation. The solution becomes

\begin{equation}
f(t,E,x)=[(1-x)e^{xsin(t)D_{\omega}}+xe^{(x-1)sin(t)D_{\omega}}]n_F(E).
\end{equation}

Following the approach in Ref. \cite{shytov} we take the Fourier transform in energy domain to find the exponent of the finite difference operator which leads to

\begin{equation}
e^{z D_{\omega}}\Phi(E)=\sum^{\infty}_{n=-\infty} J\left(2z)\right)\Phi(E-n\omega/2).
\end{equation}

Substituting this in the equation for the distribution equation, restoring dimensions and using time averaging  $\overline{J_{2n}(2asin(t))}=J^2_n(a)$ we arrive at the final general expression for the distribution function.

\begin{equation}\label{slow field function}
\bar{f}(E,x)=\left(1-\frac{x}{L}\right)\sum^{\infty}_{n>\frac{E}{\hbar \omega}}J^2_n\left(\frac{xeV}{L\hbar \omega}\right)n_F\left(E-\frac{n\hbar \omega}{2}\right)+\frac{x}{L}\sum^{\infty}_{n>\frac{E}{\hbar \omega}}J^2_n\left(\frac{(\frac{x}{L}-1)eV}{\hbar \omega}\right)n_F\left(E-\frac{n\hbar \omega}{2}\right)
\end{equation}

So we see close resemblance with Tien-Gordon theory where the probability of absorbing $n$ field quanta is also given by squared Besselfunctions. The resemblance with a dc biased wire is also visable in the pre-factors $1-x/L$ and $x/L$, which gives the number of electrons that enter position $x$ from the right and the left reservoir \cite{nagaev}.

\begin{figure}[h!]
    \centering
    \includegraphics[width=1\textwidth]{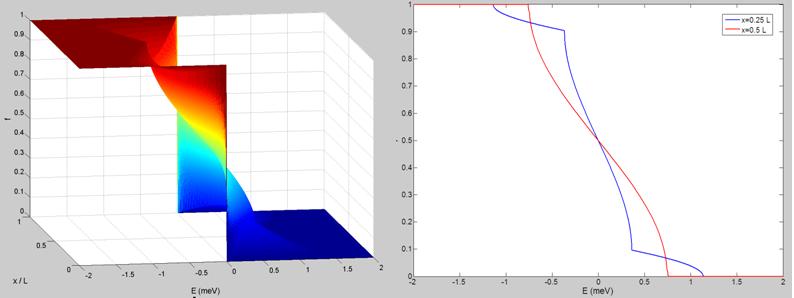}
    \caption{Left the electron energy distribution in a mesoscopic wire ac biased in the slow field, strong signal limit ($\omega \tau_D << 1$, $\hbar \omega << eV$), right with the blue line the distribution on position $x=0.25$ and with the red line the distribution on position $x=0.5$.}
    \label{figure:slowfield}
\end{figure}

When the field energy is much larger than the photon energy, $\hbar \omega<<eV$, the asymptotic form of the Bessel function at $x/\omega \approx n>>1$ may be used, which gives

\begin{equation}
\bar{f}(E,x)=\left(1-\frac{x}{L}\right)F_0(E,x/L)+\frac{x}{L} F_0(E,1-x/L)
\end{equation}

where $F_0(E,x/L)=\frac{1}{\pi}cos^{-1}(\tilde{E})$ for $|\tilde{E}|<1$ with $\tilde{E}=\frac{LE}{xeV}$. For $\tilde{E}<-1$ the occupation is one and for $\tilde{E}>1$ the occupation is zero. Figure \ref{figure:slowfield} shows the distribution in slow field, strong signal limit.

\subsection{\label{slow field}Fast field limit}

In the fast field limit the diffusion time is much larger than the reciprocal frequency of the field, $\omega \tau_D>>1$. This means that equation \ref{dimensionless qde} practically becomes time-independent, since the time-derivative is proportional to $1/\omega \tau_D$. Averaging equation \ref{dimensionless qde} over the field period, leads to an equation for the time-averaged distribution function.

\begin{equation}\label{fastfield}
\left[\frac{\partial^2}{\partial x^2}+\frac{1}{2}D^2_{\omega}\right]\bar{f}(E,x)=0
\end{equation}

In the limit $\hbar \omega<<eV$ the finite difference operator $D_{\omega}$ can be replaced by the partial energy derivative $\partial/\partial \epsilon$. This makes equation \ref{fastfield} become a Laplace equation in a two-dimensional strip defined by $0<x<1$ and $-\infty<\epsilon<\infty$. This strip can be conformally mapped by the function $w=exp[\pi i (x+i \sqrt(2)\epsilon)]$  onto the half-plane $Im w>0$ \cite{shytov} \cite{kwok}. The boundary condition on the line $Im w=0$ at zero temperature is set to be $\bar{f}_{\infty}(w)=0$ for $|Re w|<0$ and $\bar{f}_{\infty}(w)=1$ for $|Re w|>0$. The imaginary part of the analytic function gives the solution of this boundary value problem.

\begin{equation}
\bar{f}_{\infty}(w)=Im\frac{1}{\pi}ln\left(\frac{1-w}{1+w}\right)
\end{equation}

When the original dimensional units are restored the final expression for the time-averaged distribution function is

\begin{equation}
\bar{f}_{\infty}(\epsilon,x)=\frac{1}{\pi}cot^{-1}\left(\frac{sinh(\pi \sqrt{2}\epsilon/eV)}{sin(\pi x/L)}\right).
\end{equation}

\begin{figure}[h!]
    \centering
    \includegraphics[width=1\textwidth]{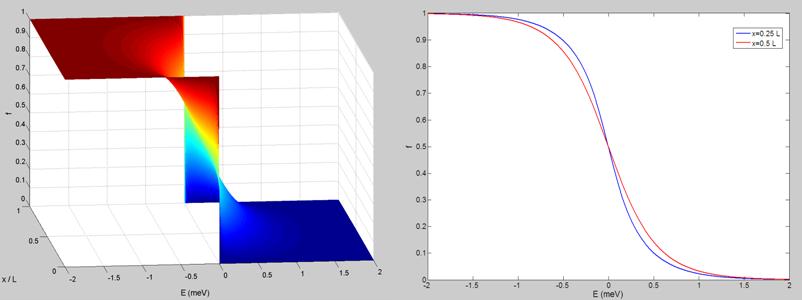}
    \caption{Left the electron energy distribution in a mesoscopic wire ac biased in the fast field, strong signal limit ($\omega \tau_D >> 1$, $\hbar \omega << eV$), right with the blue line the distribution on position $x=0.25$ and with the red line the distribution on position $x=0.5$.}
    \label{figure:fastfield}
\end{figure}

In the fast field the energy distribution does not have to go to zero at high energies. The energy gained from the field is not limited by $eV$. Instead an electron has a finite probability of oscillating several times back and forth with the field in the wire before leaving the wire, thereby gaining multiple energy quanta of the field which sum exceeds $eV$. Figure \ref{figure:fastfield} shows the electron energy distribution in the fast field, strong signal limit.

\section{\label{theory_section2}Incorporating inelastic interactions}

So far only coherent transport is considered. If the length of the wire is extended in such a way that the diffusion time becomes of the same order as the phase coherence time and energy relaxation time this simple model breaks down. Therefore this model has to be extended to account for electron-electron and electron-phonon interactions. This is done by evaluating the complete Dyson equation \ref{eq:Dysonwithinteractions}, where we isolate the collision term for the elastic impurity scattering that is treated with a relaxation time approximation in the same way as before.

\begin{equation}\label{eq:Dyson with int}
		G(1,1')=G_{0}(1,1')+i I_{im}[G(1,1')]+\frac{1}{\hbar}\int dx_2 \int dx_3 \int d\tau_2 \int d\tau_3 G_0(1,2)\Sigma(2,3)G(3,1')
\end{equation}

By substituting the equation of motion for the Green function of a free particle we obtain again two conjugate equations.

\begin{quotation}
~
\end{quotation}

\begin{eqnarray}\label{eq: diff Dyson1 int}
\nonumber
\left\{i\frac{\partial}{\partial t_{1}}-\frac{\hbar}{2m}\left(-i\nabla_{1}-\frac{e}{\hbar}A_{1}\right)^{2}\right\}G(1,1')=\delta(x_1-x_{1'})\delta(t_{1}-t_{1'})+i I_1[G(1,1')]\\
+\frac{1}{\hbar}\int d\tau \int dy \Sigma(x_1,t_1,y,\tau)G(y,\tau,x_{1'},t_{1'})
\end{eqnarray}

\begin{eqnarray}\label{eq: diff Dyson2 int}
\nonumber	\left\{-i\frac{\partial}{\partial t_{1'}}-\frac{\hbar}{2m}\left(i\nabla_{1'}-\frac{e}{\hbar}A_{1'}\right)^{2}\right\}G(1,1')=\delta(x_{1}-x_{1'})\delta(t_{1}-t_{1'})-i I_2[G(1,1')]\\
+\frac{1}{\hbar}\int d\tau \int dy G(x_1,t_1,y,\tau)\Sigma(y,\tau,x_{1'},t_{1'})
\end{eqnarray}

As before we are interested in the distribution function, so we concentrate on the lesser Green function by an analytic continuation of the above functions where we concentrate on the self energy part of the functions. The remaining part of the equations in the derivation is similar to the derivation without inelastic interactions.

\begin{equation}\label{self energy1}
I_1[G^<]=\frac{1}{\hbar}\int d\tau \int dy \left(\Sigma^r(x_1,t_1,y,\tau)G^<(y,\tau,x_{1'},t_{1'})+\Sigma^<(x_1,t_1,y,\tau)G^a(y,\tau,x_{1'},t_{1'})\right)
\end{equation}

\begin{equation}\label{self energy2}
I_2[G^<]=\frac{1}{\hbar}\int d\tau \int dy \left(G^r(x_1,t_1,y,\tau)\Sigma^<(y,\tau,x_{1'},t_{1'})+G^<(x_1,t_1,y,\tau)\Sigma^a(y,\tau,x_{1'},t_{1'})\right)
\end{equation}

These two equations are subtracted from each other.

\begin{eqnarray}
\nonumber
I[G]=\frac{1}{\hbar}\int d\tau \int dy(\Sigma^r(x_1,t_1,y,\tau)G^<(y,\tau,x_{1'},t_{1'})+\Sigma^<(x_1,t_1,y,\tau)G^a(y,\tau,x_{1'},t_{1'})\\
-G^r(x_1,t_1,y,\tau)\Sigma^<(y,\tau,x_{1'},t_{1'})-G^<(x_1,t_1,y,\tau)\Sigma^a(y,\tau,x_{1'},t_{1'}))
\end{eqnarray}

Now the following identities are introduced to gain insight in the derivation \cite{haug_jauho}.

\begin{eqnarray}
\nonumber
A^r &=&\frac{1}{2}(A^r+A^a)+1/2(A^r-A^a)\\
\nonumber
A^a &=& \frac{1}{2}(A^a+A^r)+1/2(A^a-A^r)\\
\nonumber
\Sigma &=& \frac{1}{2}(\Sigma^r+\Sigma^a)\\
\nonumber
G &=& \frac{1}{2}(G^r+G^a)\\
\nonumber
A &=& i(G^r-G^a)\\
\nonumber
\Gamma &=& i(\Sigma^r-\Sigma^a)
\end{eqnarray}

The terms are arranged so that everything is expressed in commutators and anti-commutators.

\begin{equation}
I[G^<]=\frac{1}{\hbar}\int d\tau \int dy \left([\Sigma,G^<]+[\Sigma^<,G]+\frac{1}{2}\left\{\Sigma^>,G^<\right\}-\frac{1}{2}\left\{G^>,\Sigma^<\right\}\right)
\end{equation}

To simplify the calculations we assume the scattering to be local in space, so that the integral operation over $y$ forces the integration variable towards the central space coordinate. Also we can make the assumption of weak interactions, so that we can apply the quasi-particle approximation. Because we also used a gradient expansion of the potential, the first two commutators of the above relation are second order and can be neglected. Basically this means that the density of states of the quasi-particles in the wire is not affected by the vector potential nor the interactions. 

\begin{eqnarray}\label{self energy3}
\nonumber
I[G^<]=\frac{1}{\hbar}\int d\tau(\frac{1}{2}\Sigma^>(x_1,t_1;x_1,\tau)G^<(x_1,\tau;x_{1'},t_{1'})+\frac{1}{2}G^<(x_1,t_1;x_{1'},\tau)\Sigma^>(x_{1'},\tau;x_{1'},t_{1'})\\
-\frac{1}{2}\Sigma^<(x_1,t_1;x_1,\tau)G^>(x_1,\tau;x_{1'},t_{1'})-\frac{1}{2}G^>(x_1,t_1;x_{1'},\tau)\Sigma^<(x_{1'},\tau;x_{1'},t_{1'}))
\end{eqnarray}

When we now also assume that the scattering is instantaneous, the integral over $\tau$ forces the integration variable towards the second time variable of the self energy in the product of the self energy and the Green function.

\begin{eqnarray}\label{self energy4}
\nonumber
I[G^<]=\frac{1}{\hbar}(\frac{1}{2}\Sigma^>(x_1,t_1;x_1,t_1)G^<(x_1,t_1;x_{1'},t_{1'})+\frac{1}{2}G^<(x_1,t_1;x_{1'},t_{1'})\Sigma^>(x_{1'},t_{1'};x_{1'},t_{1'})\\
-\frac{1}{2}\Sigma^<(x_1,t_1;x_1,t_1)G^>(x_1,t_1;x_{1'},t_{1'})-\frac{1}{2}G^>(x_1,t_1;x_{1'},t_{1'})\Sigma^<(x_{1'},t_{1'};x_{1'},t_{1'}))
\end{eqnarray}

As we assume a slow variation of the Green function induced by the vector potential and we assume the interactions to be weak, we can state that the effect of the self energy on time $t_1$ is the same as that at time $t_{1'}$. So the self energies at $t_1$ and $t_{1'}$ can be replaced by a single self energy $\Sigma(x_1,t_1;x_{1'},t_{1'})$.

\begin{equation}\label{self energy5}
I[G^<]=\frac{1}{\hbar}\left(\Sigma^>(x_1,t_1;x_{1'},t_{1'})G^<(x_1,t_1;x_{1'},t_{1'})-\Sigma^<(x_1,t_1;x_{1'},t_{1'})G^>(x_1,t_1;x_{1'},t_{1'})\right)
\end{equation}

By applying the Wigner transformation to this collision term, the product of the self energies with the Green functions can be interpreted as the imaginary in- and out scattering rates $i\hbar\Gamma_{e,h}$ with the electron and hole distribution \cite{hague} \cite{ferry}.

\begin{equation}\label{scattering rate}
I[f]=i\int dE' \int dp' e^{i(E't+p'r)/\hbar}(\Gamma_h(R,T,E',p')f(R,T,E',p')-\Gamma_e(R,T,E',p')f_h(R,T,E',p'))
\end{equation}

This is again multiplied by $e^{-i(Et-pr)/\hbar}$ and integrated over $t$ and $r$ leading to the final form of the total collision term due to inelastic scattering where this is multiplied by $i$ from the rest of the equation.

\begin{equation}\label{collision term}
I_{tot}[f]=\Gamma_h(R,T,E,p)f(R,T,E,p)-\Gamma_e(R,T,E,p)(1-f(R,T,E,p))
\end{equation}

The two parts of the quantum diffusion equation are again connected.

\begin{equation}
 	\left\{\frac{\partial}{\partial T}-v\left[\frac{\partial}{\partial x}-\frac{eU}{\hbar L}sin(\omega T)D_{\omega}\right]\right\} f(E,p,x,T)=I_{im}[f]+I_{tot}[f]
\end{equation}

Same procedure is followed to come to a diffusion equation as for elastic impurity scattering. The equation is divided in an even and odd part, where the elastic impurity scattering only contributes to the even part and the inelastic interactions contribute to the odd part.

\begin{equation}\label{even coll}
\frac{\partial}{\partial T}f(x,E,T)+v\left[\frac{\partial}{\partial x}-\frac{eU}{\hbar L}sin(\omega T)D_{\omega}\right]f_o(p,x,E,T)=I_{tot}[f(x,E,T)]
\end{equation}

\begin{equation}\label{odd coll}
\frac{\partial}{\partial T}f_o(p,x,E,T)+v\left[\frac{\partial}{\partial x}-\frac{eU}{\hbar L}sin(\omega T)D_{\omega}\right]f(x,E,T)=I[f_o(p,x,E,T)]
\end{equation}

Taking the same relaxation time approximation $I[f_o(p,x,E,T)]=-f_o(p,x,E,T)/\tau_{im}$ for the impurity scattering leads to the desired quantum diffusion equation.

\begin{equation}\label{qde_inel}
\left\{\frac{\partial}{\partial T}-D\left[\frac{\partial}{\partial x}-\frac{eU}{\hbar L}sin(\omega T)D_{\omega}\right]^2\right\}f(E,x,T)=I_{tot}f(E,x,T)
\end{equation}

So we see that the quantum diffusion equation \ref{quantum diffusion eq} is extended with a term that controls the in- and outscattering of quasi-particles at energy $E$ due to inelastic collisions. These inelastic collisions could be due to the interaction between two quasi-particles or due to the interaction between a quasi-particle and a phonon. In the next section we will derive expressions for these interactions.

\section{\label{scattering}Inelastic scatterering}

The main energy relaxation mechanisms are electron-electron \footnote{the electron is in fact a quasi-particle} and electron-phonon scattering and the sum of these contributions give the total interaction term.

\begin{equation}\label{tot interaction term}
I_{tot}[f]=I_{e-e}[f]+I_{e-ph}[f]
\end{equation}

Both collision terms have an inscattering and outscattering term as seen in equation \ref{collision term}. A quasi-particle with energy $E$ has an collision term

\begin{equation}\label{in-outscattering}
I_{coll}(x,E,\left\{f\right\})=I^{in}_{coll}(x,E,\left\{f\right\})-I^{out}_{coll}(x,E,\left\{f\right\})
\end{equation}

The collision terms due to electron-electron scattering and electron-phonon scattering can be calculated independently of each other. First we will tread the interaction between electrons and phonon. Subsequently we look at the interactions between electrons.

\subsection{\label{e-ph}Electron-phonon interaction}

Let's first focus on the electron-phonon interactions. To begin some assumptions have to be made. When we only want to consider acoustic phonons with a dispersion relation between energy and wave vector $\epsilon_k=\hbar s q$, with $s$ the sound velocity, the phonon temperature $T_{ph}$ has to be small compared to the Debye temperature $T_D$. Further the electronic wave functions can be approximated by plane waves, which is justified by the fact that electron-phonon coupling is only relevant for higher energies and from the dispersion relation it is seen that large wave vectors are associated with these energies. Then it is probable that the electronic mean free path is larger than $1/q$. Also the electron-phonon coupling is given by a scalar deformation potential, so only the longitudinal phonons are coupled on the electrons. The matrix element describing the interaction simplifies to $|M(q)|^2=|M_0|^2q/V$, where $|M_0|^2$ is geometry independent. This only is valid for spherical Fermi surfaces \cite{anthore}.

The transition of an electron to a state with energy $E$ can either be due to the absorption or the emission of a phonon. The same can be said of the transition out of the state with energy $E$. We can define the transition due to absorption by $W^-$ and the transition due to emission by $W^+$. Further we know that the state from which the particle departes has to be occupied and the state in which the particle arrives has to be unoccupied. The latter is a direct consequence of the fact that we look at fermions and according to the Pauli exclusion principle a state can only be occupied by a single fermion. This leads to the following collision terms \cite{rammer}.

\begin{eqnarray}\label{eph inscattering}
\nonumber
I^{in}_{eph}(x,E_k,[f])=\int dE_{k'} W^+(x,E_{k'},E_k)f(x,E_k-E_{k'})(1-f(x,E_k))n_{ph}(E_{k-k'})\\
+\int dE_{k'}W^(x,E_{k'},E_k)-f(x,E_k-E_{k'})(1-f(x,E_k))(1+n_{ph}(E_{k'-k})
\end{eqnarray}

\begin{eqnarray}\label{eph outscattering}
\nonumber
I^{out}_{eph}(x,E_k,[f])=\int dE_{k'} W^+(x,E_{k'},E_k)f(x,E_k)(1-f(x,E_k-E_{k'}))(1+n_{ph}(E_{k-k'}))\\
+\int dE_{k'}W^-(x,E_{k'},E_k)f(x,E_k)(1-f(x,E_k-E_{k'}))n_{ph}(E_{k'-k})
\end{eqnarray}

Here $n_{ph}$ represents the Bose energy distribution of the phonons, $n_{ph}(E)=(exp(E/kT)-1)^{-1}$. The transition probabilities are given by Fermi's Golden Rule \cite{rammer}.

\begin{equation}\label{golden rule}
W^{\pm}(x,E_{k'},E_k)=\frac{2 \pi}{\hbar} |\alpha_{k'-k}|^2\delta(E_{k'}-E_k \pm E_{\pm(k-k')})
\end{equation}

To obtain the collision rate at which an electron with wave vector $k$ emits or absorbs a phonon of energy $E_{|k-k'|}$ the equations \ref{eph inscattering} and \ref{eph outscattering} have to be summed over $k'$ with $E(k-k')$ fixed. A detailed derivation can be found in Ref. \cite{rammer}. 

\begin{eqnarray}\label{eph in scattering2}
\nonumber
I^{in}_{eph}(x,E,[f])=2 \pi \int d\epsilon \alpha^2F(\epsilon)f(x,E-\epsilon)(1-f(x,E))n_{ph}(\epsilon)\\
+2 \pi \int d\epsilon \alpha^2F(\epsilon)f(x,E+\epsilon)(1-f(x,E))(1+n_{ph}(\epsilon))
\end{eqnarray}

\begin{eqnarray}\label{eph out scattering2}
\nonumber
I^{out}_{eph}(x,E,[f])=2 \pi \int d\epsilon \alpha^2F(\epsilon)f(x,E)(1-f(x,E-\epsilon))(1+n_{ph}(\epsilon)\\
+2 \pi \int d\epsilon \alpha^2F(\epsilon)f(x,E)(1-f(x,E+\epsilon))n_{ph}(\epsilon)
\end{eqnarray}

The so called Eliashberg function $\alpha^2F(\epsilon)$ is dependent on the coupling between the electrons and phonons. In Ref. \cite{giazotto} this function is determined to be

\begin{eqnarray}\label{Eliashberg}
\alpha^2F(\epsilon)=\frac{|M|^2 \epsilon^2}{4 \pi^2 s^3 N(0)}, & ~ &\\
|M|^2=\frac{\pi s^3 \Sigma}{12 \zeta(5)k^5_b}.
\end{eqnarray}

Here $|M|^2$ is the matrix element depending on the defined deformation potential and $N(0)$ is the electronic density of states at Fermi level. The precise microscopic form of $|M|^2$ is dependent on the details of the lattice structure. Therefore in Ref. \cite{giazotto} they present this matrix element in terms of a measurable quantity $\Sigma$ related to the power dissipated to the lattice of volume $V$ by $P=\Sigma V T^5$. A detailed form of the electron-phonon interactions and the temperature dependence in disordered conductors can be found in Ref. \cite{sergeev}.

\subsection{\label{e-e}Electron-electron interaction}

The interaction between quasi-particles is due to the Coulomb potential of the particles. This Coulomb interaction is screened  by an effective medium build from all the electrons in the metal. Altshuler \textit{et al}. showed that multiple scattering events due to disorder in the system reduces the lifetime of the quasi-particle \cite{altshuler}. At zero temperature the lifetime of a particle obeying Fermi statistics in state $|\alpha>$ with energy $\epsilon_{\alpha}$ above Fermi level that interacts with a particle in state $|\gamma>$ with energy $\epsilon_{\gamma}$ directly follows from Fermi's Golden Rule \cite{akkermans}.

\begin{equation}\label{e-e golden rule}
\frac{1}{\tau_{\alpha}}=\frac{4 \pi}{\hbar} \sum_{\beta \gamma \delta} | \left\langle \alpha \gamma | U | \beta \delta \right\rangle |^2 \delta(\epsilon_{\alpha}+\epsilon_{\gamma}-\epsilon_{\beta}-\epsilon_{\delta})
\end{equation}

$U$ is the interaction potential from which the states $|\alpha>$ and $|\gamma>$ evolve in the states $|\beta>$ and $|\delta>$. This lifetime has to be averaged over all states having energy $\epsilon$ in order not to single out a give state.

\begin{equation}\label{e-e golden rule2}
\frac{1}{\tau_{ee}(\epsilon)}=\frac{4 \pi}{\hbar \nu_0} \sum_{\alpha \beta \gamma \delta} | \left\langle \alpha \gamma | U | \beta \delta \right\rangle |^2 \delta(\epsilon_{\alpha}+\epsilon_{\gamma}-\epsilon_{\beta}-\epsilon_{\delta})\delta(\epsilon-\epsilon_{\alpha})
\end{equation}

\begin{figure}[h!]
    \centering
    \includegraphics[width=0.75\textwidth]{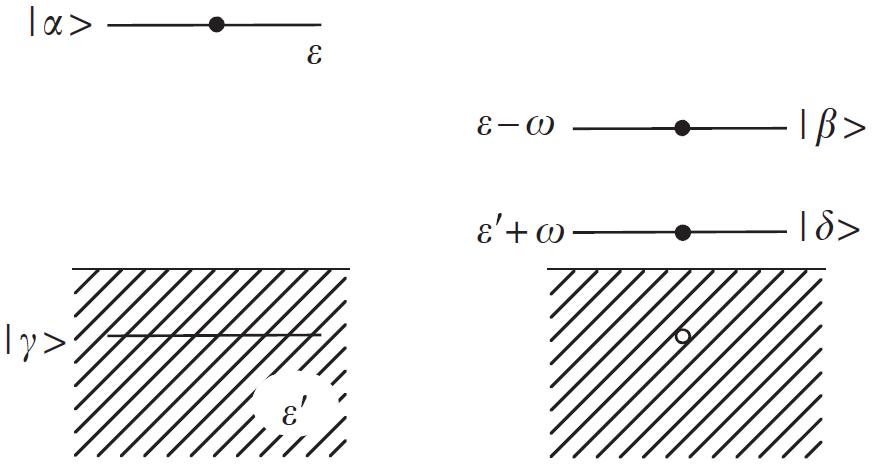}
    \caption{The energy exchange in scattering between quasi-particles. Left the initial situation, right the final situation \cite{akkermans}.}
    \label{figure:energy_exchange_scattering}
\end{figure}

When the energy of the states $|\gamma>$ are denoted by $\epsilon'$ and the energy exchange involved in the scattering is $\omega$, energy conservation leads to energies of the final states $|\beta>$ and $|\delta>$ of $\epsilon-\omega$ and $\epsilon'+\omega$. This is depicted in figure \ref{figure:energy_exchange_scattering}.

Considering all possible initial states $|\gamma>$ leads to integration over $\epsilon'$ and $\omega$.

\begin{equation}\label{e-e golden rule3}
\frac{1}{\tau_{ee}(\epsilon)}=\frac{4 \pi}{\hbar \nu_0} \int^{\epsilon}_0 d\omega \int^{0}_{-\omega}d\epsilon' \sum_{\alpha \beta \gamma \delta} | \left\langle \alpha \gamma | U | \beta \delta \right\rangle |^2 \delta(\epsilon-\epsilon_{\alpha})\delta(\epsilon'-\epsilon_{\gamma})\delta(\epsilon-\omega-\epsilon_{\beta})\delta(\epsilon'+\omega-\epsilon_{\delta})
\end{equation}

Now when the requirements of zero temperature and the Fermi statistics are dropped, this approach still holds when we include the occupation numbers of the states in the obtained result \ref{e-e golden rule3}.

\begin{eqnarray}\label{e-e scattering time}
\frac{1}{\tau_{ee}(\epsilon)}=\frac{4 \pi}{\hbar \nu_0} \int^{\epsilon}_0 d\omega \int^{0}_{-\omega}d\epsilon' (f_{\epsilon'}(1-f_{\epsilon-\omega})(1-f_{\epsilon'+\omega})+(1-f_{\epsilon'})f_{\epsilon-\omega}f_{\epsilon'+\omega})W^2(\omega)
\end{eqnarray}

Where

\begin{equation}
W^2(\omega)=\sum_{\alpha \beta \gamma \delta} | \left\langle \alpha \gamma | U | \beta \delta \right\rangle |^2 \delta(\epsilon-\epsilon_{\alpha})\delta(\epsilon'-\epsilon_{\gamma})\delta(\epsilon-\omega-\epsilon_{\beta})\delta(\epsilon'+\omega-\epsilon_{\delta}).
\end{equation}

To complete the collision term for electron-electron interactions it is convenient to let go the notation of Ref. \cite{akkermans} and proceed with the notation used for electron-phonon interactions. We define the kernel $K(\epsilon)$, which follows from $(4\pi)/(\hbar \nu_0)W^2(\omega)$. Further the collision rate can be splitted in the inscattering and outscattering term by multiplying the first part by $f$ and the second part by $1-f$.

\begin{equation}\label{e-e inscattering}
I^{in}_{ee}(x,E,[f])=\int d\epsilon \int dE' K(\epsilon)f(x,E-\epsilon)f(x,E'+\epsilon)(1-f(x,E))(1-f(x,E'))
\end{equation}

\begin{equation}\label{e-e outscattering}
I^{out}_{ee}(x,E,[f])=-\int d\epsilon \int dE' K(\epsilon)f(x,E)f(x,E')(1-f(x,E-\epsilon))(1-f(x,E'+\epsilon))
\end{equation}

In Ref. \cite{altshuler} and Ref. \cite{anthore} the matrix element of the transition in a disorded medium is calculated. Here we will not follow the complete derivation, but directly look at the result for the kernel $K(\epsilon)$.

\begin{equation}\label{ee kernel}
K(\epsilon)=\frac{\nu_F}{4\pi^4 \hbar^3}\int d\textbf{q}|U_{\epsilon/\hbar}(\textbf{q})|^2\left(\frac{D\textbf{q}^2}{D^2\textbf{q}^4+(\epsilon/\hbar)^2}\right)^2
\end{equation}

The bare Coulomb potential $U_0(\textbf{q})$ and the polarizability $\Pi(\textbf{q},\epsilon/\hbar)$ of the electron fluid determines the screened Coulomb potential $U_{\epsilon/\hbar}(\textbf{q})$ effectively experienced by the quasi-particles.

\begin{equation}\label{screened coulomb potential}
U_{\epsilon/\hbar}(\textbf{q})=\frac{U_0(\textbf{q})}{1+\Pi(\textbf{q},\epsilon/\hbar)U_0(\textbf{q})}
\end{equation}

where

\begin{equation}\label{polarizability}
\Pi(\textbf{q},\epsilon/\hbar)=\nu_F\frac{D\textbf{q}^2}{D\textbf{q}^2-i\epsilon/\hbar}.
\end{equation}

In a metal the density of states $\nu_F$ is so large (order of $10^{47} J^{-1}m^{-3}$) that the polarizability dominates the denominator in the expression of the screened Coulomb potential. Therefore equation \ref{screened coulomb potential} simplifies to

\begin{equation}\label{simplified screened coulomb potential}
U_{\epsilon/\hbar}(\textbf{q})=\frac{1}{\Pi(\textbf{q},\epsilon/\hbar)},
\end{equation}

and the total kernel becomes

\begin{equation}
K(\epsilon)=\frac{1}{4\pi^4 \nu_F \hbar^3} \int \frac{d\textbf{q}}{D^2\textbf{q}^4+(\epsilon/\hbar)^2}.
\end{equation}

If we consider a metallic wire with cross-section $S=wt$, where $w$ is the width and $t$ is the thickness of the wire, only the uniform modes in transverse dimensions contribute to $K(\epsilon)$ if the energies $\epsilon$ are smaller than $\hbar D/max(w^2,t^2)$. This leads to

\begin{equation}\label{final ee kernel}
K(\epsilon)=\left(\sqrt{2D} \pi \hbar^{3/2} \nu_F S\right)^{-1}\epsilon^{-3/2}
\end{equation}

This derivation leads to a difference with the result for the screened Coulomb interactions obtained by Kamanev and Andreev \cite{kamenevandreev}. They found $K(\epsilon)$ to be a factor 2 larger. Experiments showed that the energy dependence of the collision term is accurate, but the intensity is off. A discussion can be found in Ref. \cite{huard} and Ref. \cite{huard2}.

\section{\label{summery_chap4}Summary}

In this chapter we used the fact that the electrons involved in the ac quantum transport in a diffusive wire can be described as quasi-particles according to the Fermi liquid theory. For coherent transport the energy distribution of the quasi-particles obeys a relative simple quantum diffusion equation. The non-equilibrium in a mesoscopic, diffusive wire induced by a time-dependent field manifests itself in the energy distribution. When the length of the wire is extended, the transport becomes incoherent and the redistribution of energy among the quasi-particles has to be evaluated. For this reason the relative simple quantum diffusion equation is extended with a collision integral accounting for electron-electron and electron-phonon interactions. In the next chapter the model is evaluated using numerical calculation methods.


\chapter{\label{chapter5}Numerical results}

\section{\label{intro_sim}Introduction}

The model developed in chapter \ref{chapter4} allows the evaluation of the quasi-particle energy distribution in a mesoscopic wire ac biased with irradiation. For very short wires, where the phase coherence time and energy relaxation time exceed the diffusion time, the transport is fully coherent and the distribution function in the wire is never an equilibrium function. The non-equilibrium description is quite different in the two field limits, $\omega \tau_D<<1$ and $\omega \tau_D>>1$, as discussed in section \ref{limitcases}. In the slow field limit ($\omega \tau_D<<1$) the quasi-particle energy distribution is varying in time, following the oscillation of the field instantaneously. In the limit $eV>>\hbar \omega$ this shows close resemblance with the dc biased wire and the quasi-particle energy distribution is given by a two step function which varies in time. The fast field limit ($\omega \tau_D>>1$) is quite different. In this limit the quasi-particle energy distribution is given by a time-independent multiple step function. For energies $eV>>\hbar \omega$ the steps smooth out and a continuous function is obtained which provides a finite probability of finding a quasi-particle far from the Fermi energy.

To evaluate the slow field regime and fast field regime we can define some ratio $\hbar \omega / eV$ and vary the product $\omega \tau_D$. Shytov showed that the crossover from low-frequency behavior to high-frequency behavior occurs at $\omega \tau_D \approx 100$ \cite{shytov}. This is due to the fact that that the quasi-particle energy distribution relaxes at $t \rightarrow \infty$ as $exp(-\mu t)$, where $\mu=\pi^2/\tau_D$ is the lowest non-zero eigenvalue of the diffusion operator. It is reasonable to assume that the crossover occurs when the relaxation time is of the order of the field period, $2\pi/\omega$. So the crossover is estimated to occur at $\omega \tau_D \propto 2 \pi^3 \approx 62$, which is close to 100.

This theoretical research is done in an experimental research group. The strong connection with experimental physics leads to the desire to evaluate the model for realistic situations (THz frequencies and field amplitudes of 1-20 meV), so that when an experimental setup is realized the model can provide the understanding of the experimental results. We apply these conditions in the evaluation of equation \ref{qde_inel} using numerical calculation methods. The equation is expressed in dimensionless parameters in the same way we did for the discussed limit situations of coherent transport.

\begin{equation}\label{dimensionless qde_inel}
\left\{\frac{\partial}{\partial t}-\frac{1}{\omega \tau_D}\left[\frac{\partial}{\partial x}-sin(t)D_{\omega}\right]^2\right\}f(t,E,x)=\frac{I_{tot}f(t,E,x)}{\omega}.
\end{equation}

As explained in the previous chapter, the collision term can be neglected for fully coherent transport. The energy of the quasi-particles in the wire is only affected by photon absorption and the diffusive transport itself. This means that the neglect of the collision term is only valid for short wires. To make this somewhat more quantitative, we consider the phase coherence time of a quasi-particle. Two phase breaking mechanisms are distinguished, electron-phonon interaction and electron-electron interaction. The experimental part of the research focuses on aluminum wires with a diffusion coefficient of about 100 cm$^2$s$^{-1}$ measured at liquid helium temperatures, so that we first concentrate on this material and temperature. Above temperatures of 1 K the phase breaking mechanism is electron-phonon interaction. The phase coherence time is approximated by \cite{huard}:

\begin{equation}\label{phase coherence time ph}
\frac{1}{\tau^{(e-ph)}_{\phi}}=\frac{7 \pi \zeta(3)}{9}\frac{E^2_FN(0)k^3_b}{\hbar^3\rho s^4 k^2_F}T^3.
\end{equation}

Here $E_F$ is the Fermi energy, $N(0)$ is the density of states at Fermi energy, $\rho$ is the mass density, $s$ is the speed of sound and $k_F$ is the Fermi wave vector. The phase coherence time at 2 K, which can be achieved in a pumped liquid helium cryostat, is approximately 10 ns. This is equivalent to a wire of length $L=\sqrt{D \tau_{\phi}}=10 \mu$m. So for wires shorter than this length the transport is coherent. Since this is an approximation we decided to use in our calculations wires of maximum length of 7 $\mu$m, with a diffusion time of 5 ns, to be certain that the transport is coherent. We evaluate the quasi-particle energy distribution for coherent transport at 2 K from the slow field regime to the fast field regime. For the slow field regime $\omega \tau_D=1$ we choose a wire of 56 nm and a field frequency of 0.5 THz. In the fast field regime $\omega \tau_D=30000$ we take a wire of 7 $\mu$m and a field frequency of 2 THz. A wire of 400 nm and a field frequency of 1 THz makes the evaluation of the intermediate regime $\omega \tau_D=100$ possible. We define the ratio $\hbar \omega / eV=0.4$ for all regimes, so the field amplitude varies from 5 meV in the slow field regime to 20 meV in the fast field regime.

For extended wires, the diffusion time can exceed the energy relaxation time, so that the transport is incoherent. In this report we will focus on the fast field regime for incoherent transport. The slow field regime is already quite well understood \cite{pothier} \cite{R.Schrijvers} and thereby the frequency of the field should be extremely low to have a small product $\omega \tau_D$, where $\tau_D$ should be of the order of $\tau_E$ the energy relaxation time. 

The effect of electron-phonon interactions is evaluated at a temperature of 2 K. We determined the phase coherence time for electron-phonon interaction to be 10 ns. This will be our reference in defining the ratio between diffusion time and energy relaxation time, since the energy relaxation time is of the same order of magnitude as the coherence time. The intensity of the interaction between quasi-particle and phonon can be calculated from equation \ref{Eliashberg}. In Ref. \cite{giazotto} the quantity $\Sigma$ related to the power dissipation is given to be about 1 GWm$^{-3}$K$^{-5}$. This brings the intensity to about 2 ns$^{-1}$meV$^{-3}$, which is close to the empirical intensity of 4 ns$^{-1}$meV$^{-3}$ which followed from experiments done by Huard et al. \cite{huard}. 

Below 1 K the situation becomes a bit complicated, since aluminum is no longer a normal metal, but has experienced a phase transition to the superconducting phase. We proceed below 1 K with an undefined material with the same diffusion coefficient of 100 cm$^2$s$^{-1}$, so that we can evaluate the effect of electron-electron interactions on the energy distribution of the quasi-particles. The phase coherence time for electron-electron interactions is approximated with \cite{huard}:

\begin{equation}\label{ee time}
\frac{1}{\tau^{ee}_{\phi}}=\left(\frac{\pi k_{ee} k_b}{2 \sqrt{\hbar}}\right)^{-2/3} T^{-2/3}.
\end{equation}

Here $k_ee$ is the prefactor in the kernel of equation \ref{final ee kernel} and given by $\left(\sqrt{2D} \pi \hbar^{3/2} N(0) S\right)^{-1}$.  A temperature of 500 mK and a cross-section of the wire of 400 nm$^2$ gives an intensity of the interactions of 0.8 ns$^{-1}$meV$^{-1/2}$. The empirical intensity found by Huard et al. in silver is 0.4 ns$^{-1}$meV$^{-1/2}$, so this can be used as a realistic value. This leads to a phase coherence time of about 1 ns at 500 mK. The energy relaxation time is of the same order of magnitude.

So the purpose of this chapter is dual. First, we want to investigate the quasi-particle energy distribution for coherent transport and how the slow field regime differs from the fast field regime. Second, we want to investigate the quasi-particle energy distribution for incoherent transport in the fast field regime and how weak interactions are distinguished from strong interactions.

\section{\label{calcmethod}Calculation method}

Numerical calculation principles allow the evaluation of the quantum diffusion equation \ref{dimensionless qde_inel} \cite{dahlquist}. We use Euler's method using finite difference approximations for the space, time and energy variables. By iterating the calculation a stable solution for the quasi-particle energy distribution is obtained. This iteration is performed on the time variable, so that every time step $dt$ results in a new function which arises from the old function and the non-time operation part of the equation:

\begin{equation}\label{fnew}
f_{new}=f_{old}+\frac{dt}{\omega \tau_D}df
\end{equation}

where

\begin{equation}\label{df}
df=(D_x)^2f_{old}+2sin(m d t)D_x D_E f_{old}+sin^2(m d t)(D_E)^2 f_{old}+\tau_D(Iin-Iout).
\end{equation}

Here $D_{x,E}$ is the finite difference operator for space and energy, respectively, and $m$ is the number of iteration. For the diffusion equation without inelastic scattering the terms $Iin$ and $Iout$ in equation \ref{df} disappear. The finite difference operators are sparse matrices, which means that the percentage of zero elements greatly exceeds the percentage of non-zero elements and their distribution is such that it is advantageous to use this for a more efficient calculations. The MATLAB function \textit{sparse} provides the possibility to exploit the sparse nature of the operator. What this function does is isolate the non-zero elements, so that only these elements are used in the calculation. The MATLAB code of the simulation program can be found in the appendix.

\section{\label{sim}Simulation of realistic coherent and incoherent transport situations}

\subsection{\label{coherent_transport}Coherent transport in diffusive wires}

\begin{figure}[h!]
    \centering
    \includegraphics[width=0.8\textwidth]{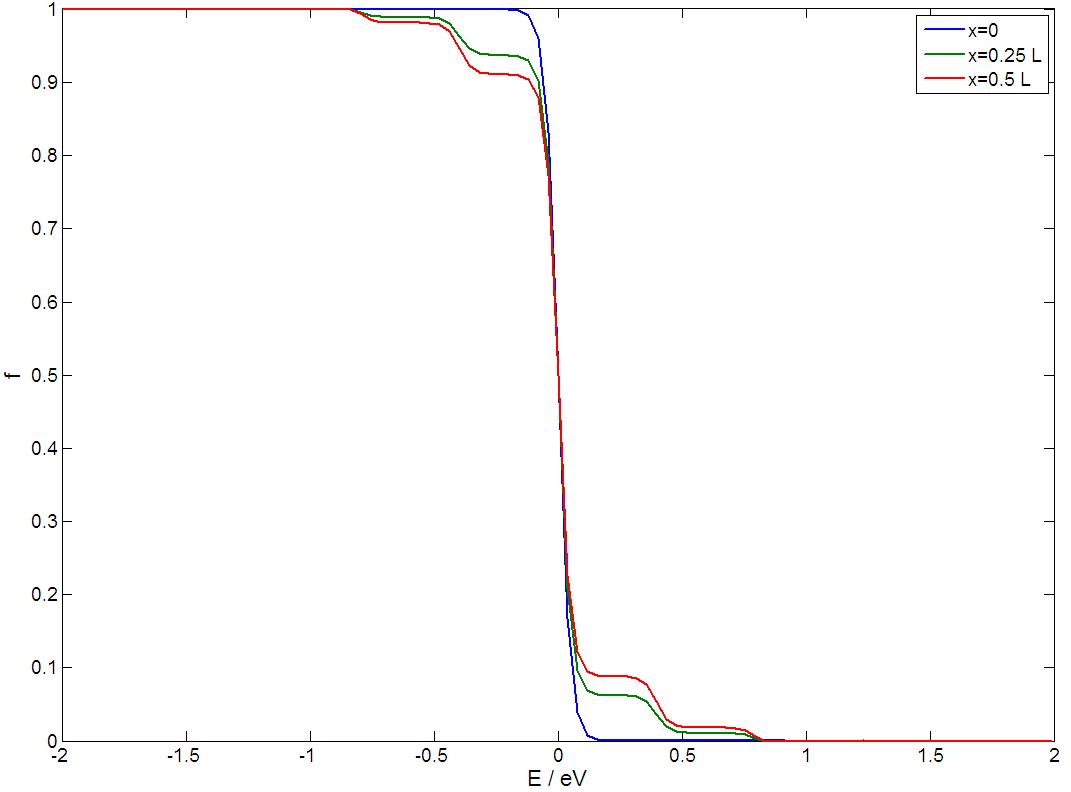}
    \caption{The quasi-particle energy distribution in the slow field regime, $\omega \tau_D=1$, and $\hbar \omega / eV=0.4$ at 3 different positions in the wire at 2 K.}
    \label{figure:timeaverslowfield}
\end{figure}

As explained in the previous section, the crossover from low-frequency behavior to high-frequency behavior occurs at $\omega \tau_D=100$. So when we want to evaluate the slow field regime it is sufficient to have a product $\omega \tau_D=1$, which is two orders of magnitude below the crossover. As said before this is based on realistic values, but throughout this section we will only work with relative values. The amplitude of the field is such that $\hbar \omega/eV=0.4$. In the slow field regime the energy distribution is highly time-dependent. The time-averaged distribution function is given in figure \ref{figure:timeaverslowfield} for three different positions in the wire. The full space dependency is shown in the three dimensional figure in appendix \ref{appendix_f}.

\begin{figure}[h!]
    \centering
    \includegraphics[width=1.0\textwidth]{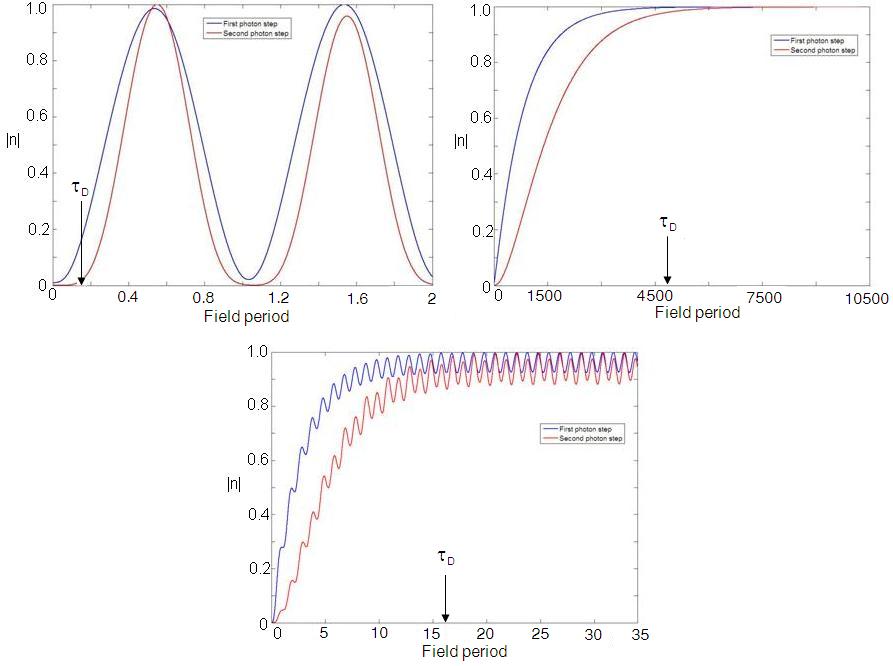}
    \caption{The time-evolution of the occupation in the photon steps for the three frequency regimes: up left the low-frequency regime $\omega \tau_D=1$, up right the high-frequency regime $\omega \tau_D=30000$ and down in the middle the crossover $\omega \tau_D=100$. The blue line gives the normalized occupation at the first photon step and the red line at the second photon step.}
    \label{figure:timedependency}
\end{figure}

The time-dependency in the slow field regime is evaluated by running the simulation during two field periods and plot the normalized occupation at the photon steps. This normalization is performed by taking the value for every iteration on the first and second photon step, determine the maximum value during the iteration process and divide the value determined in every iteration by this maximum value: $|n_i|=f_i(E+\hbar \omega/2, E+3\hbar \omega/2)/max(f_i(E+\hbar \omega/2, E+3\hbar \omega/2))$. The first photon step immediately follows the field, where the second photon step shows a slight delay as shown in the up left picture in figure \ref{figure:timedependency}.

\begin{figure}[h!]
    \centering
    \includegraphics[width=0.8\textwidth]{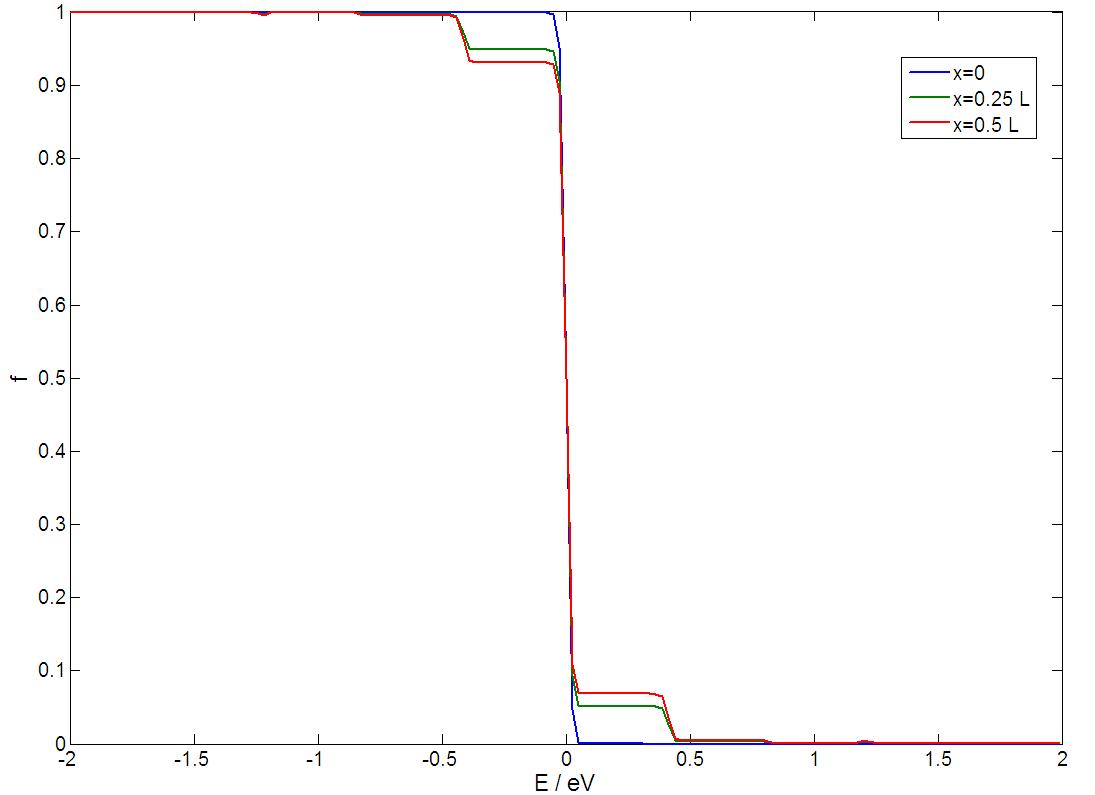}
    \caption{The quasi-particle energy distribution in the fast field regime for $\omega \tau_D=30000$ and $\hbar \omega / eV=0.4$ at 3 different positions in the wire at 2 K.}
    \label{figure:fastfieldregime}
\end{figure}

For the fast field regime we take $\omega \tau_D=30000$, two orders of magnitude above the crossover from low-frequency behavior to high-frequency behavior. The amplitude of the field is again defined so that $\hbar \omega/eV=0.4$. The simulation of this situation is time-averaged depicted in figure \ref{figure:fastfieldregime} at three different positions in the wire, where we average over a large number of periods to obtain the final time-independent distribution. The full space dependency is shown in the three dimensional figure in appendix \ref{appendix_f}. To evaluate the time-dependence in the fast field regime the normalized value of the occupation in the three photon steps in the distribution is plotted during the evolution of the function. It appears that in the fast field regime the energy distribution indeed becomes time-independent as shown in the up right picture in figure \ref{figure:timedependency} and the occupation at the photon energies is maximum when the diffusion time is reached.

The intermediate frequency regime where the crossover occurs from low-frequency behavior to high-frequency behavior is evaluated at $\omega \tau_D=100$. The time-averaged distribution function at three different positions is given in figure \ref{figure:crossover}. The full space dependency is shown in the three dimensional figure in appendix \ref{appendix_f}. The time-dependency in the distribution function is drastically decreased at the crossover as seen in figure \ref{figure:timedependency}. When the diffusion time is reached, the occupation at the photon energies is at maximum with a slight oscillatory deviation with the field period.

\begin{figure}[h!]
    \centering
    \includegraphics[width=0.8\textwidth]{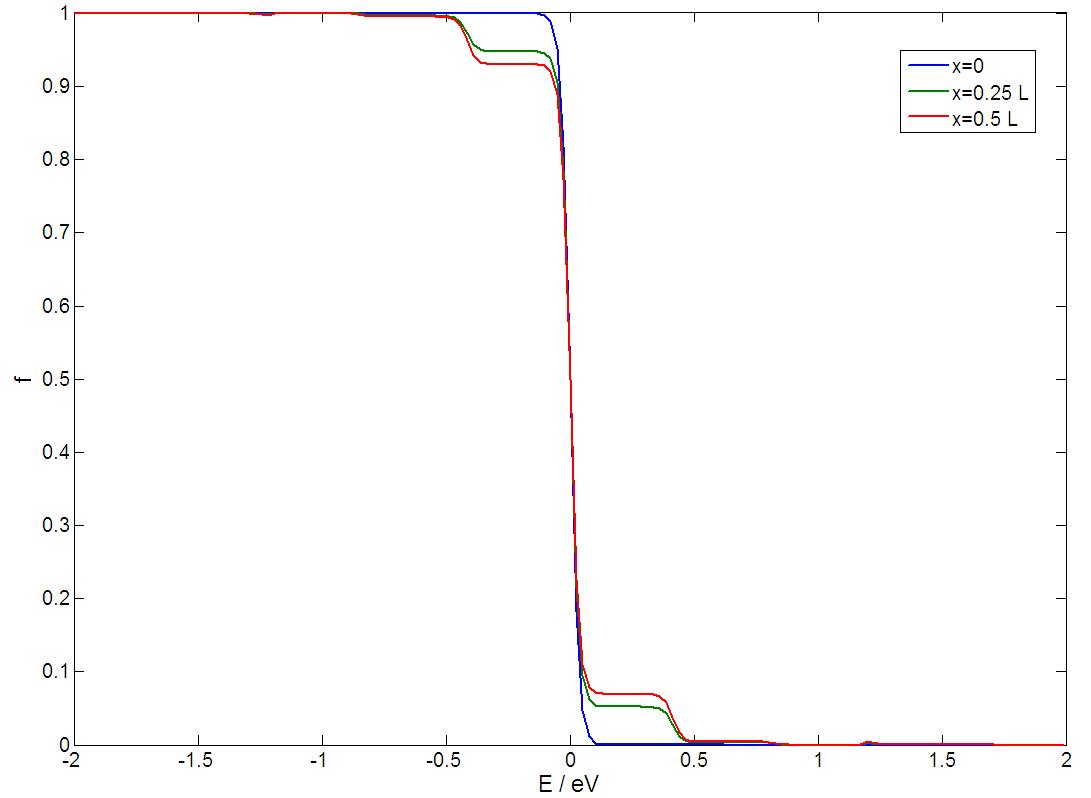}
    \caption{The quasi-particle energy distribution in the intermediate frequency regime for $\omega \tau_D=100$ and $\hbar \omega / eV=0.4$ at 3 different positions in the wire at 2 K.}
    \label{figure:crossover}
\end{figure}

\subsection{\label{incoherent_transport}Incoherent transport in diffusive wires}

As seen in the previous section the absorption of field quanta in the short wire, where the transport of quasi-particles is fully coherent, induces a staircase structure in the quasi-particle energy distribution. Now we want to investigate what happens when the wire is extended, so that photon-absorption is no longer the only mechanism that affects the energy distribution, but also interactions between quasi-particles and between quasi-particles and phonons come into play. It appears that these interactions redistribute the quasi-particles with respect to the energy, so that the occupation of the energy levels is changed with respect to the occupation in the coherent situation. The effect of the two phase breaking mechanisms is quite different. We expect that the interactions between quasi-particles cause a smearing in the staircase structure, while the interactions between quasi-particles and phonons cause the annihilation of the photon steps and finally, in the strong interaction limit, leave a Fermi function with the bath temperature. In this section we will limit ourself to the fast field regime which is, as explained in the previous section, the most interesting domain.

\begin{quotation}
~
\end{quotation}

\textbf{Electron-phonon interactions}

Since the first experiments are planned to be done at liquid helium temperatures, we first focus on the effect of electron-phonon interactions on the energy distribution of the quasi-particles. We can distinguish different interaction regimes. The weak interaction regime is found for $\tau_D \approx \tau_E$ and the strong interaction regime is found for $\tau_D >> \tau_E$.

\begin{figure}[h!]
    \centering
    \includegraphics[width=0.8\textwidth]{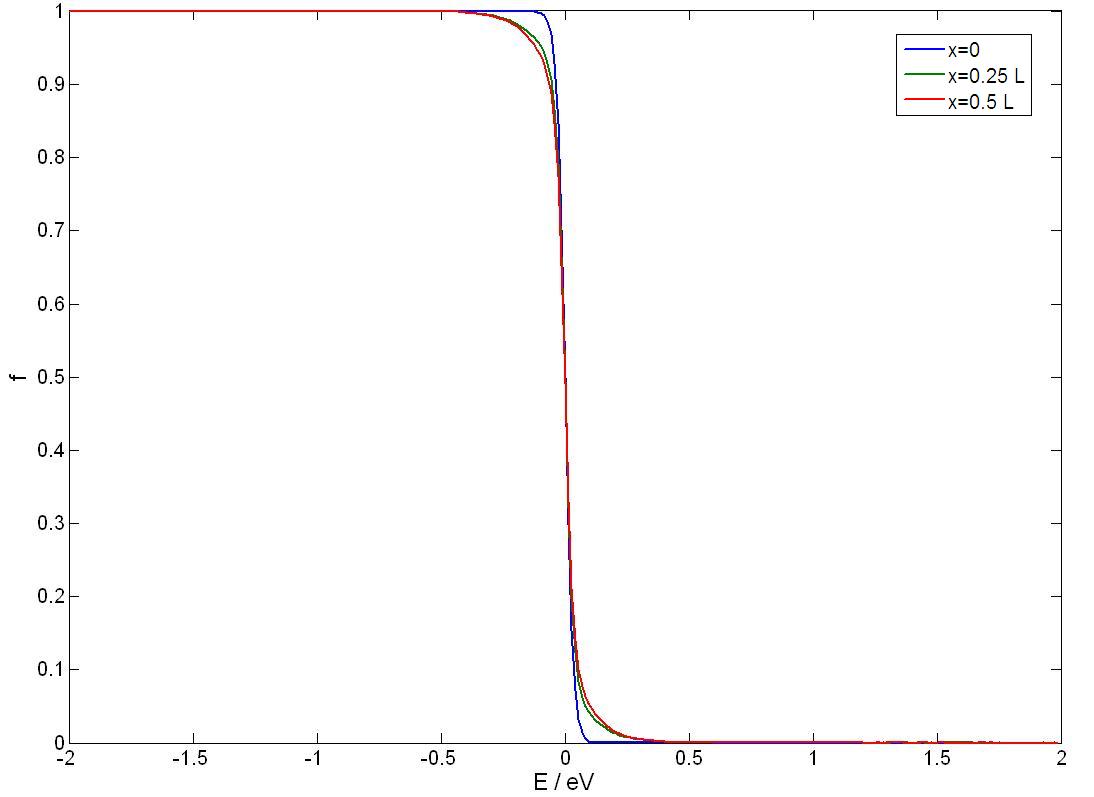}
    \caption{The quasi-particle energy distribution in the fast field, weak electron-phonon interaction with $\omega \tau_D=50000$, $\hbar \omega / eV=0.4$ and $\tau_D \approx \tau_E$ at 3 different positions in the wire at 2 K.}
    \label{figure:fastfield_weakint}
\end{figure}

Let's first look at the weak interaction regime where the diffusion time is of the order of the energy relaxation time, so we define $\omega \tau_D=50000$ and $\tau_D \approx \tau_E$. What we expect is that the energy gained from the field by a quasi-particle is redistributed, where the photon steps due to the absorption of multiple field quanta are first influenced. The result of this simulation is shown in figure \ref{figure:fastfield_weakint}. The expected disappearance of the photon steps due to the absorption of multiple field quanta is indeed observed and the transition in the first photon step is smoothed. The space dependency is shown in the three dimensional figure in appendix \ref{appendix_f}. 

\begin{figure}[h!]
    \centering
    \includegraphics[width=0.8\textwidth]{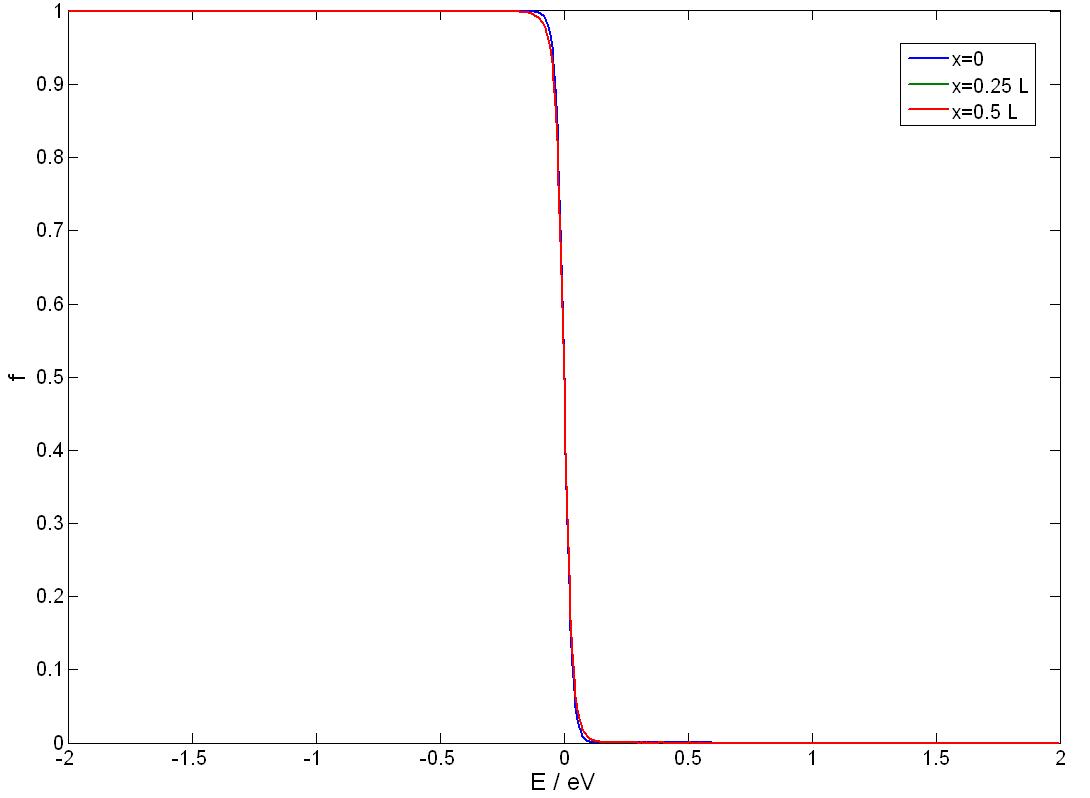}
    \caption{The quasi-particle energy distribution in the fast field, strong electron-phonon interaction with $\omega \tau_D=10^7$, $\hbar \omega / eV=0.4$ and $\tau_D \approx 200 \tau_E$ at 3 different positions in the wire at 2 K.}
    \label{figure:fastfield_strongint}
\end{figure}

When the length of the wire is increased, the diffusion time becomes much higher than the energy relaxation time. So we enter the strong interaction regime and define $\omega \tau_D=10^7$ and $\tau_D \approx 200 \tau_E$. In the fast field regime we expect a Fermi function at the bath temperature on every position in the wire. The result, shown in figure \ref{figure:fastfield_strongint}, indeed shows a Fermi function. There is a slight deviation from the Fermi function with the bath temperature.

When we look at the deviation of the calculated energy distributions from the equilibrium function at bath temperature we see what the effect of weak and strong interactions is. For weak interactions the deviation is clearly defined by the photon energy, but the photon step is smoothed. For strong interactions the deviation is no longer defined by the photon energy and the width and height of the peak is small. The height of the peak is in both situations however for energies below Fermi energy somewhat larger. This observed deviation is probably caused by the discretization of the variables and the fact that the interactions are calculated after $n$ iterations, instead for each iteration, to increase calculation speed. It is reasonable to believe that this has no physical meaning, but is just some numerical error which can be solved by solving the equations with a program written in C. This should provide a much higher calculation speed, so that the discretization can be optimized.

\begin{figure}[h!]
    \centering
    \includegraphics[width=1\textwidth]{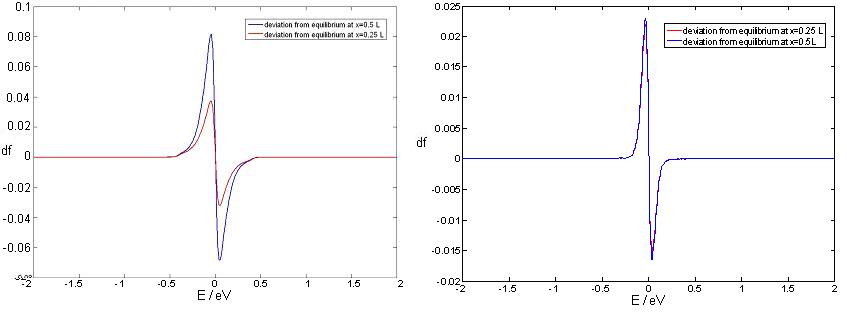}
    \caption{The deviation from the equilibrium function of the bath temperature in the fast field regime for left weak electron-phonon interactions ($\tau_D \approx \tau_E$) and right strong electron phonon interactions ($\tau_D \approx 200 \tau_E$). The red line gives the deviation in at $x=0.25 L$ and the blue line at $x=0.5 L$. For weak interactions the effect of the photon step is clearly visible at 0.4 which stems with the defined ratio $\hbar \omega /eV$. For strong interactions the width and height of this step is drastically decreased.}
    \label{figure:deviation_eph_int}
\end{figure}

\begin{quotation}
~
\end{quotation}

\textbf{Electron-electron interactions}

\begin{figure}[h!]
    \centering
    \includegraphics[width=1\textwidth]{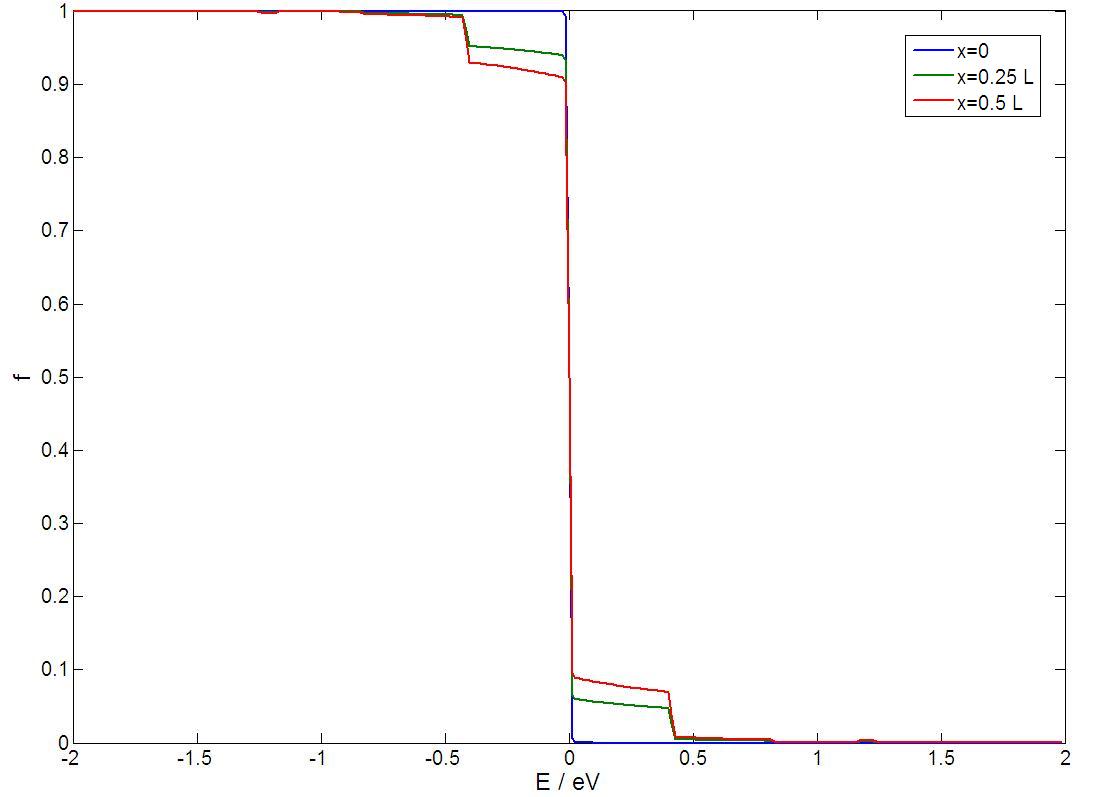}
    \caption{The quasi-particle energy distribution in the fast field, weak interaction regime with $\omega \tau_D=10000$, $\hbar \omega / eV=0.4$ and $\tau_D \approx  \tau_E$ at 3 different positions in the wire at 500 mK.}
    \label{figure:fastfield_weak_eeint}
\end{figure}

The effect of electron-electron interactions is quite different from the effect of electron-phonon interactions and dominant for lower temperature so we will evaluate this effect at a temperature of 500 mK. Lets first look at the effect of weak interactions when the diffusion time is of the order of the relaxation time, $\tau_D \approx \tau_{E}$ and $\omega \tau_D=10000$. Figure \ref{figure:fastfield_weak_eeint} shows that for weak interactions there is some smearing, but the photon steps are still good defined. The full space dependency is depicted in the three dimensional figure in appendix \ref{appendix_f}.

\begin{figure}[h!]
    \centering
    \includegraphics[width=1\textwidth]{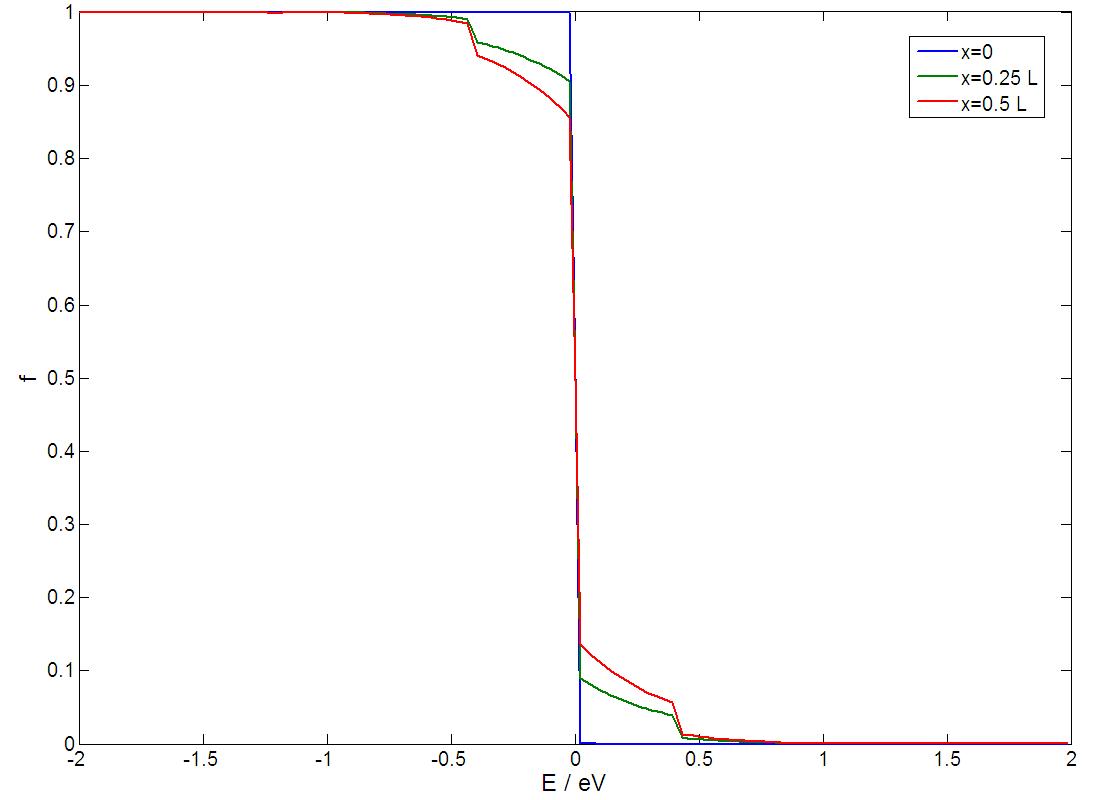}
    \caption{The quasi-particle energy distribution in the fast field, weak interaction regime with $\omega \tau_D=75000$, $\hbar \omega / eV=0.4$ and $\tau_D \approx 7.5 \tau_E$ at 3 different positions in the wire at 500 mK.}
    \label{figure:fastfield_weak_ee2int}
\end{figure}

When we increase the length of the wire, the time that the electron spends traveling through the wire increases also and the effect of the interactions becomes more significant. Figure \ref{figure:fastfield_weak_ee2int} shows that the interactions indeed are more relevant for longer wires and the smearing in the photon steps is clearly visible.

\begin{figure}[h!]
    \centering
    \includegraphics[width=1\textwidth]{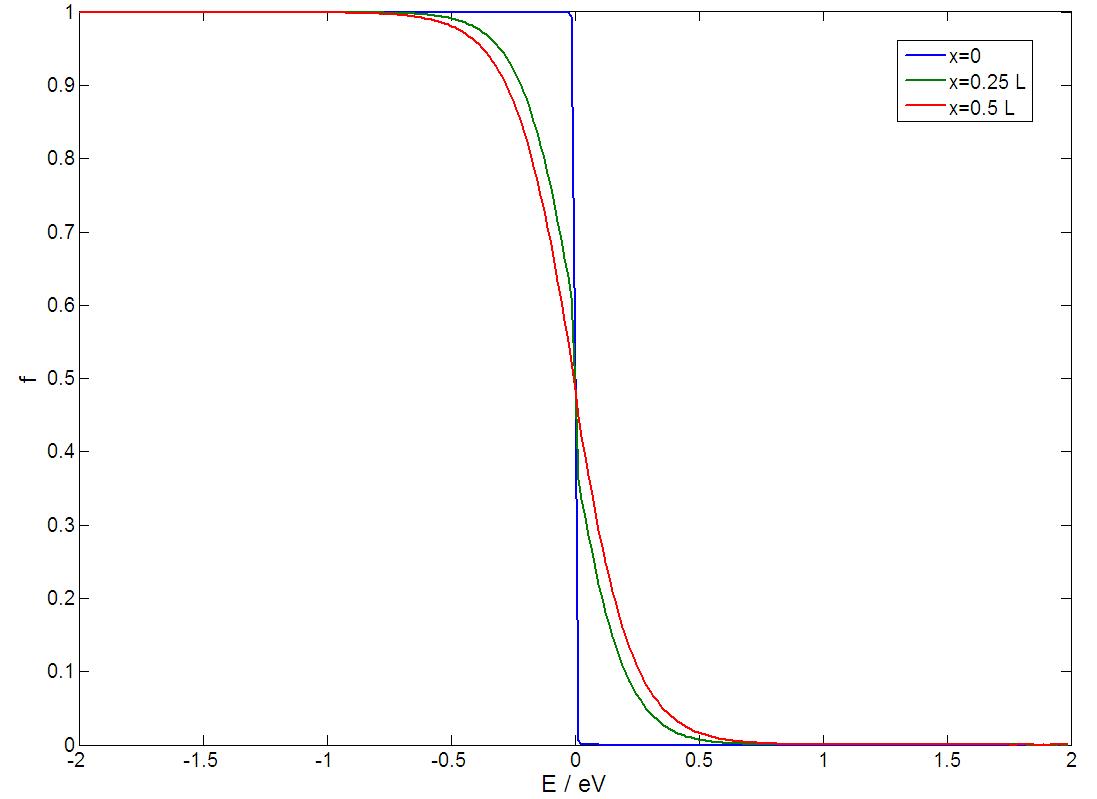}
    \caption{The quasi-particle energy distribution in the fast field, strong electron-electron interaction regime with $\omega \tau_D=2000000$, $\hbar \omega / eV=0.4$ and $\tau_D \approx 200 \tau_E$ at 3 different positions in the wire at 500 mK.}
    \label{figure:fastfield_strong_eeint}
\end{figure}

\begin{figure}[h!]
    \centering
    \includegraphics[width=1\textwidth]{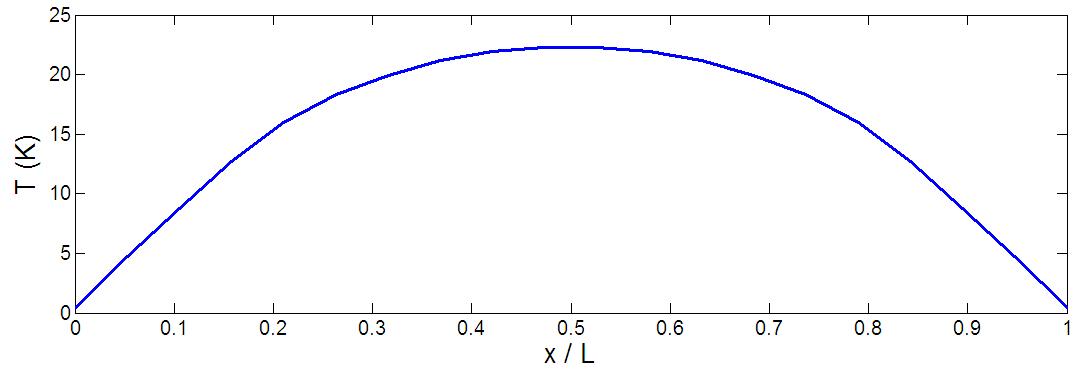}
    \caption{The effective temperature profile across the wire in the case of strong interactions.}
    \label{figure:Tprofile_strong_eeint}
\end{figure}

By further increasing the length of the wire, the energy relaxation rate becomes dominant with respect to the diffusion time. The energy gained from the electric field is redistributed in a Fermi function with an effective temperature, so that an effective temperature profile arises across the wire analogously to the dc biased macroscopic wire evaluated in the introduction. In figure \ref{figure:fastfield_strong_eeint} three Fermi functions are given at different positions in the wire for $\omega \tau_D=2000000$ and $\tau_D \approx 200 \tau_E$ at 500 mK. Figure \ref{figure:Tprofile_strong_eeint} shows the effective temperature profile across the wire obtained by fitting the energy distribution on every position in the wire to a Fermi function using a least square method. The distribution function on every position in the wire is depicted in the three dimensional figure in appendix \ref{appendix_f}.

\begin{figure}[h!]
    \centering
    \includegraphics[width=1\textwidth]{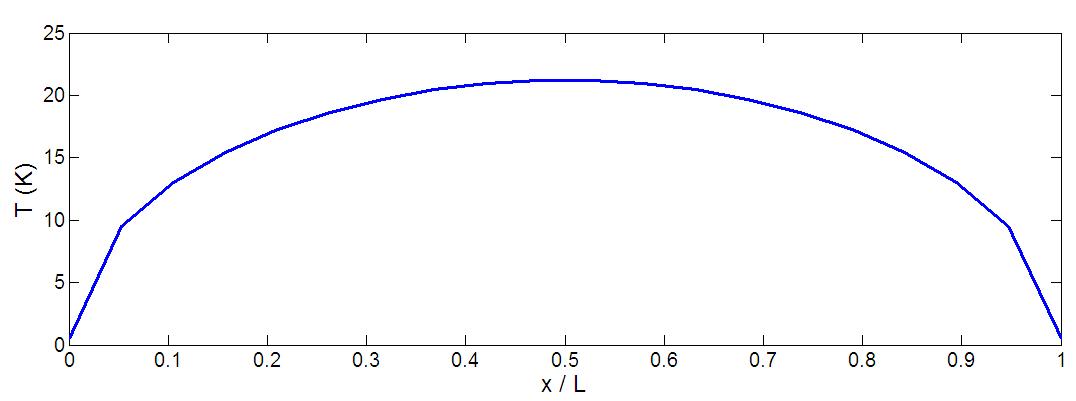}
    \caption{The theoretical predicion of the effective temperature profile across the wire in the case of strong interactions with $V=\hbar \omega / e$.}
    \label{figure:theor_Tprof}
\end{figure}

Figure \ref{figure:theor_Tprof} shows the theoretical prediction for the effective temperature profile given by $T_e(x)=\sqrt{T^2+\frac{x}{L}\left(1-\frac{x}{L}\right)3/\pi^2\left(e/k_b\right)^2V^2}$ when the voltage across the wire is taken to be $V=\hbar \omega / e$. It shows resemblance with the effective temperature profile resulting from the simulation. It seems plausible to say that the observed difference is due to numerical inaccuracy. Another possible reason for this observed deviation could come from the fact that we look at the fast field regime and the position of photon absorption is responsible for the difference. However, a closed statement on this calls for further study. 

\begin{figure}[h!]
    \centering
    \includegraphics[width=1\textwidth]{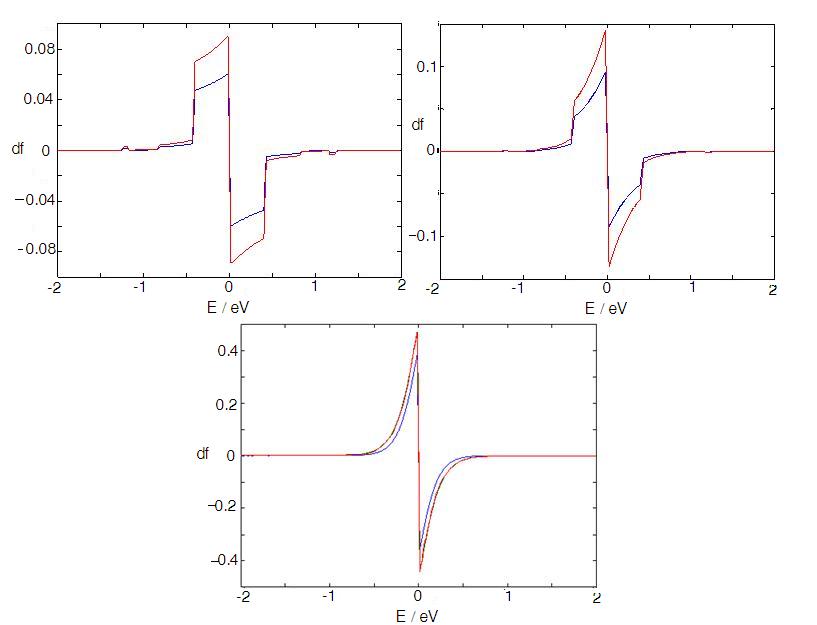}
    \caption{The deviation from a equilibrium function at bath temperature for up left weak interactions ($\tau_D \approx \tau_E$), up right weak interactions ($\tau_D \approx 7.5 \tau_E$) and down in the middle strong interactions ($\tau_D \approx 200 \tau_E$). The red line gives the deviation at $x=0.25 L$ and the blue line at $x=0.5 L$. The interactions cause a smearing ending in a local equilibrium with an effective temperature.}
    \label{figure:deviation_ee_int}
\end{figure}

In figure \ref{figure:deviation_ee_int} the deviation from the equilibrium function at bath temperature is shown. It appears that the smearing induced for weak interactions, $\tau_D \approx \tau_E$, causes the smooth deviation of two subtracted Fermi functions at different temperature for strong interactions, $\tau_D >> \tau_E$.

\chapter{\label{chapter6}Probing the quasi-particles energy distribution}

\section{\label{probing eed1}System description}

Obviously it is desirable to be able to obtain experimentally the quasi-particle energy distribution in a diffusive wire subject to high frequency irradiation to test the model. To this end a system is designed which should provide this possibility. An aluminum diffusive wire is connected to large aluminum reservoirs where the quasi-particle can relax to equilibrium. 

These equilibrium reservoirs are designed in such a way that they also function as antenna for the high frequency radiation. This means that the thickness of the metal reservoirs should be larger than the penetration depth of the radiation. The penetration depth can be calculated by $\delta=\sqrt{\frac{2}{\omega \mu \sigma}}$ \cite{lee}. Here $\omega$ is the radial frequency, $\mu$ is the permeability and $\sigma$ is the conductivity. The penetration depth of radiation with a frequency of 1 THz is in aluminum approximately 82 nm \cite{lee}. Based on this number the reservoirs are designed 100 nm thick.

The antenna design is not the subject of this research and since it is a very sophisticated field of science, we did not spend time on calculating field profiles. Instead we used a bow-tie antenna design, so that imperfections in the design matter less due to the broadband character. The bow-tie antenna is designed in such a way that it is self-complementary \cite{stutzman}, by designing the triangles of the bow-tie with a 45 degrees angle with respect to the wire \cite{tomioka}. In this way the system is frequency independent and we are not limited by a capacitive effect. The length of the triangular sides of the antenna are chosen in multiples of the wavelength of irradiation. By choosing the frequency of irradiation we take into account that for pronounced photon steps in the energy distribution the photon energy $\hbar \omega$ should exceed the thermal broadening $k_b T$.

\begin{figure}[h!]
    \centering
    \includegraphics[width=1\textwidth]{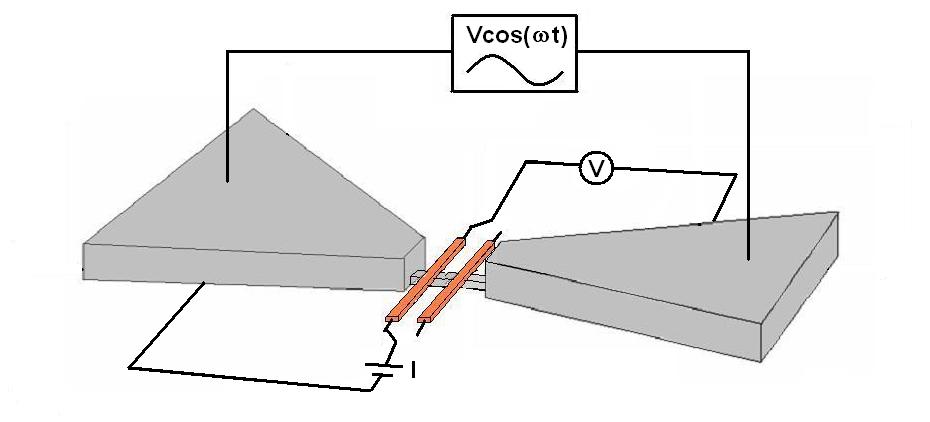}
    \caption{A schematic overview of the system existing of an antenna coupled on an aluminum wire. The niobium probes on the wire enable the differential conductance measurements by setting a current through the NIS junction and measure the voltage drop. The THz radiation is represented by $Vcos(\omega t)$.}
    \label{figure:measure_setup}
\end{figure}

We use the same probing method as Pothier et al. did for their experiments on the quasi-particle distribution function in a dc biased wire. On top of the aluminum wire, an insulating layer is positioned whereon superconducting probes of niobium are placed perpendicular, so that when polarized light is used the probes are not affected. In this way NIS junctions are formed and from differential conductance measurement the quasi-particle energy distribution can be obtained. The photon energy should not exceed the gap energy of the used superconductor, otherwise cooper pairs will be broken and trouble the probing of the energy distribution. Figure \ref{figure:measure_setup} shows a schematic overview of the system where two probes are connected on the wire with one connection to set the current through the NIS junction and one connection to measure the voltage drop over the junction. The THz radiation coupled by the antennas causing the ac bias on the aluminum wire is given by $Vcos(\omega t)$.

First measurement were planned at liquid helium temperature (pumped 2 K - unpumped 4.2 K). The phase coherence time can be calculated using equation \ref{phase coherence time ph} since the dominant phase breaking mechanism is electron-phonon interaction. This approximation is used to define the length of the wires ranging from 500 nm for coherent transport to 100 $\mu$m for fully incoherent transport. When we compare these values with the values stated in the introduction of the previous chapter, we see that we are not measuring in the slow field regime. This has its origin in the fact that we are interested in fields with THz frequencies.

\section{\label{probing eed2}Fabrication}

\begin{figure}[h!]
    \centering
    \includegraphics[width=0.5\textwidth]{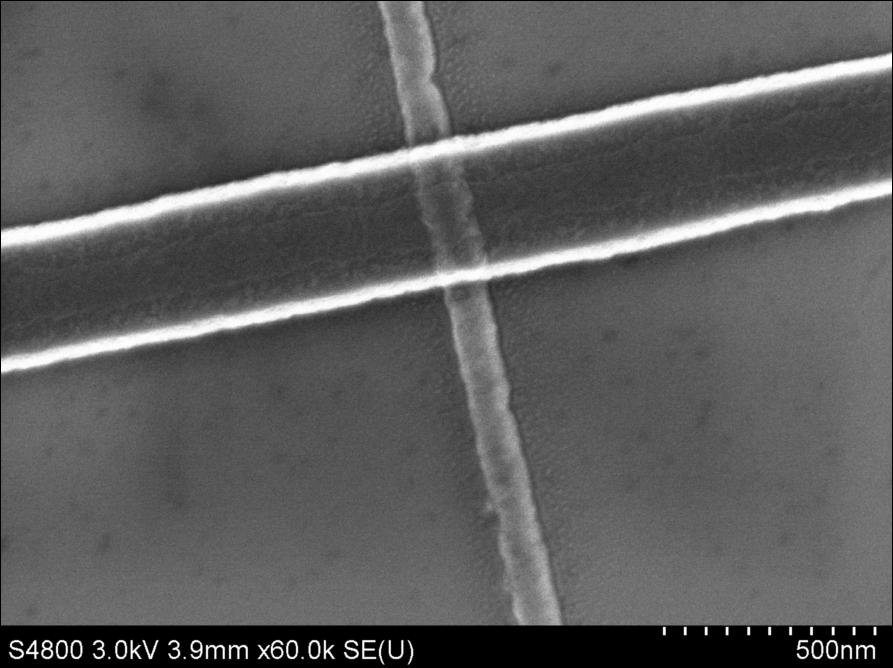}
    \caption{A SEM picture of a NIS junction. Vertical the aluminum wire and horizontal the niobium probe.}
    \label{figure:junction}
\end{figure}

The fabrication of the samples is not yet proven to be fully successful. The samples to test the NIS junctions however are functioning well enough to conclude that with that recipe the NIS junctions provide the ability to probe the distribution function in a wire driven out of equilibrium. Figure \ref{figure:junction} shows a scanning electron microscope (SEM) picture of a NIS junction.

The fabrication of the structures described is done in three steps, in which the nanowires, the antenna and the tunneling probes are defined. The different structures are aligned by means of markers. To define the Al nanowires a double resist layer is patternd with electron beam lithography. The Pmma 950k/Pmma 495k resist is developed in MIBK:IPA for 60 seconds and rinsed for 30 seconds in IPA. A 20 nm thick Al film is evaporated at a rate of 1 angstrom/sec at a pressure of 1e-7 mbar, and lift off is done in hot acetone. The procedure for the definition of the antennas is similar to the nanowires, except for a cleaning step prior to the deposition of the antennas, and an Al film which is 90 nm thick instead of 20 nm. To fabricate the tunneling probes the sample is cleaned for six minutes in an Argon plasma, after which it is oxidized for 40 minutes in a pure Oxygen atmosphere of 1 mbar. A 80 nm thick Niobium film is sputtered in situ, at a pressure of 8e-3 mbar and a rate of 1 nm/sec. An etch mask is created using SAL resist and electron beam lithography, using MF-322 for developing. The Niobium is subsequently etched in an SF$_6$/O$_2$ plasma for 5 minutes with end-point detection. The remaining resist is removed in PRS 3000 resist stripper and the sample is cleaned in acetone. 

\begin{figure}[h!]
    \centering
    \includegraphics[width=1\textwidth]{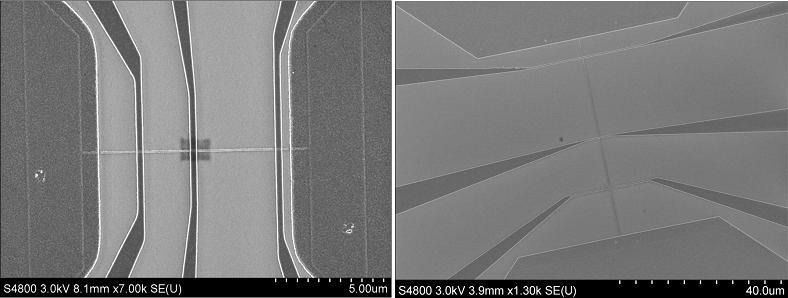}
    \caption{SEM pictures of two samples with wires of different length.}
    \label{figure:antenna_wires}
\end{figure}

Figure \ref{figure:antenna_wires} shows SEM pictures of two samples with wires of different length. Measurements on most recent samples showed an improvement of performance. The outlook for experimental results becomes more promising.

\section{\label{probing eed4}Measurements on NIS junctions}

\subsection{\label{dIdVsection}Differential conductance of a NIS junction}

The energy distribution function of the quasi-particles in the wire can be obtained from conductance measurements of the NIS junctions of the superconducting wires on top of the mesoscopic wire of interest. In appendix \ref{appendix_dIdV} the differential conductance of a NIS junction is derived. The differential conductance is a convolution of the distribution function in the normal wire and the density of states in the superconducting probe. So the unknown distribution function can be obtained by deconvolve the differential conductance with the BCS density of states.

\begin{equation}\label{dIdV_NIS}
\frac{dI}{dV}=-\frac{1}{R_t}\int n_{BCS}(E) f'_x(E-eV)dE
\end{equation}

\begin{figure}[h!]
    \centering
    \includegraphics[width=0.9\textwidth]{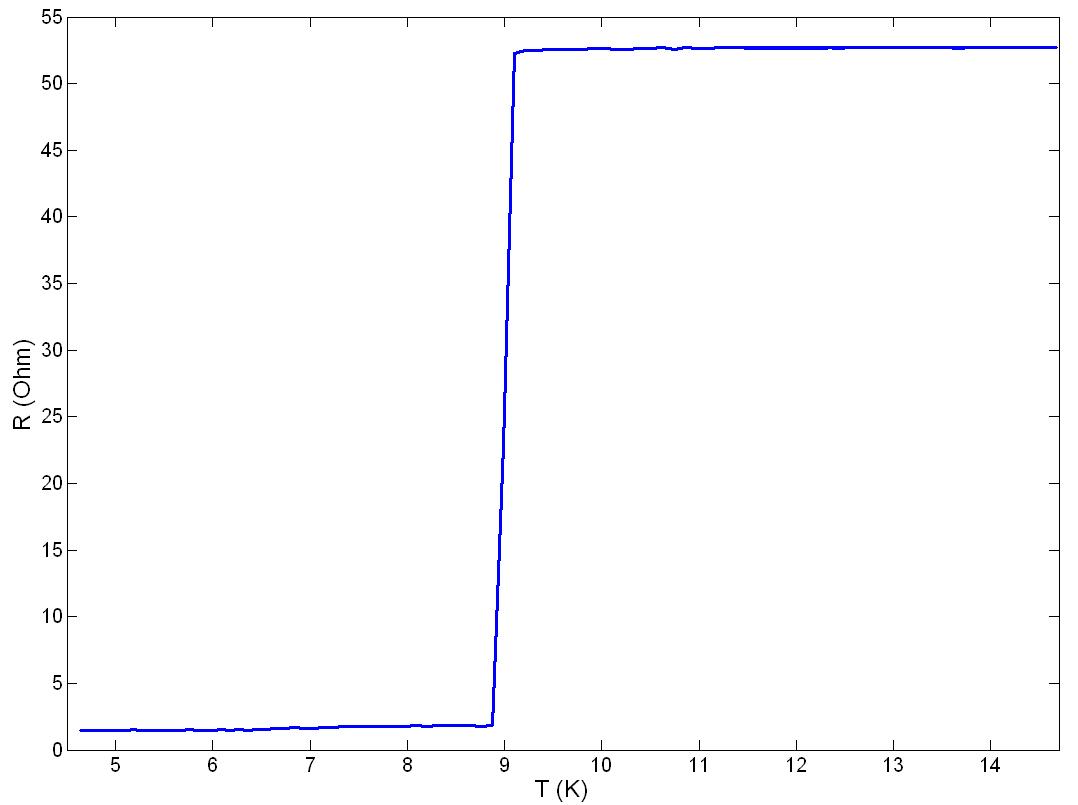}
    \caption{Measured phase transition of the niobium material used for the probes.}
    \label{figure:RT}
\end{figure}

The differential conductance of the NIS junction is only usefull for probing the energy distribution when the material used for the superconductor is indeed superconducting at liquid helium temperatures. The phase transition of niobium is measured by doing a RT-measurement using a dipstick. The sample is mounted in the vacuum tube of the dipstick with a heating resistance connected. By applying a current to this heating resistance the temperature of the sample is increased from 4.2 K to the desired temperature above the critical temperature. So when simultaneously the resistance of a niobium wire is measured using a four point measurement we can obtain the RT-characteristic. The four point measurement is done by setting a current bias to the niobium wire using a current source and measure the voltage drop across the wire. It appears that the niobium indeed becomes superconducting at the expected critical temperature of 9 K. A result of a RT-measurement is shown in figure \ref{figure:RT}.

\begin{figure}[h!]
    \centering
    \includegraphics[width=1.0\textwidth]{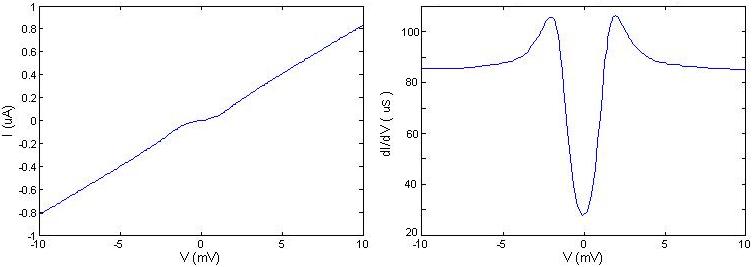}
    \caption{Measured IV and dI/dV of a NIS junction which is a convolution of the quasi-particle energy distribution in the normal wire and the density of states of the superconducting probe.}
    \label{figure:dIdV}
\end{figure}

To test the NIS junctions, differential conductance measurements are performed for wires (not connected to equilibrium reservoirs) with the superconducting wires on top forming the NIS junction. The measurements are performed at liquid helium temperature (4.2 K) using a dipstick, where again the sample is mounted in a vacuum tube. A current source is used to apply a current bias to the NIS junction. By measuring the voltage drop over the junction in a four points measurement setup, the IV characteristic is obtained. Using a lock-in amplifier the differential conductance of this characteristic is determined. The distribution function can be obtained from such measurements on the NIS junction because of the non-linear IV behavior. Figure \ref{figure:dIdV} shows a measurement of the IV-curve and the dI/dV-curve of a NIS junction.

The deconvolution is executed using a steepest descent method. This method is commonly known for the use in minimizing functions \cite{meza}. This is used to deduce the distribution function from the differential conductance using equation \ref{dIdV_NIS}. First the effective density of states of the superconductor in the NIS junction is deduced from a $dI/dV$ measurement on a wire in equilibrium, so that fits for the energy gap, tunneling resistance and electron temperature can be implemented. Then an initial distribution function can be chosen in such a way that the calculation converges in a relative short time \cite{gueron}. The initial distribution function is used to calculate the differential conductance and this is compared to the measured differential conductance for a wire out of equilibrium. If the difference is equal or smaller than the desired precision the deconvolution is completed and the initial distribution is the distribution of the electrons in the wire out of equilibrium. If the difference between the two $dI/dV$'s is larger than the desired precision a new distribution is calculated using the square deviation:

\begin{equation}\label{chi}
\chi^2=\sum_k\left(\left.\frac{\partial I}{\partial V}\right|^k_{calc}-\left. \frac{\partial I}{\partial V}\right|^k_{meas}\right)^2
\end{equation}

The occupation probability at each energy in the distribution is incremented by the partial derivative of the square deviation with respect to the occupation factor $f^k$ at that energy:

\begin{eqnarray}\label{iterate_f}
\nonumber
f^k_{it+1} & = & f^k_{it}+\lambda \frac{\partial \chi^2}{\partial f^k}\\
& = & f^k_{it}+\lambda'n'_{BCS}\left(\left.\frac{\partial I}{\partial V}\right|^k_{calc}-\left. \frac{\partial I}{\partial V}\right|^k_{meas}\right)
\end{eqnarray}

This calculation is iterated until the desired precision is achieved. The procedure is illustrated in figure \ref{figure:deconv_scheme}. The deconvolution is executed in MATLAB. The script that is used is given in appendix \ref{appendix_deconv_script}.

\begin{figure}[h!]
    \centering
    \includegraphics[width=0.95\textwidth]{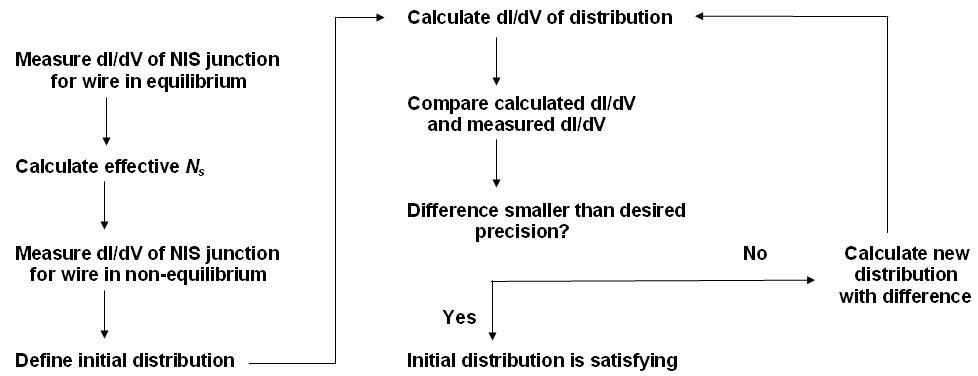}
    \caption{Scheme of the deconvolution procedure.}
    \label{figure:deconv_scheme}
\end{figure}

\subsection{\label{expecteddIdV}NIS differential conductance for ac biased wire}

\begin{figure}[h!]
    \centering
    \includegraphics[width=0.95\textwidth]{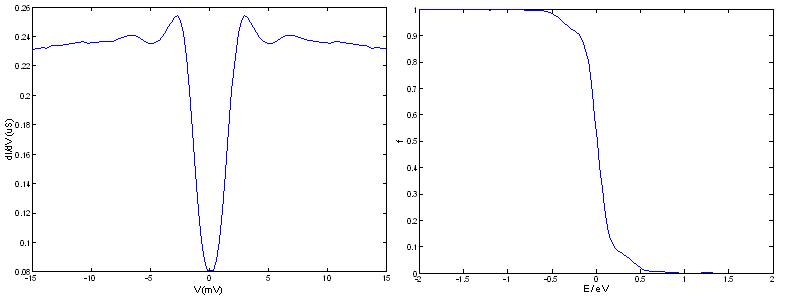}
    \caption{The calculated differential conductance left for the distribution function right.}
    \label{figure:f_and_dIdV}
\end{figure}

The fabrication of samples with wires connected to antennas and superconducting probes on top of the wires is not yet proven to be successful. Therefore the proposed model of chapter \ref{chapter4} cannot yet be verified. We can however calculate what the expected differential conductance measurements will look like when we measure it while driving the wire out of equilibrium with a time-dependent electric field. We do this by taking a calculated distribution function from the model and calculated dI/dV with equation \ref{dIdV_NIS}. This is done for the fast field regime where $\omega \tau_D=30000$ and $\hbar \omega /eV=0.4$ at a temperature of 2 K. So we look at a coherent transport situation comparable to that discussed in chapter \ref{chapter5}, where we take into account that the photon energy should be below the gap energy of niobium. The distribution function from the model and the calculated differential conductance of the NIS junction on the wire are given in figure \ref{figure:f_and_dIdV}.

The experiments with the described system will proceed and hopefully lead to a satisfying result that can be used to verify the theoretical model.


\chapter{Conclusion and discussion}

\section{\label{conclusion}Conclusion}
In this project we studied the ac quantum transport in a quasi-one dimensional, normal metal wire, where the transport is diffusive, connected between equilibrium reservoirs. For coherent transport, where the phase of a charge carrier is preserved, photon absorption and the diffusive character of the transport influence the energy distribution of the quasi-particles inside the wire. When the diffusion time, i.e. the time that a quasi-particle spends in the wire, exceeds the energy relaxation time, the mutual interaction of quasi-particles and the interaction between quasi-particles and phonons causes incoherent transport and influences the energy distribution. 

Often scattering theory is used to describe the transport of charge carriers through nano-structures. However, in our situation many processes involving the energy of the charge carriers come into play in the scattering region, i.e. the diffusive wire. Therefore we studied the effect of an ac bias applied to a diffusive wire by looking at the energy distribution of the quasi-particles inside the wire. 

Previous work on this subject was independently done by R. Schrijvers \cite{R.Schrijvers} and A.V. Shytov \cite{shytov}.      R. Schrijvers approached the situation by applying Tien-Gordon theory to the reservoirs and calculated the energy distribution in the wire with a semi-classical diffusion equation, which is a Boltzmann equation extended for inelastic interactions. He concluded that in this way only the slow field limit without inelastic interactions is adequately described. This is due to the fact that the assumption is made that the path traveled by the charge carrier in the wire is of no influence on the energy distribution. Also Tien-Gordon theory assumes averaging over time, which troubles the correct evaluation of the collision integral.

A.V. Shytov calculated the energy distribution in a diffusive wire where the energy relaxation time exceeds the diffusion time with a quantum diffusion equation. This approach seems valid in all frequency regimes, but only for coherent transport.

Stimulated by these approaches we derived from Green function formalism a quantum diffusion equation equivalent to that of Shytov and we did a second derivation from the full Dyson equation to extend this model to account for inelastic scattering processes. This approach seems successful, as the quasi-particle energy distribution can be calculated in every frequency regime and the limit situations of strong interaction processes provide the correct distribution. This approach is justified by the Landau theory of Fermi liquids. This states that there is an one-to-one correspondence between the states of a non-interacting particle system and the states of an interacting particle system provided that the excitations are near the Fermi level. This means that the excited states of an interacting system are labeled with the same quantum numbers as those of a non-interacting system. The interactions have the effect that the electrons are treated as quasi-particles, particles which are closely related to their environment.

For wires where the energy relaxation time exceeds the diffusion time, the transport is fully elastic and the energy distribution is calculated by a quantum diffusion equation without an inelastic collision term. The distribution function behaves differently in the two field limits, $\omega \tau_D <<1$ and $\omega \tau_D >>1$. For the slow field, strong signal limit ($\omega \tau_D << 1$, $\hbar \omega /eV << 1$) the dc situation is approached and the energy distribution is given by a time-varying two step function. In the fast field, strong signal limit ($\omega \tau_D >> 1$, $\hbar \omega /eV << 1$) a quasi-particle can oscillate multiple times with the field in the wire before leaving the wire, therefore gaining more energy quanta. This results in a time-independent electron energy distribution which does not go to zero at higher energies. 

Numerical simulations for a finite ratio $\hbar \omega / eV$ shows the photon steps in the energy distribution. In the slow field regime the distribution is highly time-dependent. The energy distribution directly follows the field and the photon steps oscillate from zero photon absorption to maximum photon absorption. In the fast field limit the energy distribution becomes completely time-independent. The maximum photon absorption is reached when the diffusion time is exceeded. In the crossover the two effects are both observed. There is a slight oscillation around the maximum photon absorption value of the fast field regime.

When the energy relaxation time becomes comparable to the diffusion time, the transport is no longer coherent and scattering is inelastic. Numerical simulations show how the energy exchange processes of mutual quasi-particles and between quasi-particles and phonons influence the energy distribution in the fast field regime, which is of interest for experimental situations. The interaction between quasi-particles and phonons annihilates first the photon steps in the distribution. In the strong electron-phonon interaction limit the Fermi function at bath temperature is found on every position in the wire.

The interaction between quasi-particles is quite different. It causes a smearing in the photon steps. In the strong electron-electron interaction limit a local equilibrium is reached on every position in the wire. The photon absorption, diffusive transport and the interactions cause an effective temperature profile across the wire. The effective temperature profile is determined by fitting the distribution from the simulation on every position with a Fermi function. The obtained temperature profile deviates from the calculated profile. It is not yet fully understood whether this is caused by a numerical error or that it is caused by some physical effect, like for instance the position in the wire that the photon absorption takes place.

So the complicated interplay between the effect of photon absorption, diffusive transport and inelastic scattering on the  quasi-particle energy distribution seems to be accurately described by our model.

\section{\label{discussion}Discussion}

The model developed in this work is not yet verified by experiments. The fabrication of the required samples is not yet proven to be completely succesfull due to the failure of fabrication apparatus, however the outlook is promising and the experimental work will be continued by members of this research group.

To calculate the collision term, the interactions are assumed to be instantaneous and local. An experiment can verify this assumption, so it is desirable to proceed with the experimental part of this project. The comparison between experiment and theoretical model has to provide a full insight in the ac quantum transport in diffusive quasi-one dimensional wires and how the non-equilibrium is shown in the quantum statistics of the quasi-particles in the wire.

While our MATLAB code seems to calculate the quasi-particle energy distribution influenced by photon absorption, diffusive transport and inelastic collisions in a correct manner, it also appears that the MATLAB code is not very efficient. A program written in C should in principle work more efficient. This provides the opportunity to optimize the discretization of the variables, so that a more accurate result is obtained.


\chapter*{\label{acknowledgements}Acknowledgements}

The phrase \textit{Tempus fugit} is the first thing that comes to mind when I look back at the time spent in this group working on my master thesis project. It was certainly an interesting and challenging time. The project did not go as planned. One of the main goals was to experimentally obtain data on the subject. However, reality seems to play tricks on you when you want to control things. Apparatus failure caused such delay in the fabrication of the samples that I did not get to experience the beauty of  experimental success. This pushed me further to theoretical research and I have to say, I didn't mind. It was really intriguing how the physics on small scale revealed itself to me by doing the math and combining concepts. In addition I learned different experimental techniques in preparation of the big experiment which didn't come. 

I would like to thank prof. Teun Klapwijk for giving me the opportunity of doing this challenging work. It really gave me the chance to evolve in different disciplines. I also would like to express my gratitude towards my daily supervisors, Nathan and Rik. There were times that I was not certain whether everything would work out, but they gave me the confidence that leaded to this satisfying result and educated me in the necessary basics of experimental and theoretical research.

Furthermore, it was just a fun time. I really experienced to be a part of the group. I especially want to thank Nathan, Rik, Eduard and David for the interesting conversations on everything and nothing. Also I want to thank Reinier for his company during the whole year in the huge students office. I want to thank the whole group for this fun time and for their part in my education. I am confident that the gained experience in this group will be of great help in the continuation of my career.

\clearpage{\pagestyle{empty}}
 \bibliographystyle{unsrt}
  \bibliography{references}



\appendix

\chapter{\label{appendix_shotnoise}Shot noise}

Shot noise, the phenomenon that arises because of charge quantization, provides information about the statistics of charge carriers involved in quantum transport. In chapter \ref{chapter2} the mean square fluctuations in the occupation of incident, reflected and transmitted state are derived to be:

\begin{eqnarray}\label{mean sq fluctuations2}
\left\langle \delta n_T \delta n_T\right\rangle&=&-TRf^2\\
\left\langle (\delta n_T)^2\right\rangle&=&Tf(1-Tf)\\
\left\langle (\delta n_R)^2\right\rangle&=&Rf(1-Rf)
\end{eqnarray}

It appears that at zero temperature, when the distribution is given by a step function at chemical potential, the shot noise disappears for full reflectance or full transmittance. At finite temperatures the mean square fluctuations fluctuate like the incident state with occupation $f$.

Now when we proceed with this simplified model of a single incident charge carrier to investigate the fluctuations in the current, we can consider a perfect conductor with three separated channels. One for the incident state, one for the reflected state and one for the transmitted state. The carriers move in one directions with a velocity $v(E)$ dependent on the energy of the charge carrier. For an energy interval $dE$, the incident current is given by $dI_{in}=ev(E)d\rho(E)$. $\rho(E)$ is the energy dependent density of carriers per unit length. It is given by $\rho(E)=n_{in}(E)\nu(E)dE$, where $\nu(E)$ is the density of states per unit length. In a perfect conductor $\nu(E)=(2\pi \hbar v(E))^{-1}$. This leads to $dI_{in}=e(2\pi \hbar)^{-1}n_{in}(E)dE$. For the transmitted and reflected channel  the same procedure can be followed leading to $dI_{T}=e(2\pi \hbar)^{-1}n_{T}(E)dE$ and $dI_{R}=e(2\pi \hbar)^{-1}n_{R}(E)dE$. Integrating gives the expressions for the current. When the occupation numbers vary slowly in time, the derivation can be expanded by just taking the occupation not only energy-dependent, but also time-dependent.

\begin{eqnarray}\label{current}
I_{in}(E,t)&=&\frac{e}{2 \pi \hbar}\int n_{in}(E,t)dE\\
I_{T}(E,t)&=&\frac{e}{2 \pi \hbar}\int n_{T}(E,t)dE\\
I_{R}(E,t)&=&\frac{e}{2 \pi \hbar}\int n_{R}(E,t)dE
\end{eqnarray}

For low frequency fluctuations in the current, these expressions can be Fourier transformed giving the frequency dependent current.

\begin{eqnarray}\label{current}
I_{in}(\omega)&=&\frac{e}{2 \pi \hbar}\int n_{in}(E+\hbar \omega)dE\\
I_{T}(\omega)&=&\frac{e}{2 \pi \hbar}\int n_{T}(E+\hbar \omega)dE\\
I_{R}(\omega)&=&\frac{e}{2 \pi \hbar}\int n_{R}(E+\hbar \omega)dE
\end{eqnarray}

The fluctuations in current and occupation number are directly related. The current noise power is in the zero frequency limit $S_{II}=e(2\pi \hbar)^{-1}\int S_{nn}(E)dE$. From the current it is seen that the charge carriers arrive at a rate of $dE/(2\pi \hbar)$ in each energy interval. This contributes to the noise with the mean square fluctuations of the relevant state. Therefore $S_{nn}(E)=1/(\pi \hbar)\left\langle \delta n \delta n\right\rangle$. Substitution in the current noise power relations leads to

\begin{eqnarray}\label{current noise power}
S_{I_{in}I_{in}}&=& 2\frac{e^2}{2 \pi \hbar} \int f(1-f)dE\\
S_{I_{T}I_{T}}&=& 2\frac{e^2}{2 \pi \hbar} \int Tf(1-Tf)dE\\
S_{I_{R}I_{R}}&=& 2\frac{e^2}{2 \pi \hbar} \int Rf(1-Rf)dE\\
S_{I_{T}I_{R}}&=&-2\frac{e^2}{2 \pi \hbar} \int TRf^2dE.
\end{eqnarray}

So here we see explicitly the earlier found conclusion for the fluctuations that shot noise disappears for full transparency or full reflectance at zero temperature. At finite temperature the distribution function is thermally broadened and the shot noise will not disappear due to thermal fluctuations. When now the system under consideration is extended to a situation where a multi-channel scatterer is placed between two terminals with respectively distribution $f_L$ and $f_R$ the noise power is given by

\begin{equation}\label{noise power}
S=\frac{e^2}{\pi \hbar} \sum_{n} \int dE{T_n(E)[f_L(1-f_L)+f_R(1-f_R)]+T_n(E)[1-T_n(E)](f_L-f_R)^2}.
\end{equation}

The scale of the energy dependence of the transmission coefficients is usual much bigger than the thermal and bias energy. Therefore these quantities can be taken in equation \ref{noise power} at Fermi energy. Then the noise power becomes

\begin{equation}\label{noise power2}
S=\frac{e^2}{\pi \hbar}[2k_bT\sum_{n} T_n^2+eVcoth\left(\frac{eV}{2k_bT}\right)\sum_n T_n(1-T_n)]
\end{equation}

\chapter{\label{appendix_sim}MATLAB code of the simulation program}

\section{\label{coh_sim}Script for the simulation of coherent transport}

\footnotesize
\begin{verbatim}



%==========================================================================
%CLEAN UP, FUNDAMENTALS
%==========================================================================
clear all;
tic;

%fundamental constants
e = 1.602e-19;
hbar = 6.63e-34/2/3.141592; %6.6e-16; %
kb = 1.38e-23; % 8.6e-5;

%==========================================================================
%PARAMETERS
%==========================================================================

%general
saveall=0;
saverepeat=1250;

%time discretization
Nt0=7;
dt=2*pi/Nt0;
int_method='euler';
Nt=5000*Nt0;%1000000;

%energy discretization
T=2;
omega=0.3e12*2*3.141592;
V=3e-3; 
limit_E=2; %limit in multiples of eV
hw=hbar*omega/e/V;
Nw=16; %must be even! 
dE=hw/Nw;
NE=ceil(limit_E/dE);
E=-NE*dE:dE:(NE-1)*dE+dE/2;
NE=length(E);

kT=kb*T/e/V;

%space discretization
D_method='lagrange_2';
Nx=100;
x = linspace(0,1,Nx);
dx = 1/(Nx-1);
%tauD= 1e-9;
%z=omega*tauD;
z=30000;

fprintf('homega/eV= %f kT/eV= %f dE=%f Emax= %f omega tau= %f \n',hw,kT,dE,NE*dE,z);

%==========================================================================
%INITIALIZATION
%==========================================================================

%INITIAL AND BOUNDARY CONDITIONS
Fl=1./(exp((E)/kb/T*e*V)+1);
Fr=1./(exp((E)/kb/T*e*V)+1);
Fold=[ones(Nx-1,1)*Fl; Fr];
Feq=[ones(Nx-1,1)*Fl; Fr];
Fave=[ones(Nx-1,1)*Fl; Fr];

%MATRICES
D_x=dx1(Nx,1,dx,D_method);
D_xx=dx2(Nx,1,dx,D_method);

D_E=dE1(1,NE,Nw,dE)';
D_EE=dE2(1,NE,Nw,dE)';

first_step=[ ];
second_step=[];
third_step=[];

%==========================================================================
%INTEGRATION
%==========================================================================

switch lower(int_method)
    case 'euler'
        %euler
        for m=1:Nt
            w=sin(m*dt);
            dF=D_xx*Fold+2*w*D_x*Fold*D_E+w^2*Fold*D_EE;
            Fnew=Fold+dF*dt/z;
            Fnew([1 Nx],:)=[Fl;Fr];
            Fold=Fnew;
            Fave=((m-1)*Fave+Fold)/m;
            first_step=[first_step Fnew(round([Nx/2])',NE/2+Nw/2)];
            second_step=[second_step Fnew(round([Nx/2])',NE/2+3*Nw/2)];
            third_step=[third_step Fnew(round([Nx/2])',NE/2+5*Nw/2)];
            if mod(m,5000)==0 
                for o=1:Nx
                    for p=1:NE
                        if Fnew(o,p)>1
                            Fnew(o,p)=1;
                        end
                        if Fnew(o,p)<0
                            Fnew(o,p)=0;
                        end
                    end
                end
            end
            if mod(m,1000)==0 %saverepeat
                clc;
                fprintf('iteratie %i, time %f',m,toc);%round(m/Nt0*100)
                %tic;
%                 hold on;
                plot(E,Fnew(round([1 Nx/4 Nx/2])',:));
                pause(.2);
                if saveall
                save(['Fnew' num2str(m)],'Fnew');
                end
            end
        end
 end
 
\end{verbatim}

\section{\label{op_sim}Functions for the used operators}

\textbf{First spatial derivative}

\begin{verbatim}

function S=dx1(Nx,NE,h,method)

switch lower(method)
    case {'lagrange_1'}
        %lagrange 1st order
        r=2;
        w1=[-3 4 -1];
        w3=[-1 0 1];
        n=1;
    case 'lagrange_2'
        %lagrange
        r=12;
        w1=[-25 48 -36 16 -3];
        w2=[-3 -10 18 -6 1];
        w3=[1 -8 0 8 -1];
        n=2;
    case 'least'
        %least squares
        r=70;
        w1=[-54 13 40 27 -26];
        w2=[-34 3 20 17 -6];
        w3=[-2 -1 0 1 2]*7;
        n=2;
end

%the boundaries of the matrix
base0=1:NE;
v0=ones(1,NE);

%first boundary
array1=[];
column1=[];
values1=[];
for k=1:length(w1)
    array1=[array1 base0];
    column1=[column1 base0+(k-1)*NE];
    values1=[values1 w1(k)*v0];
end
column1=[column1 NE*Nx-column1+1];
array1=[array1 NE*Nx-array1+1];
values1=[values1 values1];

%second boundary (if needed)
array2=[];
column2=[];
values2=[];
if n==2
for k=1:length(w2)
    array2=[array2 base0+NE];
    column2=[column2 base0+(k-1)*NE];
    values2=[values2 w2(k)*v0];
end
column2=[column2 NE*Nx-column2+1];
array2=[array2 NE*Nx-array2+1];
values2=[values2 values2];
end

%central part
if n==1
       base1=1:NE*(Nx-2);
v1=ones(1,NE*(Nx-2));
else
       base1=1:NE*(Nx-4);
v1=ones(1,NE*(Nx-4));
end
array3=[];
column3=[];
values3=[];
for k=1:length(w3)
    array3=[array3 base1+n*NE];
    column3=[column3 base1+(k-1)*NE];
    values3=[values3 w3(k)*v1];
end

%the total matrix
S=sparse([array1 array2 array3],[column1 column2 column3],[values1 values2 values3])/r/h;

\end{verbatim}

\textbf{Second spatial derivative}

\begin{verbatim}

function S=dx2(Nx,NE,h,method)

switch lower(method)
    case {'lagrange_1'}
        %lagrange 1st order
        r=1;
        w1=[2 -5 4 -1];
        w3=[1 -2 1];
        n=1;
    case 'lagrange_2'
        %lagrange
        r=12;
        w1=[35 -104 114 -56 11];
        w2=[10 -15 -4 14 -6 1];
        w3=[-1 16 -30 16 -1];
        n=2;
    case 'least'
        %least squares
        r=14;
        w1=[9 -15 -2 13 -5]*2;
        w2=[11 -16 -4 12 -3];
        w3=[2 -1 -2 -1 2]*2;
        n=2;
end

%the boundaries of the matrix
base0=1:NE;
v0=ones(1,NE);

%first boundary
array1=[];
column1=[];
values1=[];
for k=1:length(w1)
    array1=[array1 base0];
    column1=[column1 base0+(k-1)*NE];
    values1=[values1 w1(k)*v0];
end
column1=[column1 NE*Nx-column1+1];
array1=[array1 NE*Nx-array1+1];
values1=[values1 values1];

%second boundary (if needed)
array2=[];
column2=[];
values2=[];
if n==2
    for k=1:length(w2)
        array2=[array2 base0+NE];
        column2=[column2 base0+(k-1)*NE];
        values2=[values2 w2(k)*v0];
    end
    column2=[column2 NE*Nx-column2+1];
    array2=[array2 NE*Nx-array2+1];
    values2=[values2 values2];
end

%central part
if n==1
    base1=1:NE*(Nx-2);
    v1=ones(1,NE*(Nx-2));
else
    base1=1:NE*(Nx-4);
    v1=ones(1,NE*(Nx-4));
end
array3=[];
column3=[];
values3=[];
for k=1:length(w3)
    array3=[array3 base1+n*NE];
    column3=[column3 base1+(k-1)*NE];
    values3=[values3 w3(k)*v1];
end

%the total matrix
S=sparse([array1 array2 array3],[column1 column2 column3],[values1 values2 values3])/r/h/h;

\end{verbatim}

\textbf{First energy derivative}

\begin{verbatim}

function S=dE1(Nx,NE,Nw,h)

%boundaries
array_0=[1:Nw NE-[1:Nw]+1];
column_0=[ones(1,Nw) ones(1,Nw)*NE];
values_0=[-ones(1,Nw) ones(1,Nw)];

%central part
base=1:NE-Nw;
v0=ones(1,NE-Nw);

array_1=[base+Nw base];
column_1=[base base+Nw];
values_1=[-v0 v0];

array=[array_0 array_1];
column=[column_0 column_1];
values=[values_0 values_1];

for k=1:Nx-1
    array=[array array_0+NE*k array_1+NE*k];
    column=[column column_0+NE*k column_1+NE*k];
    values=[values values_0 values_1];
end

%the total matrix
S=-sparse(array,column,values)/2/h/Nw;

\end{verbatim}

\textbf{Second energy derivative}

\begin{verbatim}

function S=dE2(Nx,NE,Nw,h)

%boundaries
array_0=[1:2*Nw 2:2*Nw 1:2*Nw];
array_0=[array_0 NE+1-array_0];
column_0=[ones(1,2*Nw) 2:2*Nw 2*Nw+1:4*Nw];
column_0=[column_0 NE+1-column_0];
values_0=[-1 ones(1,2*Nw-1) -2*ones(1,2*Nw-1) ones(1,2*Nw)];
values_0=[values_0 values_0];

%central part
base=1:NE-2*Nw; %4*Nw
v0=ones(1,NE-2*Nw); %4*Nw

array_1=[base+1*Nw base+1*Nw base+1*Nw]; %2*Nw ; 2*Nw ; 2*Nw
column_1=[base base+1*Nw base+2*Nw]; %2*Nw ; 4*Nw
values_1=[v0 -2*v0 v0];

array=[array_0 array_1];
column=[column_0 column_1];
values=[values_0 values_1];

for k=1:Nx-1
    array=[array array_0+NE*k array_1+NE*k];
    column=[column column_0+NE*k column_1+NE*k];
    values=[values values_0 values_1];
end

%the total matrix
S=sparse(array,column,values)/(2*Nw*h)^2;

\end{verbatim}

\section{\label{incoh_sim}Script for the simulation of incoherent transport}

\begin{verbatim}

%==========================================================================
%CLEAN UP, FUNDAMENTALS
%==========================================================================
clear all;
tic;

%fundamental constants
e = 1.602e-19;
hbar = 6.63e-34/2/3.141592; %6.6e-16; 
kb = 1.38e-23; % 8.6e-5;

%==========================================================================
%PARAMETERS
%==========================================================================

%general
saveall=0;
saverepeat=1250;

%time discretization
Nt0=1.1;
dt=2*pi/Nt0;
Nt=79000*Nt0;%20000;

%energy discretization
T=0.5;
omega=1.6e12*2*3.141592;
V=16e-3; 
limit_E=2; %limit in multiples of eV
hw=hbar*omega/e/V;
Nw=50; %must be even! 
dE=hw/Nw;
NE=ceil(limit_E/dE);
E=-NE*dE:dE:(NE-1)*dE+dE/2;
NE=length(E);

kT=kb*T/e/V;

%interaction parameters
int_mech='eph'; %interaction mechansim, ee for electron-electron, eph for electron phonon

rho=2.7e3;
EF=12/V; %J
dos=2e47*e*V;
s=6.42e3;
kf=1.75e10;
D=100e14; %nm^2s^-1
dos=2e47;
S=20e-9*20e-9;
sigma=1e9;

Ke=(sqrt(2*D)*pi*hbar^(3/2)*dos*S)^(-1);

NE_ee=NE;
E_ee=linspace(10,14,NE_ee)/V;
dE_ee=4*e/NE_ee/V;

NE_ph=NE;
E_ph=linspace(-2,2,NE_ph);
dE_ph=4*e/NE_ph;

kph=sigma/24/zeta(5)/dos/kb^5*(e*V)^2*dE_ph;
%kph=4e12*1e-3/V; %has to be expressed in V

n_ph=(1./(exp(abs(E_ph)/kb/2*e*V)-1))';

%space discretization
D_method='lagrange_2';
Nx=20;
x = linspace(0,1,Nx);
dx = 1/(Nx-1);
%tauD= 1e-9;
%z=omega*tauD;
z=2000000;

fprintf('homega/eV= %f kT/eV= %f dE=%f Emax= %f omega tau= %f \n',hw,kT,dE,NE*dE,z);

%==========================================================================
%INITIALIZATION
%==========================================================================

%INITIAL AND BOUNDARY CONDITIONS
Fl=1./(exp((E)/kb/T*e*V)+1);
Fr=1./(exp((E)/kb/T*e*V)+1);
Fold=[ones(Nx-1,1)*Fl; Fr];
Feq=[ones(Nx-1,1)*Fl; Fr];
Fave=[ones(Nx-1,1)*Fl; Fr];

%MATRICES
D_x=dx1(Nx,1,dx,D_method);
D_xx=dx2(Nx,1,dx,D_method);

D_E=dE1(1,NE,Nw,dE)';
D_EE=dE2(1,NE,Nw,dE)';

Iin=0;
Iout=0;
h=400;

first_step=[ ];
second_step=[];
third_step=[];

%==========================================================================
%INTEGRATION
%==========================================================================

switch lower(int_mech)
    case 'ee'
        %euler
        for m=1:Nt
            w=sin(m*dt);
            dF=D_xx*Fold+2*w*D_x*Fold*D_E+w^2*Fold*D_EE+Iin*(z/omega)-Iout*(z/omega);
            Fnew=Fold+dF*dt/z;
            Fnew([1 Nx],:)=[Fl;Fr];
            Fold=Fnew;
            Fave=((m-1)*Fave+Fold)/m;
            first_step=[first_step Fnew(round([Nx/2])',NE/2+Nw/2)];
            second_step=[second_step Fnew(round([Nx/2])',NE/2+3*Nw/2)];
            third_step=[third_step Fnew(round([Nx/2])',NE/2+5*Nw/2)];
             if mod(m,50)==0 
                for o=1:Nx
                    for p=1:NE
                        if Fnew(o,p)>1
                            Fnew(o,p)=1;
                        end
                        if Fnew(o,p)<0
                            Fnew(o,p)=0;
                        end
                    end
                end
            end
            if mod(m,h+1)==0
                for o=1:Nx
                    for p=1:NE
                        if Fnew(o,p)>1
                            Fnew(o,p)=1;
                        end
                        if Fnew(o,p)<0
                            Fnew(o,p)=0;
                        end
                    end
                end
            end
            if mod(m,h-1)==0
                for o=1:Nx
                    for p=1:NE
                        if Fnew(o,p)>1
                            Fnew(o,p)=1;
                        end
                        if Fnew(o,p)<0
                            Fnew(o,p)=0;
                        end
                    end
                end
            end
            if mod(m,800)==0
                Fnew2=Fnew;
                Fnew1=Fnew;
                [q,r]=size(Fnew);
                g1=diag(ones(r-1,1),1);
                g2=diag(ones(r-1,1),-1);
                Fnew1=circshift(Fnew1,[0 1])+[Fnew(:,1), (zeros(NE-1,Nx))'];
                b=[];
                a=[];
               for j=1:Nx;
                    k1=Fnew1(j,:);
                    k2=Fnew2(j,:);
 
                    for i=1:NE/2-1
                        h1=Fnew1(j,:)*g1;
                        h1=h1+[ones(1,1)', zeros(NE-1,1)'];
                        k1=[k1' h1']';
                        Fnew1(j,:)=h1;
                        h2=Fnew2(j,:)*g2;
                        k2=[h2' k2']';
                        Fnew2(j,:)=h2;
                    end
                    k=[k2' k1']';
                    E2=((E_ee).^(-3/2))';
                    W=(1-fliplr(k'))*Fnew(j,:)';
                    W2=Ke*E2.*W;
                    %In=W2'*fliplr(k')*dE;
                    Out=W2'*(1-k)*dE;
                    In=fliplr(Out);  
                    b=[b Out'];
                    a=[a In'];
                end
               Iout=(Fnew.*b');
               Iin=((1-Fnew).*a');
            end
            if mod(m,1000)==0 %saverepeat
                clc;
                fprintf('iteratie %i, time %f',m,toc);%round(m/Nt0*100)
                %tic;
                plot(E,Fnew(round([1 Nx/4 Nx/2])',:));
                pause(.2);
                if saveall
                save(['Fnew' num2str(m)],'Fnew');
                end
            end
        end
    case 'eph'
            %euler
        for m=1:Nt
            w=sin(m*dt);
            dF=D_xx*Fold+2*w*D_x*Fold*D_E+w^2*Fold*D_EE+Iin*(z/omega)-Iout*(z/omega);
            Fnew=Fold+dF*dt/z;
            Fnew([1 Nx],:)=[Fl;Fr];
            Fold=Fnew;
            first_step=[first_step Fnew(round([Nx/2])',NE/2+Nw/2)];
            second_step=[second_step Fnew(round([Nx/2])',NE/2+3*Nw/2)];
            third_step=[third_step Fnew(round([Nx/2])',NE/2+5*Nw/2)];
            if mod(m,50)==0 
                for o=1:Nx
                    for p=1:NE
                        if Fnew(o,p)>1
                            Fnew(o,p)=1;
                        end
                        if Fnew(o,p)<0
                            Fnew(o,p)=0;
                        end
                    end
                end
            end
            if mod(m,h+1)==0
                for o=1:Nx
                    for p=1:NE
                        if Fnew(o,p)>1
                            Fnew(o,p)=1;
                        end
                        if Fnew(o,p)<0
                            Fnew(o,p)=0;
                        end
                    end
                end
            end
            if mod(m,h-1)==0
                for o=1:Nx
                    for p=1:NE
                        if Fnew(o,p)>1
                            Fnew(o,p)=1;
                        end
                        if Fnew(o,p)<0
                            Fnew(o,p)=0;
                        end
                    end
                end
            end
            Fave=((m-1)*Fave+Fold)/m;
            if mod(m,400)==0
                h=m;
                Fnew2=Fnew;
                Fnew1=Fnew;
                [q,r]=size(Fnew);
                g1=diag(ones(r-1,1),1);
                g2=diag(ones(r-1,1),-1);
                Fnew1=circshift(Fnew1,[0 1])+[Fnew(:,1), (zeros(NE-1,Nx))'];
                b=[];
                a=[];
                for j=1:Nx; 
                    k1=Fnew1(j,:);
                    k2=Fnew2(j,:);
                    for i=1:NE/2-1
                        h1=Fnew1(j,:)*g1;
                        h1=h1+[ones(1,1)', zeros(NE-1,1)'];
                        k1=[k1' h1']';
                        Fnew1(j,:)=h1;
                        h2=Fnew2(j,:)*g2;
                        k2=[h2' k2']';
                        Fnew2(j,:)=h2;
                    end
                    k=[k2' k1']';
                    E2=(E.^2)';
                    W=E2.*(n_ph+heaviside(E_ph)');
                    Out=kph*(1-(k'))*W*dE;
%                     In=kph*fliplr(k')*W*dE;
                    In=fliplr(Out')';
                    b=[b Out];
                    a=[a In];
                end
                Iout=Fnew.*b';
                Iin=(1-Fnew).*a';
            end
            if mod(m,1000)==0 %saverepeat
                clc;
                fprintf('iteratie %i, time %f',m,toc);%round(m/Nt0*100)
                %tic;
                plot(E,Fnew(round([1 Nx/4 Nx/2])',:));
                pause(.2);
                if saveall
                save(['Fnew' num2str(m)],'Fnew');
                end
            end
        end
end

\end{verbatim}

\chapter{\label{appendix_f}Space dependency in the distribution function}

\begin{figure}[h!]
    \centering
    \includegraphics[width=0.85\textwidth]{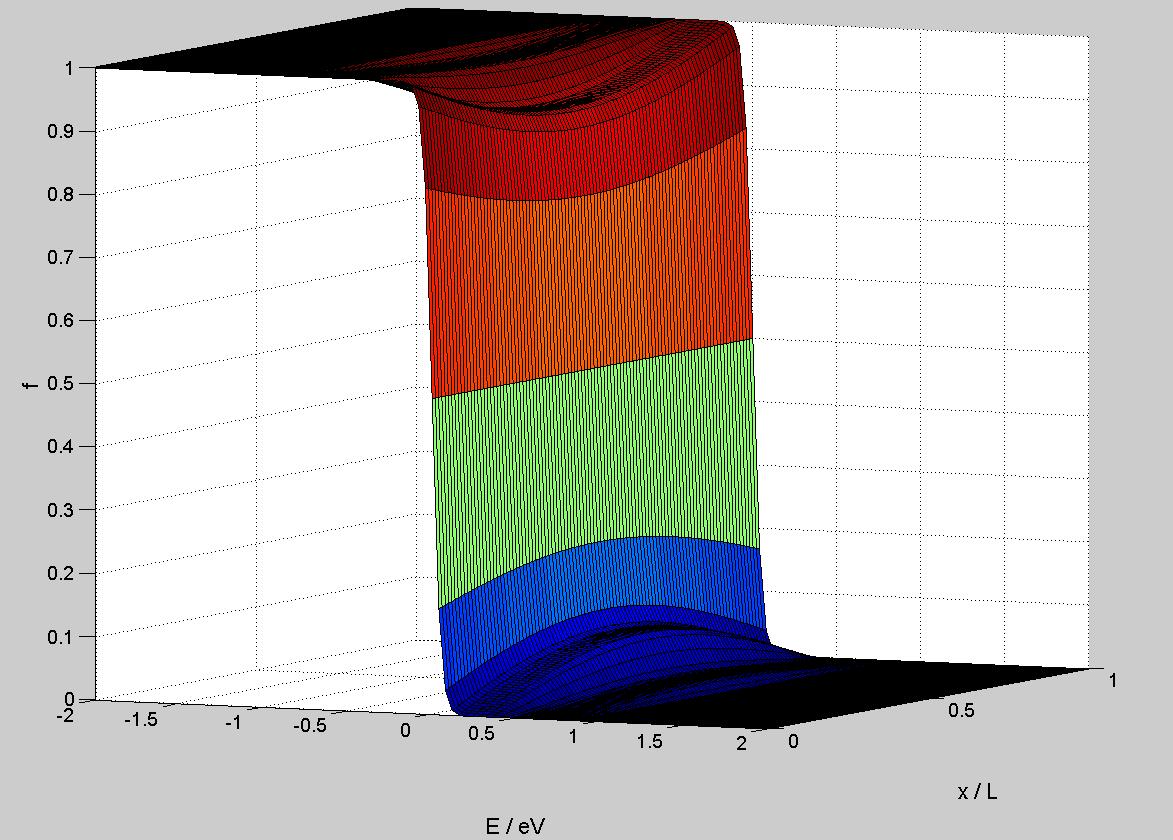}
    \caption{The quasi-particle energy distribution in the slow field regime, $\omega \tau_D=1$, and $\hbar \omega / eV=0.4$ at all positions in the wire at 2 K.}
    \label{figure:3D_slowfield}
\end{figure}

\begin{figure}[h!]
    \centering
    \includegraphics[width=0.8\textwidth]{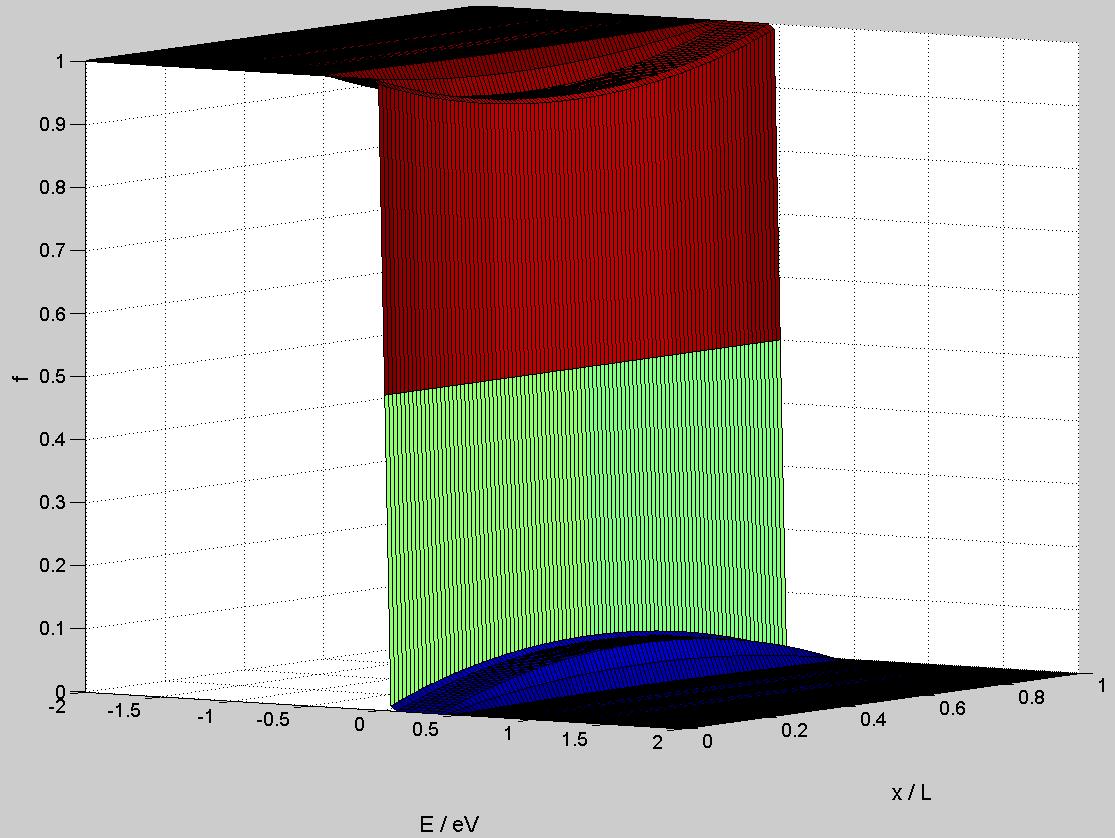}
    \caption{The quasi-particle energy distribution in the fast field regime, $\omega \tau_D=30000$, and $\hbar \omega / eV=0.4$ at all positions in the wire at 2 K.}
    \label{figure:3D_slowfield}
\end{figure}

\begin{figure}[h!]
    \centering
    \includegraphics[width=0.8\textwidth]{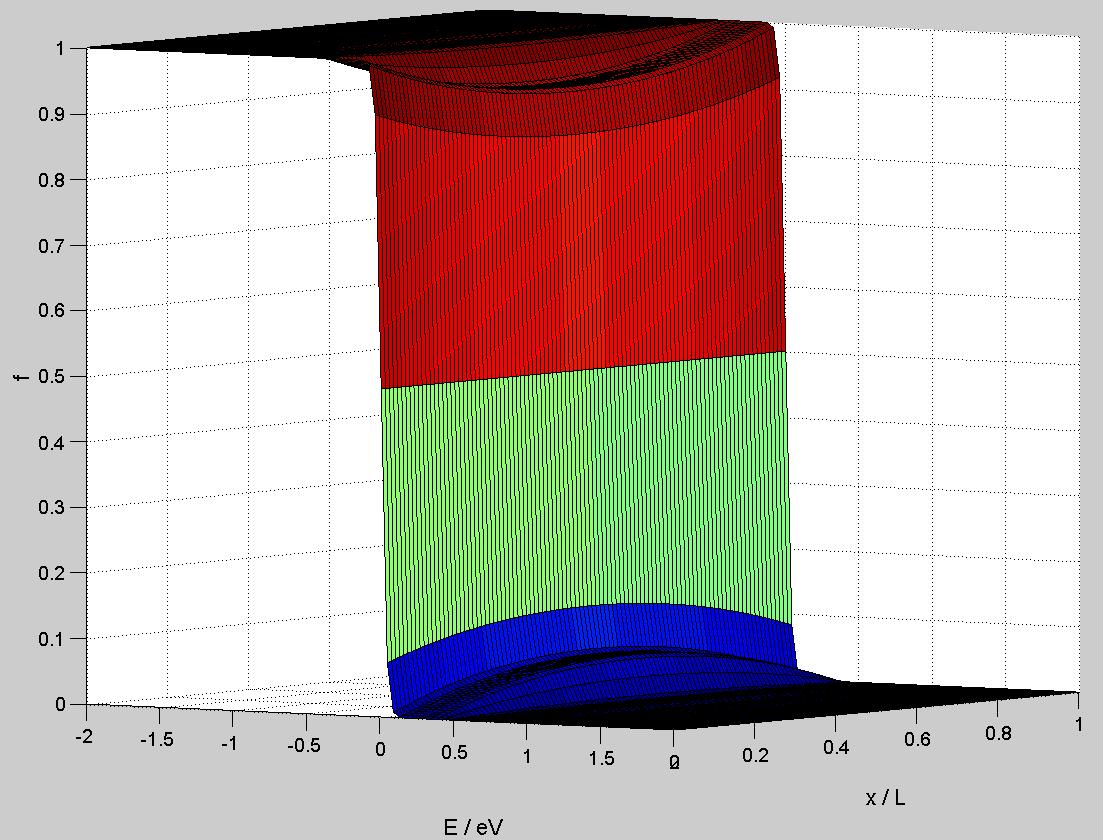}
    \caption{The quasi-particle energy distribution in the intermediate regime, $\omega \tau_D=100$, and $\hbar \omega / eV=0.4$ at all positions in the wire at 2 K.}
    \label{figure:3D_intermediate}
\end{figure}

\begin{figure}[h!]
    \centering
    \includegraphics[width=0.8\textwidth]{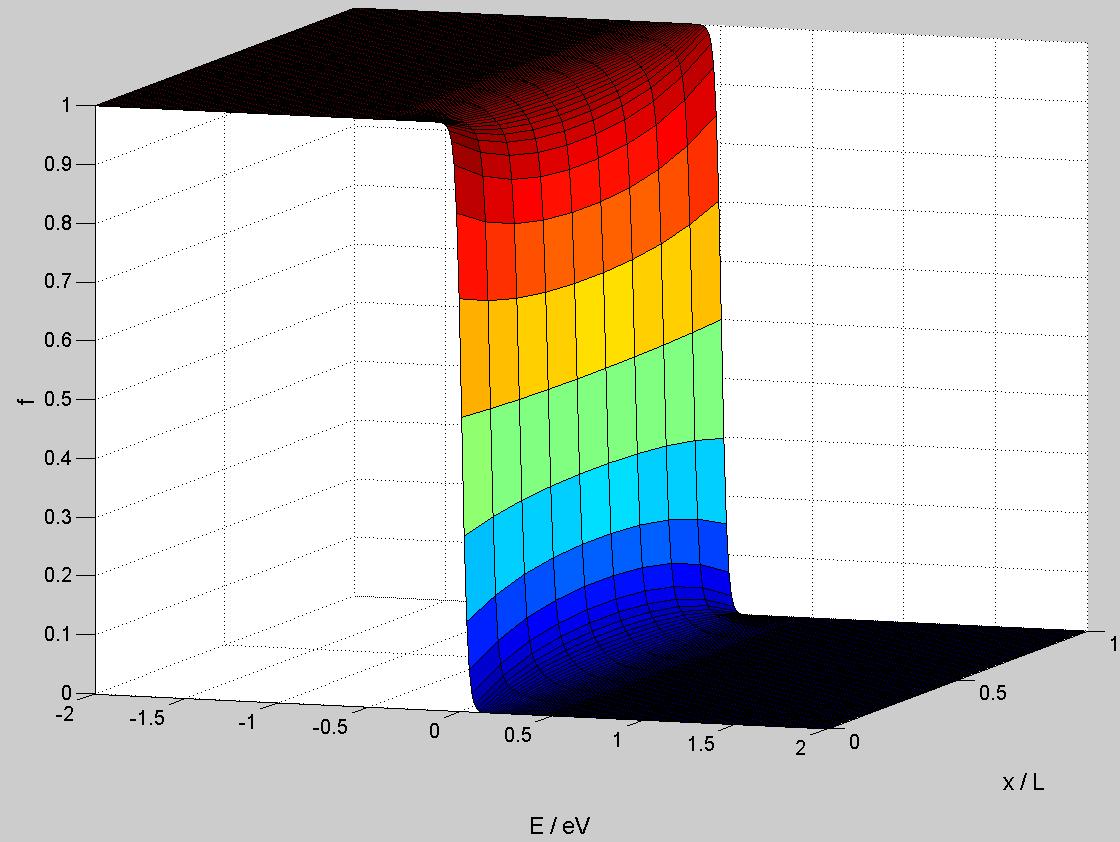}
    \caption{The quasi-particle energy distribution in the fast field, weak electron-phonon interaction regime, $\omega \tau_D=50000$, $\hbar \omega / eV=0.4$ and $\tau_D \approx \tau_E$ at all positions in the wire at 2 K.}
    \label{figure:3D_fastfield_weakint}
\end{figure}

\begin{figure}[h!]
    \centering
    \includegraphics[width=0.8\textwidth]{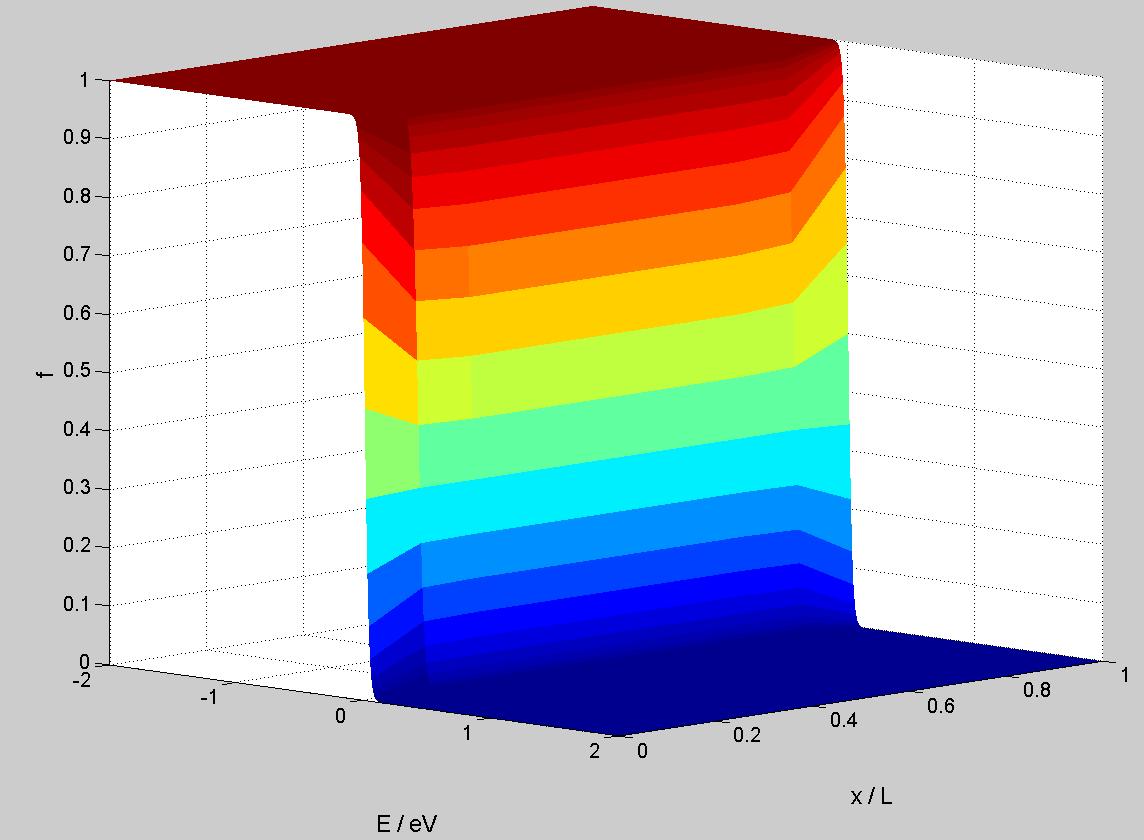}
    \caption{The quasi-particle energy distribution in the fast field, strong electron-phonon interaction regime, $\omega \tau_D=10^7$, $\hbar \omega / eV=0.4$ and $\tau_D \approx 200 \tau_E$ at all positions in the wire at 2 K.}
    \label{figure:3D-fastfield_strongint}
\end{figure}

\begin{figure}[h!]
    \centering
    \includegraphics[width=0.8\textwidth]{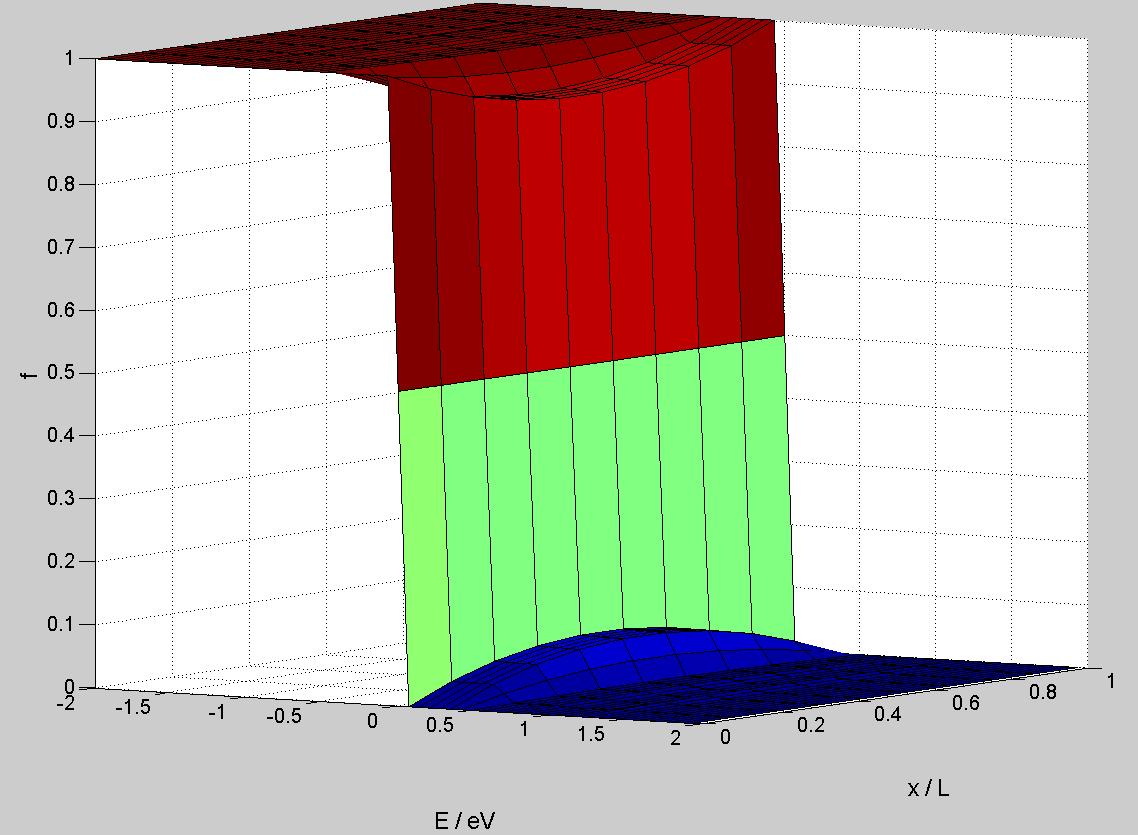}
    \caption{The quasi-particle energy distribution in the fast field, weak electron-electron interaction regime, $\omega \tau_D=10000$, $\hbar \omega / eV=0.4$ and $\tau_D \approx \tau_E$ at all positions in the wire at 500 mK.}
    \label{figure:3D_fastfield_weak_eeint}
\end{figure}

\begin{figure}[h!]
    \centering
    \includegraphics[width=0.8\textwidth]{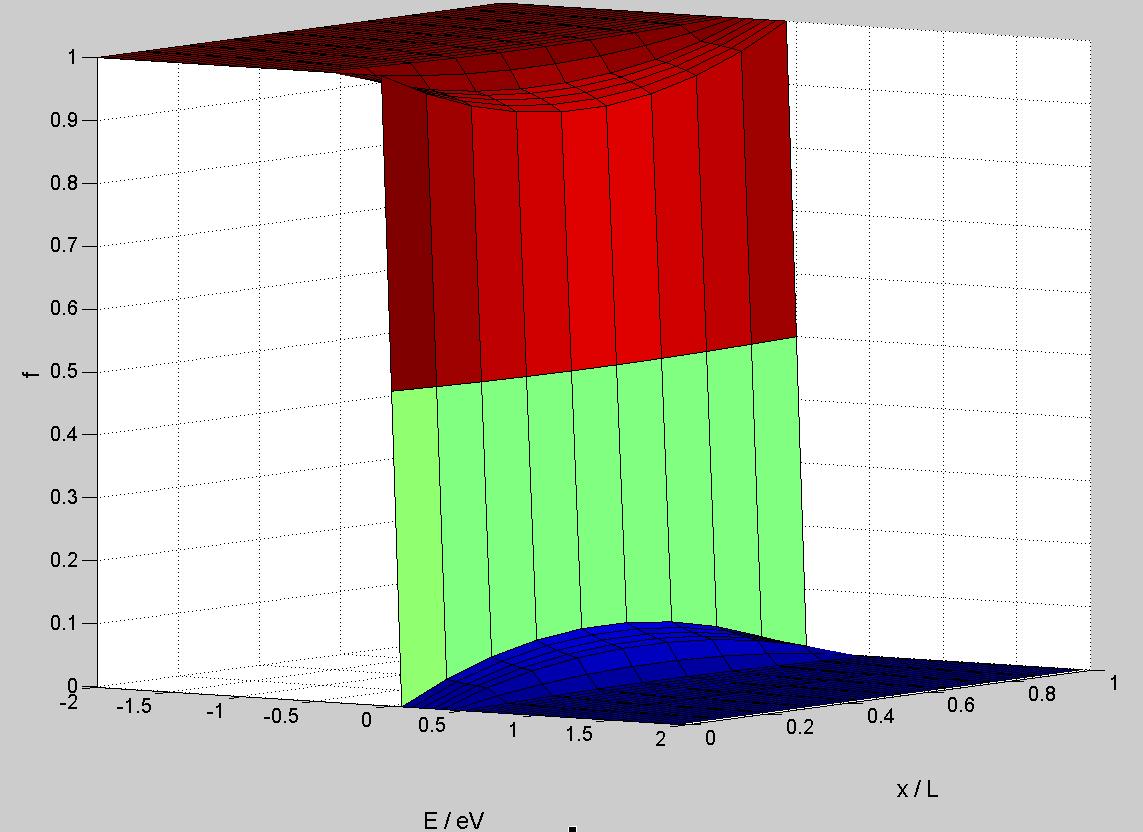}
    \caption{The quasi-particle energy distribution in the fast field, weak electron-electron interaction regime, $\omega \tau_D=75000$, $\hbar \omega / eV=0.4$ and $\tau_D \approx 7.5 \tau_E$ at all positions in the wire at 500 mK.}
    \label{figure:3D_fastfield_weak_ee2int}
\end{figure}

\begin{figure}[h!]
    \centering
    \includegraphics[width=0.8\textwidth]{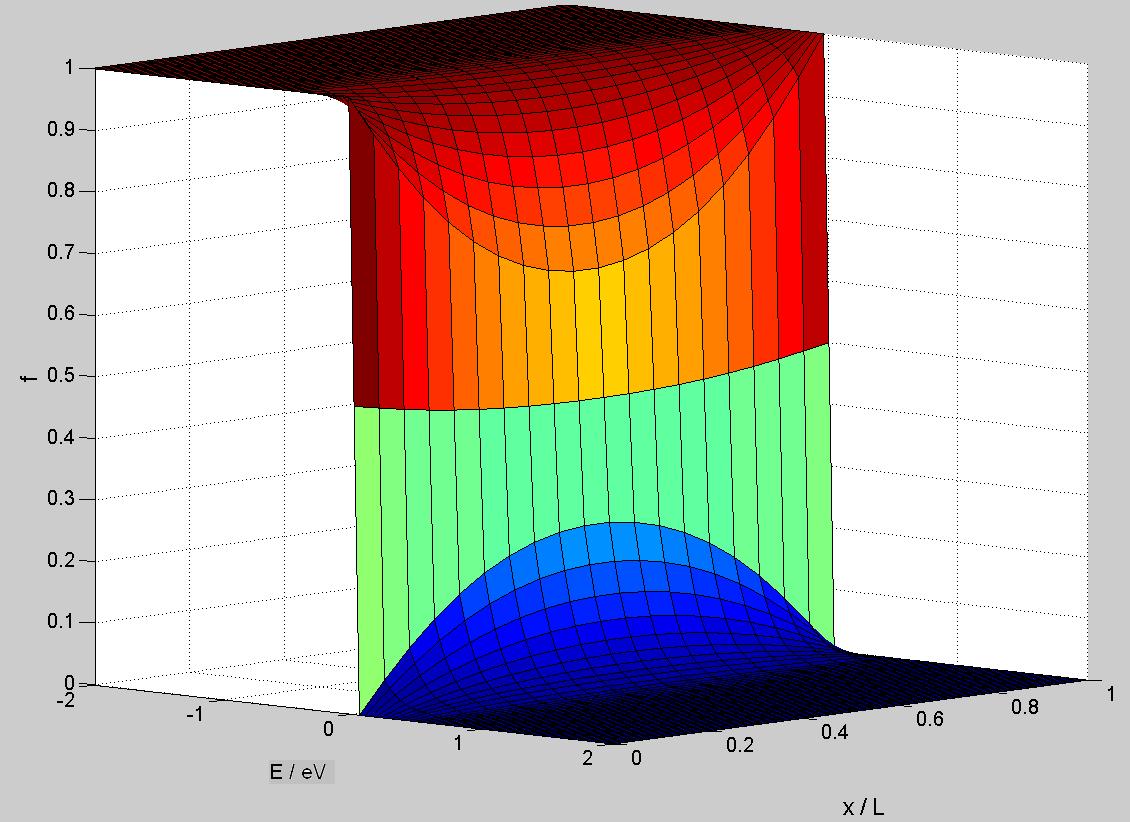}
    \caption{The quasi-particle energy distribution in the fast field, strong electron-electron interaction regime, $\omega \tau_D=2000000$, $\hbar \omega / eV=0.4$ and $\tau_D \approx 200 \tau_E$ at all positions in the wire at 500 mK.}
    \label{figure:3D_fastfield_strong_eeint}
\end{figure}

\chapter{\label{appendix_dIdV}Differential conductance of a NIS junction}

To obtain the electron energy distribution function on a certain position in the wire, superconducting probes are used that are positioned orthogonally on top of the wire with an insulating layer in between. Now by applying a current to this NIS junction we can measure the reciprocal value of the differential conductance. This differential conductance is a convolution of the density of states of the used superconductor and the probed energy distribution function. This can be calculated from Fermi's golden rule when the tunnel matrix elements are considered nearly constant for all energy states consider in the measurement. Because the energy distribution is probed with a superconductor the tunneling current is elastic, as there is no energy dissipation in a superconductor. So from the golden rule the tunneling rate from the wire to the superconductor and from the superconductor to the wire become respectively

\begin{equation}
\Gamma_{x\rightarrow p}(E)= \frac{1}{e^2 R_t}n_x(E)f_x(E)n_p(E+eV)(1-f_p(E+eV))
\end{equation}

\begin{equation}
\Gamma_{p\rightarrow x}(E)=\frac{1}{e^2 R_t}n_p(E+eV)f_p(E+eV)n_x(1-f_x(E)).
\end{equation}

Here $e$ is the elementary charge, $R_t$ is the tunnel resistance, $n_x$ and $n_p$ are respectively the density of states in the wire and in the superconductor and $f_x$ and $f_p$ are the distribution functions in respectively the wire and the superconductor. The current across the junction is calculated from these tunnel rates.

\begin{equation}
I(V)=e\int\left(\Gamma_{x\rightarrow p}(E)-\Gamma_{p\rightarrow x}(E))\right)dE
\end{equation}

When the tunnel rates are implemented in the formula above and the density of states of the superconductor is taken to be the BCS density of states $n_{BCS}(E)=\Re(|E|/\sqrt{E^2-\Delta^2})$ and as the wire is a metal the density of states in the wire is taken to be flat, we arrive at the following expression.

\begin{equation}
I(V)=\frac{1}{eR_t}\int n_{BCS}(E+eV)\left(f_x(E)-f_p(E+eV)\right) dE
\end{equation}

A variable change of $E\rightarrow E-eV$ and taking the derivative with respect to $V$ leads to an expression for the differential conductance of the NIS junction.

\begin{equation}
\frac{dI}{dV}=-\frac{1}{R_t}\int n_{BCS}(E) f'_x(E-eV)dE
\end{equation}

The distribution function shows up explicitly by integration by part and using the fact that the BCS density of states is even.

\begin{equation}
R_t\frac{dI}{dV}=1-\int n'_{BCS}(eV-E)f_x(E)dE \equiv 1-n'_{BCS} \ast f_x(eV)
\end{equation}

In practice the singular behaviour of the density of states of the superconductor is less sharp than the BCS theory predicts. To avoid problems with this aspect, the effective density of states of the superconductor in the NIS junction can be probed first for an unbiased wire. Since in this situation the energy distribution of the electrons is just a quasi-equilibrium Fermi function, the effective density of states is obtained by a deconvolution of the differential conductance of the NIS junction and the distribution function. This is basically finding a fit for the gap energy $\Delta$, the tunneling resistance $R_t$ and the electron temperature $T$.

\chapter{\label{appendix_deconv_script}MATLAB script for deconvolution}

\begin{verbatim}

clear all;
dVdI = importdata('testdatadIdV.txt');
dIdV1 = (1./dVdI);
m=length(dIdV1);
dIdV = [dIdV1' dIdV1(m)]';
kb =  8.6e-6;%1.38*10^-23;
T= 4.2;
e = 1.602e-19;
NE = length(dIdV);
NV = length(dIdV);
d = 1.3*10^(-3);%*e;
E = linspace(-10*10^(-3),10*10^(-3),NE);%*e;
dE = 20e-3/NE;%*e;
dV = 20e-3/NV;
V = linspace(-10*10^(-3),10*10^(-3),NV);%*e;
Rt = 1;%;1.2*10^4;
V2=1e-3;

for i=1:length(V)
    for j = 1:length(E)
        f(i,j) = 1/Rt*(exp((E(j)-V(i))/kb/T)./((exp((E(j)-V(i))/kb/T)+1).^2*kb*T));
    end
end
 
Ns=f\dIdV/dE;
Ns(1:5)=Ns(6);
Ns(246:m+1)=Ns(245);
 
for i = 1:length(V)
    dIdV99(i) = 1/Rt.*(exp((E-V(i))/kb/T)./((exp((E-V(i))/kb/T)+1).^2*kb*T)*Ns)*dE;
end

dIdV99(1:2)=dIdV99(3);
dIdV99(250:m+1)=dIdV99(249);

Diff_n=diag(ones(NE-1,1),1)-diag(ones(NE-1,1),-1);
Diff_n(1:2,1)=[1 ;1];
Diff_n(NE-1:NE,NE)=[1 ;1];

DnBCS=Diff_n*Ns/dE;
DnBCS(1:2)=DnBCS(3);
DnBCS(251)=DnBCS(250);

E=E';

nBCS=real(abs(E)./(E.^2-d^2).^(1/2));

fold=1./(exp(E/kb/T)+1);
Diff_f=-diag(ones(NE-1,1),1)+diag(ones(NE-1,1),-1);
 Diff_f(1:2,1)=[1 ;1];
 Diff_f(NE-1:NE,NE)=[1 ;-1];
dfold=Diff_f*fold/dV;

[m,n]=size(dIdV);

epsilon=3.9e-14*ones(m,1);

chi2=((dIdV99-dIdV').^2)';
 
 while sum(chi2 > epsilon) > 0
     fprintf('q %i \n',sum(chi2 > epsilon));
     fnew=fold+5e0*(DnBCS.*chi2.^(1/2));
     dfnew=Diff_f*fnew/dV;
     g1=diag(ones(m-1,1),1);
     g2=diag(ones(m-1,1),-1);
     dfnew1=dfnew;
     dfnew2=dfnew;
     dfnew1=circshift(dfnew1,[0 1])+[dfnew(:,1), (zeros(n-1,m))'];
     k1=dfnew1;
     k2=dfnew2;
     for i=1:NE/2-1
         h1=g1*dfnew1;
         h1=h1+[ones(1,1)', zeros(NE-1,1)']';
         k1=[k1 h1];
         dfnew1=h1;
         h2=g2*dfnew2;
         k2=[h2 k2];
         dfnew2=h2;
     end
     dfnew_m=[k2 k1];
     q=1/Rt*Ns'*dfnew_m;
     g=dIdV(1)/q(1);
     q=1/Rt*Ns'*dfnew_m*g;
     chi2=((q-dIdV').^2)';
     fold=fnew;
 end

   
plot(E,fnew)

\end{verbatim}

\end{document}